\documentclass[11pt]{article}

\usepackage[margin=1in]{geometry}
\usepackage[T1]{fontenc}
\usepackage[utf8]{inputenc}
\usepackage{lmodern}
\usepackage{microtype}
\usepackage{graphicx}
\usepackage{booktabs}
\usepackage{longtable}
\usepackage{array}
\usepackage{multirow}
\usepackage{tabularx}
\usepackage{threeparttable}
\usepackage{caption}
\usepackage{subcaption}
\usepackage{float}
\usepackage{placeins}
\usepackage{amsmath}
\usepackage{amssymb}
\usepackage{siunitx}
\usepackage{enumitem}
\usepackage{natbib}
\usepackage[hidelinks]{hyperref}
\usepackage{xurl}
\usepackage{xspace}
\usepackage{authblk}
\usepackage{fancyhdr}
\usepackage{lastpage}
\usepackage{xcolor}
\usepackage{pdflscape}

\hypersetup{
  pdftitle={From differential expression to diagnostic transfer: a nine-cohort audit of transcriptomic reproducibility across three cancers},
  pdfauthor={Athanasios Angelakis},
  pdfsubject={Cross-cohort reproducibility and locked diagnostic transfer of differential-expression signatures},
  pdfkeywords={differential expression, reproducibility, diagnostic transfer, replication, gene ranking, cancer transcriptomics, microarray}
}

\setlist{nosep,leftmargin=*}
\newcommand{\pdac}{pancreatic ductal adenocarcinoma\xspace}

\newcommand{\logfc}{$\log_2$ fold change\xspace}
\newcommand{\abslogfc}{$|\log_2\mathrm{FC}|$\xspace}
\newcommand{\rede}{REDE\xspace}
\newcommand{\redetwofold}{REDE-2Fold\xspace}

\title{\textbf{REDE: A Quantitative Framework for Differential-Expression Reproducibility and Diagnostic Transfer Across Nine Cohorts and Three Cancers}}

\author[1,2]{Athanasios Angelakis\thanks{Correspondence: \href{mailto:ath.angelakis@gmail.com}{athanasios.angelakis@unibw.de}}}
\affil[1]{BioML Lab, Research Institute CODE, Universit\"at der Bundeswehr M\"unchen, Neubiberg, Germany}
\affil[2]{Epidemiology and Data Science, Amsterdam UMC, University of Amsterdam, Amsterdam, Netherlands}
\date{}

\begin{document}
\maketitle

\begin{abstract}
\textbf{Background:} Differential expression analysis provides valid evidence of association within an analyzed cohort, but its thresholded output is frequently presented as a transferable gene signature or diagnostic biomarker panel. We tested which layers of differential-expression evidence reproduce across independent cohorts and whether discovery-derived gene panels retain locked tumor-versus-non-tumor classification performance.

\textbf{Methods:} Nine public microarray cohorts were organized into fixed discovery, validation, and external-test experiments for pancreatic ductal adenocarcinoma, breast cancer, and lung cancer. Differential expression was estimated independently with the same patient-adjusted limma-trend model. Reproducibility was evaluated for DEG burden, exact membership, top-rank overlap, signed effects, prespecified gene confirmation, and Hallmark pathways. Each discovery cohort was also divided once into two patient-level folds; genes selected in both folds with concordant direction formed a \redetwofold set. For diagnostic transfer, four training-only feature sets were compared: all discovery DEGs, the 19 strongest discovery DEGs, all \redetwofold genes, and the 19 strongest \redetwofold genes. L2-regularized logistic models, scaling, hyperparameters, gene orientation, and decision thresholds were learned only from discovery data and transferred unchanged to validation and test.

\textbf{Results:} Broad DEG-list confirmation in both independent cohorts ranged from 15.5\% to 39.5\%, increasing to 50.1\% to 84.3\% for large effects. \redetwofold enriched confirmation relative to full discovery lists but performed similarly to matched-size full-cohort rankings. Pathway replication ranged from 52.2\% to 88.9\%. Compact 19-gene panels retained high ranking discrimination: validation and test ROC-AUCs were 0.875 and 0.897 for conventional DE and 0.835 and 0.869 for \redetwofold in pancreatic cancer; 0.850 and 0.989 versus 0.905 and 0.987 in breast cancer; and 0.999 and 0.996 for both strategies in lung cancer. However, locked operating points were substantially less stable. Breast all-gene panels had test ROC-AUCs of 0.989 to 0.994 but zero specificity, pancreatic compact panels had test ROC-AUCs of 0.869 to 0.897 but sensitivities of 0.067 to 0.200, and lung compact panels had validation ROC-AUC of 0.999 but specificity of 0.091.

\textbf{Conclusions:} Transcriptomic reproducibility is hierarchical from thresholded membership through effect, pathway, discrimination, and operating-point transfer. Strong external ROC-AUC does not establish a transferable diagnostic rule, and strong differential expression does not by itself establish biomarker utility. The expanded seven-level \rede framework separates these claims, while \redetwofold provides a minimum internal feature-stability procedure that strengthens but does not replace locked independent evaluation.
\end{abstract}


\section{Introduction}

Differential gene expression is one of the most common starting points for molecular discovery. A standard analysis estimates a condition-associated effect for every gene, corrects the resulting $P$ values for multiplicity, and reports the genes that cross an adjusted significance threshold. Within the analyzed cohort, this is a coherent inferential procedure. The scientific language attached to the output often goes further. A thresholded list may be presented as a disease signature, a biomarker set, a mechanistic shortlist, or a stable prioritization of targets even when no independent cohort has tested whether the same genes retain their membership, direction, magnitude, or rank.

We term this the \emph{single-cohort fallacy}: the inference that a statistically significant gene list from one cohort represents a stable, transferable molecular signature or biomarker panel without independent evidence of rank, direction, or pathway stability.

The distinction is fundamental because a DEG call is not a direct property of a gene. It is the result of a gene-specific effect estimate combined with variance, sample size, tissue composition, measurement platform, preprocessing, annotation, model specification, and a chosen decision boundary. Binary thresholding compresses these continuous quantities into a yes-or-no label. Two genes with nearly identical evidence can fall on opposite sides of the threshold, and the same underlying biological response can produce very different list lengths when cohort size or within-pair variability changes. In the present study, for example, the breast discovery cohort classified 5,796 genes as significant, whereas the breast validation cohort classified only 1,724 under the same statistical rule. Ranking adds another layer of sensitivity because the extreme top of a genome-wide list depends on small differences among many correlated test statistics.

Concerns about the repeatability and transportability of expression signatures are longstanding. Published microarray analyses have often been difficult to reproduce computationally, and prognostic gene lists can change markedly across resamples or cohorts even when predictive performance appears similar \citep{ioannidis2009repeatability,eindor2005outcome,eindor2006thousands,venet2011random}. Conversely, controlled inter-laboratory studies have shown that broad expression measurements and fold-change patterns can be reproducible across platforms \citep{shi2006maqc,larkin2005independence}. These observations are not contradictory. Here, \emph{repeatability} denotes obtaining the same result from the same data and computational workflow, whereas \emph{reproducibility} denotes stability across independent datasets addressing the same biological contrast. Reproducibility is therefore not a single property: expression values, signed effects, thresholded sets, top ranks, and biological programs are different estimands and need not transfer equally well.

The same distinction applies when DEGs are converted into diagnostic models. A model may preserve case-control ranking, reflected by ROC-AUC, while its predicted probabilities or fixed decision threshold shift across platforms and cohorts. Discrimination, calibration, sensitivity, specificity, and predictive values are therefore separate properties, and all model-building decisions must be protected from independent validation data \citep{pepe2003statistical,collins2015tripod,steyerberg2019clinical,wolff2019probast}. Internal cross-validation can estimate development performance, but only a locked independent evaluation tests transport of the complete diagnostic rule.

Most cross-cohort DEG comparisons focus on one disease, one pair of datasets, or one overlap statistic. Such designs cannot establish whether an observed pattern is disease-specific or reflects a more general property of transcriptome-wide inference. They also frequently conflate three questions: whether a gene is independently significant after a new genome-wide search, whether a prespecified discovery gene confirms under restricted multiplicity correction, and whether its direction and magnitude are sufficiently stable to support biological prioritization. A rigorous audit should separate these questions and should evaluate the full hierarchy from binary calls to pathway-level organization.

We therefore analyzed nine independent tumor and matched non-tumor microarray cohorts spanning three biologically distinct settings: \pdac, breast cancer, and lung cancer. Within each cancer, one cohort was fixed as discovery, one as validation, and one as external test. The paired contrast, statistical model, significance definitions, rank metric, confirmation rules, and enrichment procedure were held constant. Cohorts were never pooled, and no cross-cohort batch correction was used. This design produced three parallel replication experiments and nine within-cancer cohort comparisons across 441 patients.

Our objective was not to discredit differential expression. Instead, we asked what level of claim is justified by a single-cohort result and how the answer changes when independent evidence is evaluated at multiple layers. We quantified variation in DEG burden, exact list overlap, top-k rank overlap, genome-wide effect correlation, effect-direction concordance, confirmatory replication of prespecified discovery genes, genome-wide replication, and Hallmark pathway transfer. We additionally introduced \redetwofold, a bio-machine-learning adaptation of two-fold cross-validation in which the discovery cohort is divided once, differential expression is performed independently in both folds, and only same-direction genes selected in both folds are retained. Finally, we tested the downstream claim most often implied by the term biomarker: whether all-gene and compact 19-gene panels selected exclusively in discovery retain discrimination and a locked classification threshold in two independent cohorts. The resulting synthesis defines a hierarchy from within-cohort association to diagnostic transfer and motivates \rede, a practical seven-level framework for matching molecular claims to the evidence actually evaluated.

\section{Results}

\subsection{Nine scale-audited paired cohorts formed three independent replication experiments}

The differential-expression reproducibility study included 882 expression profiles from 441 patients with complete paired tumor and non-tumor tissue. The PDAC experiment comprised 36 pairs from GSE15471, 16 pairs from GSE16515, and 45 pairs from GSE28735. The breast-cancer experiment comprised 148 pairs from GSE70947, 43 pairs from GSE15852, and 25 pairs from GSE109169. The lung-cancer experiment comprised 57 pairs from GSE32863, 44 pairs from GSE18842, and 27 pairs from GSE7670. Within each cancer, the gene space was fixed to the intersection available across its three platforms: 18,034 genes for PDAC, 7,701 for breast cancer, and 10,549 for lung cancer (Table~\ref{tab:cohorts}).

The locked tissue-classification extension used all labeled samples available in the processed matrices, totaling 902 profiles from 461 patients. Eight cohorts contained the same complete pairs used for differential expression. GSE16515 additionally contained 20 tumor-only patients, so its diagnostic evaluation included 36 tumor and 16 non-tumor samples rather than only the 16 complete pairs. These additional samples contributed only to locked validation and did not enter gene selection, model tuning, or threshold determination.

The preprocessing audit showed that six cohorts already had plausible log2-scale submitted values and were preserved. GSE16515, GSE15852, and GSE7670 contained nonnegative intensity-scale values and were transformed with $\log_2(x+1)$. Probe identifiers were mapped to platform gene symbols, multiple probes mapping to one symbol were aggregated by their median, and technical replicates in GSE15471 were averaged within patient and tissue after scale correction. The final matrices retained their submitted within-cohort normalization. No cross-cohort scaling, z-score standardization, batch correction, missing-value imputation, or pooled modeling was performed.

\begin{table}[H]
\centering
\caption{Cohorts, preprocessing, and differential-expression burden.}
\label{tab:cohorts}
\scriptsize
\resizebox{0.98\textwidth}{!}{%
\begin{tabular}{lllrrrrllrr}
\toprule
Cancer & Role & GEO & Pairs & Samples & Genes & Platform & Scale action & FDR DEGs, n (\%) & Large-effect DEGs, n (\%) \\
\midrule
PDAC & Discovery & GSE15471 & 36 & 72 & 18,034 & GPL570 & Preserve log2 & 13,038 (72.3) & 952 (5.3) \\
PDAC & Validation & GSE16515 & 16 & 32 & 18,034 & GPL570 & $\log_2(x+1)$ & 4,590 (25.5) & 971 (5.4) \\
PDAC & External test & GSE28735 & 45 & 90 & 18,034 & GPL6244 & Preserve log2 & 8,046 (44.6) & 390 (2.2) \\
\addlinespace
Breast & Discovery & GSE70947 & 148 & 296 & 7,701 & GPL13607 & Preserve log2 & 5,796 (75.3) & 819 (10.6) \\
Breast & Validation & GSE15852 & 43 & 86 & 7,701 & GPL96 & $\log_2(x+1)$ & 1,724 (22.4) & 161 (2.1) \\
Breast & External test & GSE109169 & 25 & 50 & 7,701 & GPL5175 & Preserve log2 & 4,005 (52.0) & 489 (6.3) \\
\addlinespace
Lung & Discovery & GSE32863 & 57 & 114 & 10,549 & GPL6884 & Preserve log2 & 6,394 (60.6) & 681 (6.5) \\
Lung & Validation & GSE18842 & 44 & 88 & 10,549 & GPL570 & Preserve log2 & 6,868 (65.1) & 1,424 (13.5) \\
Lung & External test & GSE7670 & 27 & 54 & 10,549 & GPL96 & $\log_2(x+1)$ & 3,835 (36.4) & 1,018 (9.7) \\
\bottomrule
\end{tabular}%
}
\parbox{0.98\textwidth}{\footnotesize \vspace{0.3em}FDR DEGs were defined by Benjamini-Hochberg adjusted $P<0.05$. Large-effect DEGs additionally required \abslogfc at least 1. Percentages use the cancer-specific common-gene space. Only complete matched pairs entered the primary analyses.}
\end{table}

\subsection{Identical statistical rules produced markedly different DEG burdens}

\begin{figure}[H]
\centering
\includegraphics[width=0.98\textwidth]{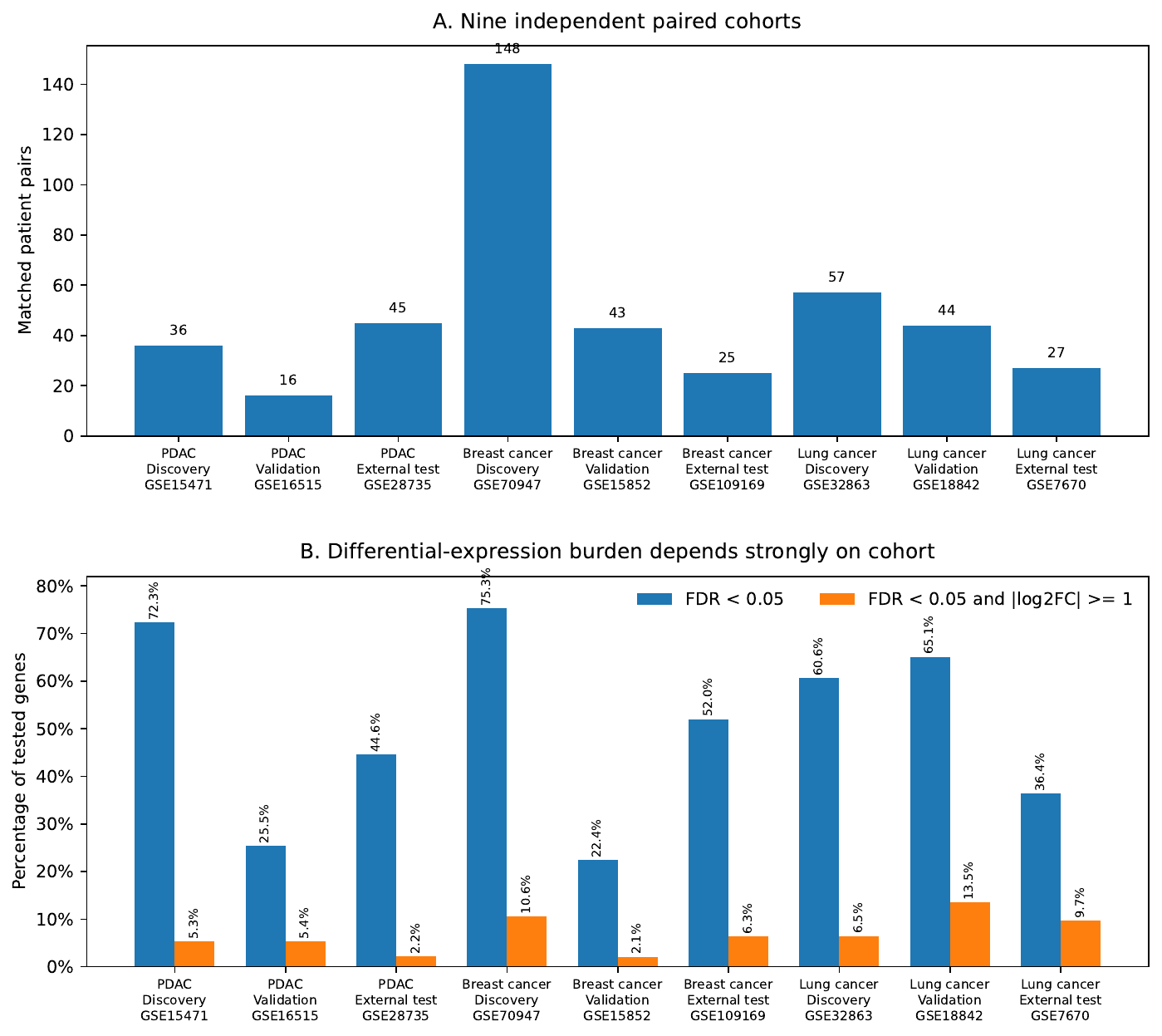}
\caption{\textbf{Study design and differential-expression burden across nine paired cohorts.} (A) Number of complete matched patient pairs in each fixed discovery, validation, and external-test cohort. (B) Percentage of the cancer-specific common-gene space meeting the primary FDR definition and the secondary large-effect definition. Identical paired analysis rules produced broad FDR lists covering 22.4\% to 75.3\% of the tested transcriptome.}
\label{fig:burden}
\end{figure}

Despite using the same paired design and FDR threshold, the fraction of the tested transcriptome classified as differentially expressed varied from 22.4\% to 75.3\% (Figure~\ref{fig:burden}). The smallest broad list was the breast validation cohort, with 1,724 genes, whereas the largest was the PDAC discovery cohort, with 13,038 genes. This 3.4-fold range in transcriptome proportion occurred even though each experiment compared tumor with patient-matched non-tumor tissue. A researcher selecting the breast validation cohort would conclude that approximately one fifth of the tested transcriptome was altered, whereas one selecting the breast discovery cohort would conclude that three quarters was altered. Both statements accurately describe their analyzed cohort, but neither is automatically a disease-level property.

The addition of a descriptive magnitude criterion, \abslogfc at least 1, reduced every list but did not make list size uniform. Large-effect calls ranged from 2.1\% to 13.5\% of the tested gene space. In PDAC, discovery and validation had nearly identical large-effect counts, 952 and 971, despite a 2.8-fold difference in their broad FDR counts. In breast cancer, the 148-pair discovery cohort identified 819 large effects, whereas validation identified 161. In lung cancer, the largest large-effect list arose in validation rather than discovery. Thus, FDR-list length was not a direct proxy for the number of large biological effects, and an effect filter did not erase cohort dependence.

\subsection{Exact FDR-list agreement was moderate and disease dependent}

Pairwise Jaccard indices for broad FDR lists ranged from 0.229 to 0.547 across the nine within-cancer cohort comparisons (Figure~\ref{fig:hierarchy}; Supplementary Table~\ref{tab:s_pairwise}). PDAC values were 0.252, 0.403, and 0.357; breast-cancer values were 0.249, 0.505, and 0.229; and lung-cancer values were 0.547, 0.425, and 0.419. The median across all comparisons was 0.403. Lung cancer therefore showed the strongest overall exact-list agreement, but even its highest value implied that almost half of the genes in the union were not shared.

Three-way membership reinforced the cohort dependence of binary calls. The number of genes significant in all three cohorts was 2,631 in PDAC, 993 in breast cancer, and 2,591 in lung cancer. These counts represented 20.2\%, 17.1\%, and 40.5\% of the respective discovery FDR lists. At the same time, substantial cohort-specific membership remained. In PDAC, 6,068 discovery DEGs were unique to discovery. In breast cancer, 2,002 genes were unique to discovery and 643 were unique to the external test. In lung cancer, 1,246 were unique to discovery and 1,606 were unique to validation. A gene could therefore have a reproducible directional effect without crossing the same significance boundary in all three datasets.

The overlap coefficient was consistently higher than the Jaccard index because a large fraction of a smaller list often appeared within a larger list. This distinction prevents a misleading interpretation of asymmetric recovery: a small cohort can recover most of its discoveries inside a larger list while the larger list still contains thousands of additional calls. Exact list reproducibility therefore cannot be summarized by one overlap percentage without specifying the denominator.

\subsection{Top-ranked genes were the least stable layer of evidence}

\begin{figure}[H]
\centering
\includegraphics[width=\textwidth]{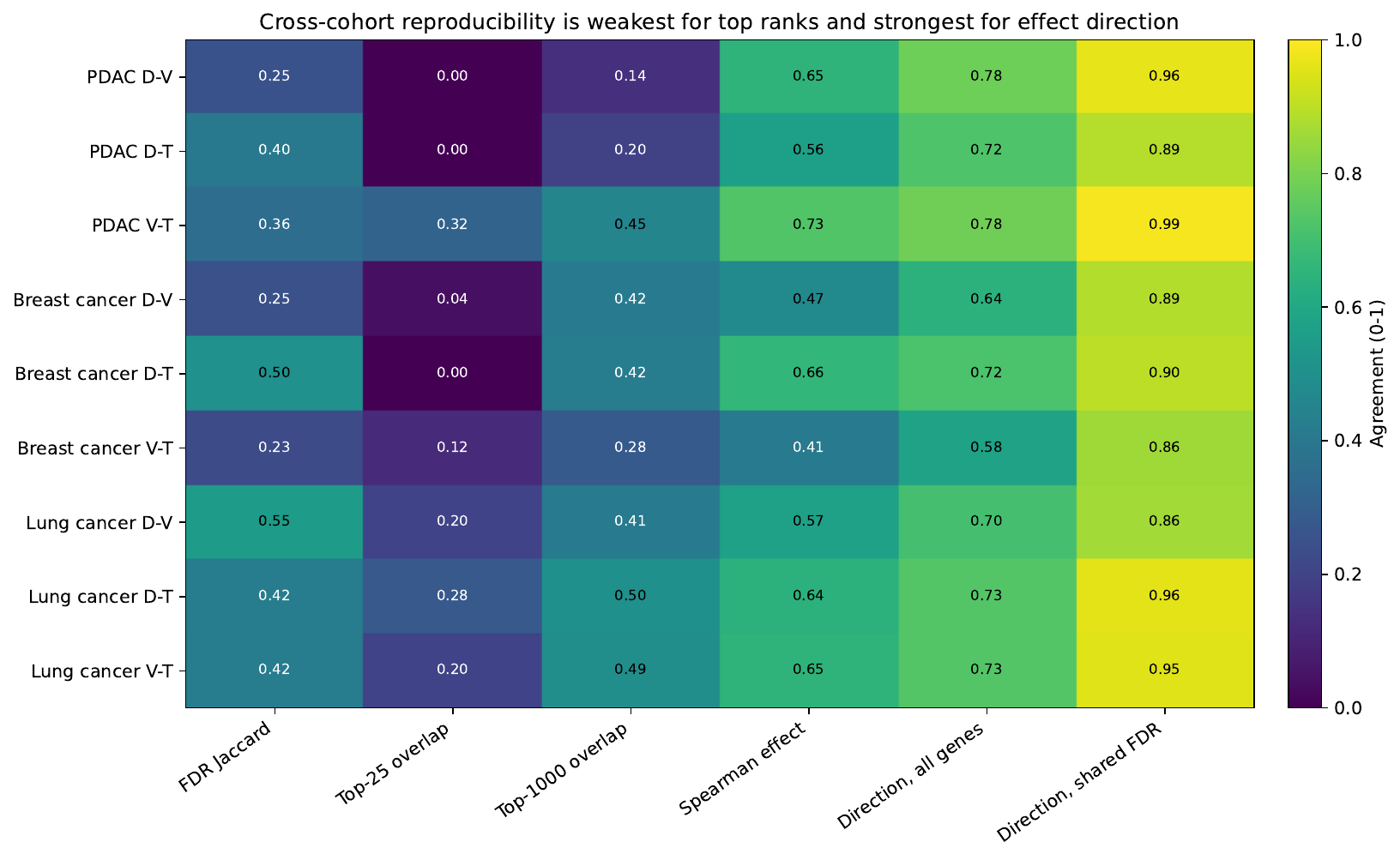}
\caption{\textbf{A reproducibility hierarchy across three cancers.} Each row is a within-cancer cohort comparison: discovery versus validation, discovery versus external test, or validation versus external test. Columns show exact FDR-list Jaccard agreement, overlap of the top 25 and top 1,000 genes ranked by absolute moderated t-statistic, Spearman correlation of genome-wide \logfc values, direction concordance across all genes, and direction concordance among genes significant in both cohorts. Extreme ranks were generally less stable than signed effects and shared-DEG direction.}
\label{fig:hierarchy}
\end{figure}

\begin{figure}[H]
\centering
\includegraphics[width=\textwidth]{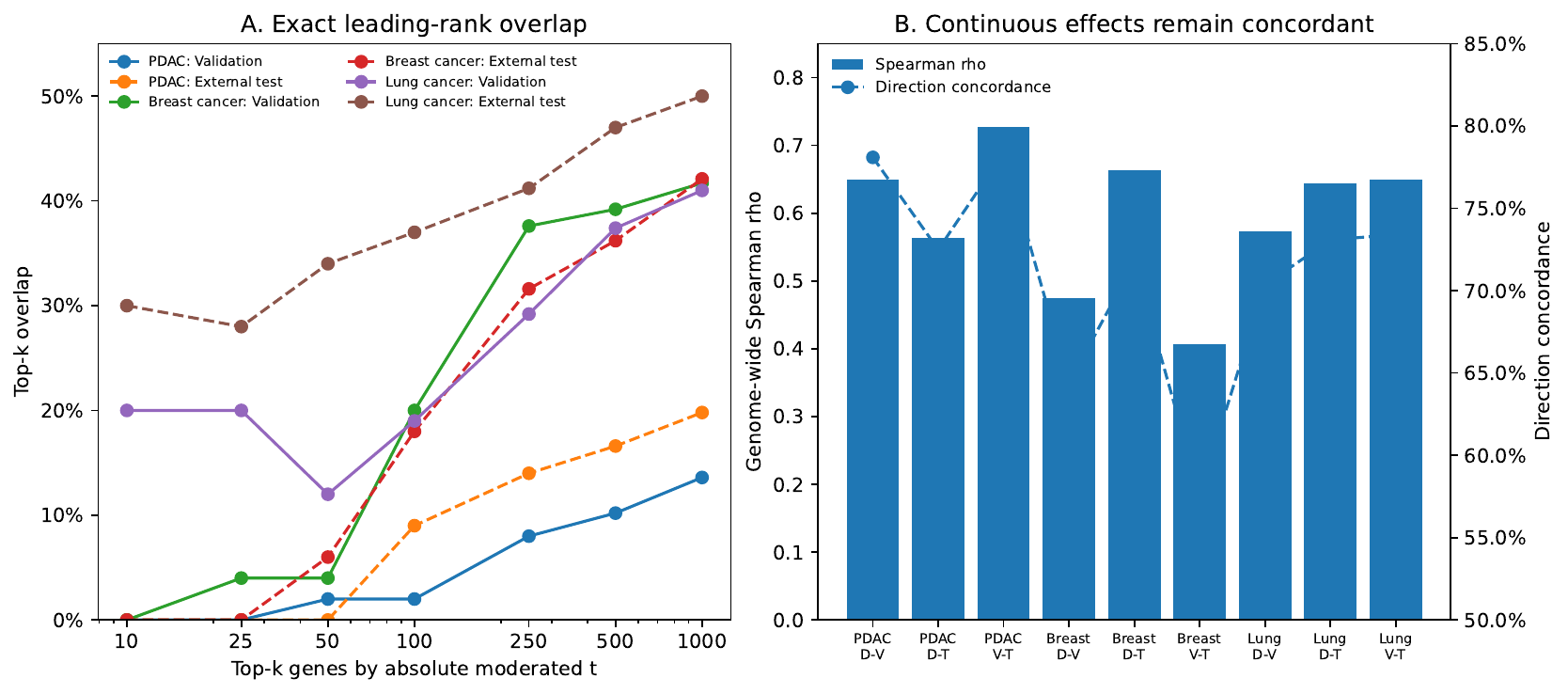}
\caption{\textbf{Rank and continuous-effect reproducibility.} (A) Shared fraction among the top $k$ genes ranked by absolute moderated t-statistic for each discovery-to-independent comparison. (B) Genome-wide Spearman correlation of $\log_2$ fold changes and all-gene effect-direction concordance for all nine within-cancer cohort pairs. D, discovery; V, validation; T, external test.}
\label{fig:rankeffects}
\end{figure}

Rank overlap was substantially weaker than shared effect direction and was highly heterogeneous across diseases (Figures~\ref{fig:hierarchy} and~\ref{fig:rankeffects}). For discovery versus validation or external test, the fraction shared among the top 25 genes ranged from 0\% to 28\%. Even in the most favorable comparison, fewer than one third of the top 25 genes were shared, and several comparisons had no overlap. PDAC discovery shared none of its top 25 genes with either independent cohort. Breast discovery shared one gene with validation and none with external test. Lung discovery showed higher but still incomplete agreement, sharing 20\% with validation and 28\% with external test. This absence of stable top-ranked candidates is consequential because functional follow-up commonly begins with precisely these extreme ranks.

Overlap increased as the list expanded, but the top rankings remained non-interchangeable. At $k=1{,}000$, discovery shared 13.6\% and 19.8\% with the two PDAC independent cohorts, 41.7\% and 42.1\% with the two breast cohorts, and 41.0\% and 50.0\% with the two lung cohorts. The cross-cancer range was therefore 13.6\% to 50.0\%. These overlaps were above the random expectation determined by the cancer-specific gene-space size, but enrichment above chance does not imply that the selected top genes are equivalent or that a short mechanistic shortlist would transfer.

The leading discovery genes also emphasized different biological compartments by cancer. The PDAC ranking was dominated by extracellular-matrix and stromal-associated genes such as \textit{COL10A1}, \textit{SULF1}, \textit{INHBA}, \textit{THBS2}, and \textit{FN1}. The breast discovery ranking included proliferation and cell-cycle genes such as \textit{EZH2}, \textit{RAD54L}, \textit{KIAA0101}, \textit{BUB1B}, and \textit{TYMS}. The lung discovery ranking was led by genes including \textit{FABP4}, \textit{FMO2}, \textit{FAM107A}, \textit{CLEC3B}, and \textit{CAV1}. These rankings were biologically interpretable, but the cross-cohort results show that interpretability within one cohort is not evidence of rank stability.

\subsection{Signed effects were more transferable than binary membership or extreme rank}

Across all genes, Spearman correlations of cohort-specific \logfc values ranged from 0.407 to 0.727, with a median of 0.644. Genome-wide direction concordance ranged from 57.6\% to 78.4\%. The lowest values occurred between the breast validation and external-test cohorts, whereas the highest effect correlation occurred between PDAC validation and external test. These values demonstrate neither perfect agreement nor random behavior. Instead, the global tumor-associated expression field was partly conserved while individual estimates shifted in magnitude and rank.

Agreement strengthened when attention was restricted to genes carrying stronger evidence. Among genes significant in both members of a cohort pair, direction concordance ranged from 85.6\% to 98.7\%, with a median of 90.2\%. The corresponding ranges were 88.8\% to 98.7\% for PDAC, 85.6\% to 90.2\% for breast cancer, and 86.2\% to 95.9\% for lung cancer. Among genes meeting the large-effect definition in both cohorts, direction concordance approached unity in PDAC and remained very high in the other cancers. The principal instability was therefore often whether a gene crossed a threshold or occupied an extreme rank, not whether independently detected effects pointed in opposite directions.

This separation is central to interpretation. A moderate Jaccard index can coexist with strong directional agreement because thousands of genes near the list boundary preserve their sign but differ in standard error or adjusted $P$ value. Conversely, a high direction-concordance value among shared DEGs is conditional on both cohorts selecting the gene and should not be mistaken for high recall of the entire discovery list. The metrics answer different scientific questions and need to be reported together.

\subsection{Prespecified confirmation exposed a strong effect-magnitude gradient}

\begin{figure}[H]
\centering
\includegraphics[width=0.96\textwidth]{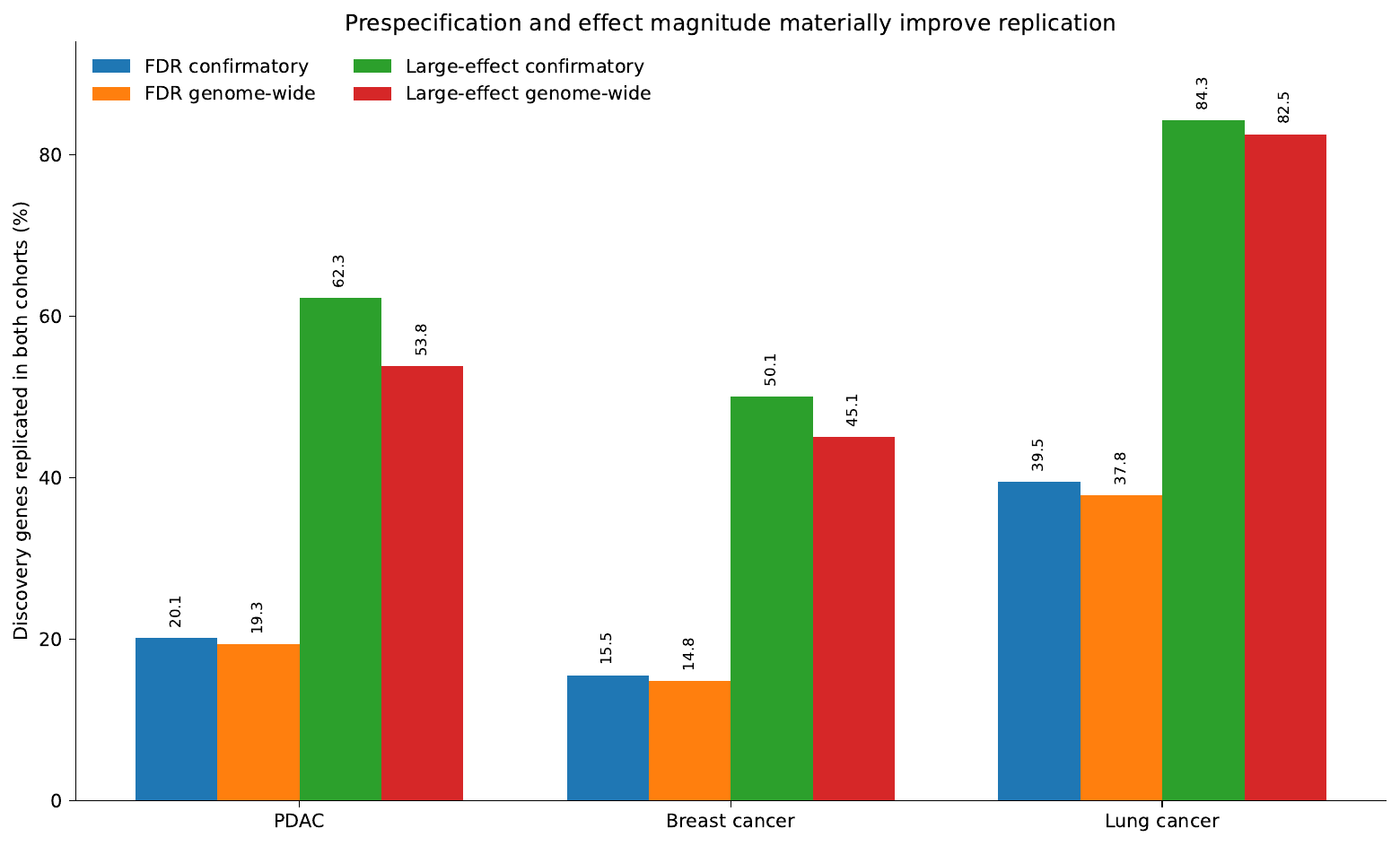}
\caption{\textbf{Independent replication of discovery-defined gene families.} Bars show the percentage of discovery genes replicated in both validation and external test. Confirmatory replication applied FDR correction across the prespecified discovery family in each independent cohort and required concordant direction. Genome-wide replication required each gene to pass the independent cohort's full genome-wide FDR threshold with concordant direction. Restricting discovery to large effects increased replication in all three cancers.}
\label{fig:replication}
\end{figure}

We next treated each discovery list as a prespecified hypothesis family and applied Benjamini-Hochberg correction only across those genes in each independent cohort, requiring the same effect direction as discovery. Under this confirmatory definition, 20.1\% of PDAC discovery FDR genes, 15.5\% of breast discovery FDR genes, and 39.5\% of lung discovery FDR genes confirmed in both validation and external test (Figure~\ref{fig:replication}). The stricter requirement that each gene also pass the independent cohort's full genome-wide FDR threshold yielded similar but slightly lower values of 19.3\%, 14.8\%, and 37.8\%.

Replication rose sharply when the discovery family was restricted to genes with \abslogfc at least 1. Confirmatory replication in both independent cohorts reached 62.3\% in PDAC, 50.1\% in breast cancer, and 84.3\% in lung cancer. Under the genome-wide independent threshold, the corresponding values were 53.8\%, 45.1\%, and 82.5\%. This pattern was consistent in all three diseases and under both confirmation definitions. Large discovery effects were therefore substantially more transferable than the broad FDR list, although they were not universally reproducible.

The confirmatory and genome-wide results should not be conflated with exact three-way intersection. Confirmatory testing asks whether a declared discovery family receives independent evidence after multiplicity correction appropriate to that family. Genome-wide replication asks whether the same gene survives a new transcriptome-wide search. Exact large-effect intersection additionally requires the effect to exceed the magnitude threshold in every cohort. These are increasingly restrictive and scientifically distinct estimands.

\subsection{REDE-2Fold enriched external confirmation but matched full-cohort evidence ranking}

We next evaluated a practical minimum for settings in which only one transcriptomic dataset is available. \redetwofold applies a technique directly analogous to running a machine-learning model under two-fold cross-validation, but transfers the validation target from predictive performance to differential-expression feature selection. The discovery patients were randomly divided once into two approximately equal and non-overlapping folds using fixed seed 3, while each matched tumor and normal or non-tumor pair remained together. No repeated resampling was performed. The two folds contained 18 and 18 PDAC pairs, 74 and 74 breast-cancer pairs, and 29 and 28 lung-cancer pairs.

Independent paired analyses identified 12,178 and 6,890 FDR genes in the two PDAC folds, 5,135 and 4,963 in the breast folds, and 5,099 and 5,092 in the lung folds. Requiring FDR below 0.05 in both folds and concordant direction retained 6,257 PDAC genes, 4,511 breast genes, and 4,238 lung genes. The corresponding between-fold Jaccard indices were 0.489, 0.807, and 0.712. Direction concordance among genes significant in both folds exceeded 99.9\% in every cancer, confirming that the filter primarily removed genes whose statistical support, rather than effect sign, depended on the patient subset.

The frozen \redetwofold sets confirmed in both independent cohorts at 25.9\% for PDAC, 19.8\% for breast cancer, and 50.2\% for lung cancer, compared with 20.1\%, 15.5\%, and 39.5\% for the complete discovery FDR lists. The absolute gains were therefore 5.7, 4.3, and 10.7 percentage points, respectively. However, full-discovery lists restricted to the same number of genes by absolute moderated t-statistic confirmed at 25.9\%, 19.9\%, and 50.5\%, and matched-size absolute-\logfc lists confirmed at 33.3\%, 19.9\%, and 49.4\%. Thus, the cross-validated intersection consistently enriched external confirmation relative to an unfiltered discovery list, but did not provide a material advantage over selecting an equally sized set with the strongest full-cohort evidence.

Effect magnitude remained complementary to internal cross-validation. Restricting the same-direction fold intersection to genes with \abslogfc at least 1 in both folds produced 536 PDAC, 719 breast, and 580 lung genes. These sets confirmed in both independent cohorts at 65.7\%, 51.9\%, and 86.0\%, modestly above the 62.3\%, 50.1\%, and 84.3\% obtained from the full-discovery large-effect lists. \redetwofold therefore provides an explicit internal stability certificate and a practical bio-machine-learning filter when no external cohort is initially available, but its output must still be distinguished from external replication.

\begin{figure}[H]
\centering
\includegraphics[width=\textwidth]{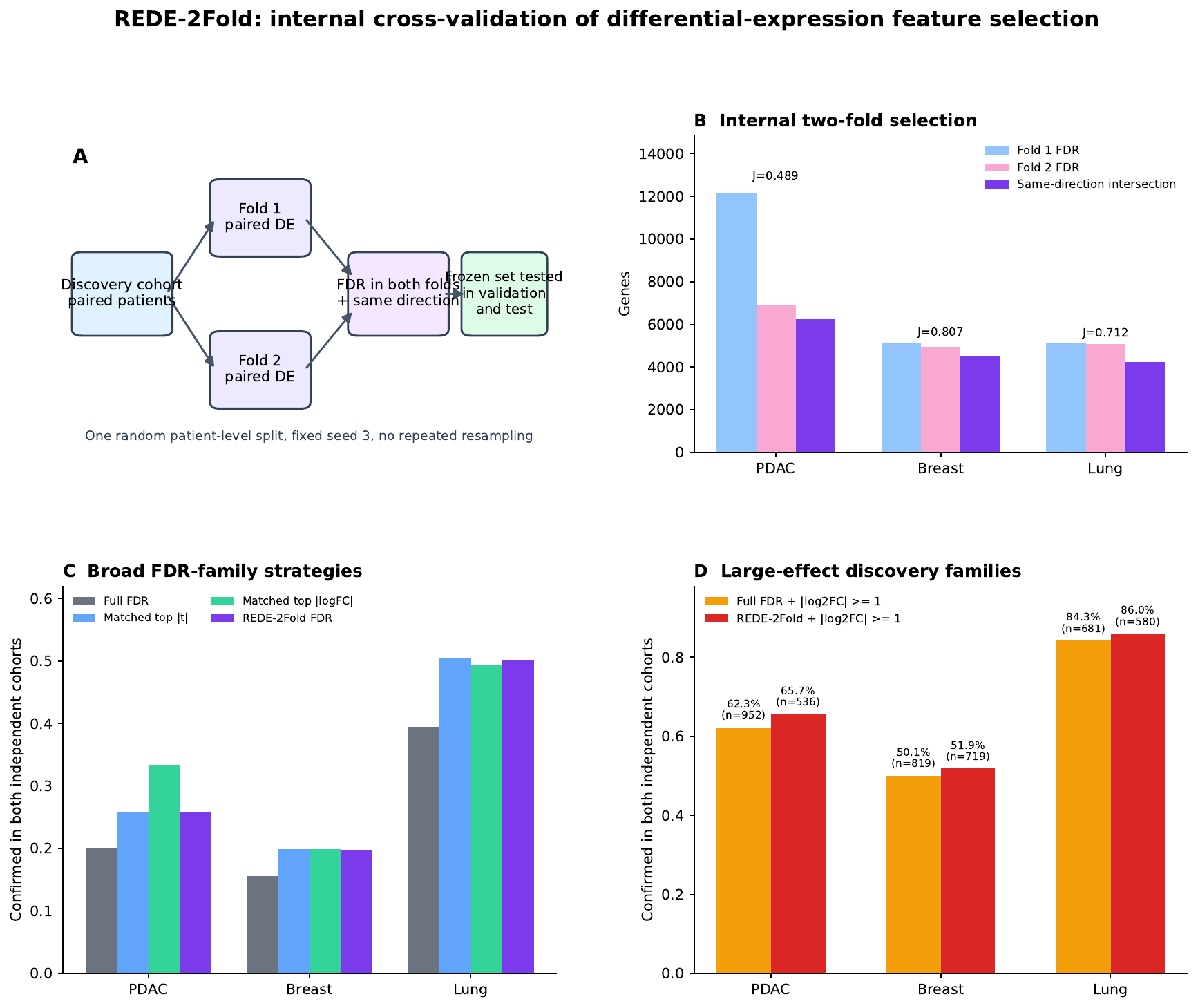}
\caption{\textbf{REDE-2Fold cross-validation of differential-expression feature selection.} (A) The discovery cohort was divided once at patient level, paired differential expression was conducted independently in each fold, and same-direction FDR genes were frozen before external evaluation. (B) Fold-specific FDR counts, same-direction intersections, and FDR-set Jaccard indices. (C) Confirmation in both independent cohorts for the full discovery FDR list, size-matched full-cohort rankings, and the \redetwofold FDR intersection. The cross-validated set improved over the full FDR list but was similar to matched-size controls. (D) Large-effect filtering remained strongly associated with transferability and produced a modest additional gain when combined with \redetwofold.}
\label{fig:rede2fold}
\end{figure}

\subsection{Pathway replication exceeded broad gene-level confirmation in every cancer}

Hallmark preranked gene set enrichment analysis provided the most consistent compression of the disease-associated transcriptome. PDAC discovery contained 36 significant Hallmark programs, of which 32 replicated in both independent cohorts with concordant normalized-enrichment-score direction, a rate of 88.9\%. Breast discovery contained 23 significant programs, of which 12 replicated in both, a rate of 52.2\%. Lung discovery contained 32 significant programs, of which 22 replicated in both, a rate of 68.8\%. In each cancer, pathway replication exceeded confirmatory replication of the broad discovery FDR list.

\begin{figure}[H]
\centering
\includegraphics[width=\textwidth]{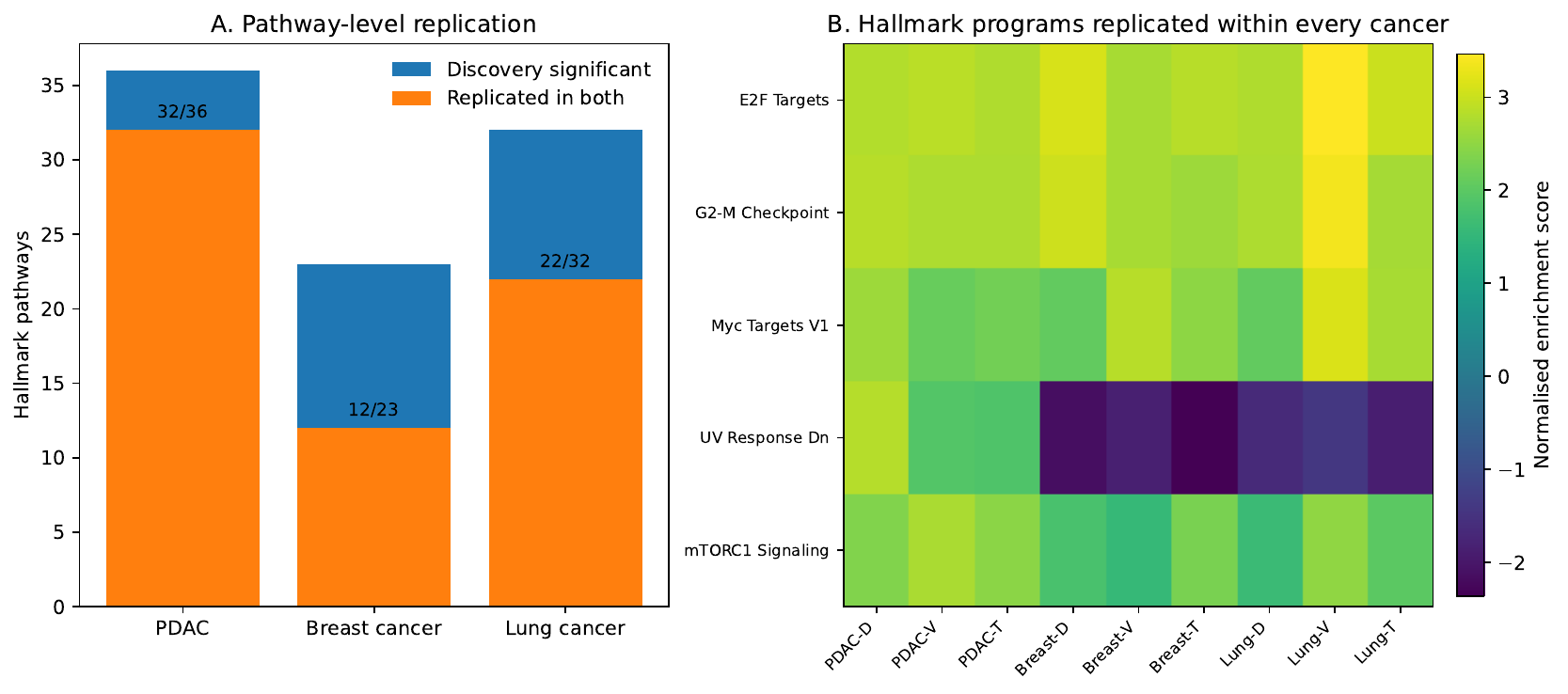}
\caption{\textbf{Pathway-level reproducibility.} (A) Discovery-significant Hallmark pathways and the subset significant with concordant direction in both independent cohorts. (B) Normalized enrichment scores for the five Hallmark programs that replicated within all three cohorts of every cancer. D, discovery; V, validation; T, external test. Disease-specific pathway heatmaps are provided in Supplementary Figure~\ref{fig:s_gsea}.}
\label{fig:pathways}
\end{figure}

Five Hallmark programs replicated within all three cohorts of all three cancers: E2F targets, G2-M Checkpoint, Myc Targets V1, UV Response Dn, and mTORC1 Signaling (Figure~\ref{fig:pathways}). E2F targets, G2-M Checkpoint, Myc Targets V1, and mTORC1 Signaling were positively enriched in every cohort, indicating a common proliferative and growth-associated axis. UV Response Dn was also reproducible within every cancer, but its enrichment direction differed: positive in PDAC and negative in breast and lung cancer. Replication of a pathway name therefore does not guarantee a universal direction across diseases.

\begin{center}
\fbox{\begin{minipage}{0.92\textwidth}
\textbf{A reproducible core of cancer transcriptomics.} E2F Targets, G2-M Checkpoint, Myc Targets V1, UV Response Dn, and mTORC1 Signaling were significant with internally concordant direction in all three cohorts of each cancer. Together, they define a core proliferative and growth-associated transcriptomic axis that remained detectable across nine independent cohorts, while the disease-specific direction of UV Response Dn illustrates why pathway names and enrichment directions must both be reported.
\end{minipage}}
\end{center}

Other programs further demonstrated cancer-specific organization. Interferon-alpha and interferon-gamma responses were positive across PDAC and breast cohorts but negative across lung cohorts. TNF-alpha signaling via NF-kB and KRAS signaling up were positive in PDAC but negative in breast and lung cancer. Glycolysis was broadly positive, with the exception of a weak negative estimate in the breast validation cohort. The stable object was therefore not one pan-cancer gene list. It was a mixture of shared core programs and disease-specific directional states.

\subsection{Locked diagnostic transfer separated discrimination from operating-point stability}

We next tested whether training-derived DEGs functioned as transferable tumor-versus-non-tumor tissue markers. Four feature strategies were frozen within each discovery cohort: the complete discovery FDR list, the top 19 discovery DEGs ranked by absolute moderated t-statistic, the complete \redetwofold intersection, and the top 19 \redetwofold genes ranked by the weaker fold-specific absolute t-statistic. The compact conventional and \redetwofold panels shared 13 of 19 genes in PDAC, 13 of 19 in breast cancer, and all 19 in lung cancer. Validation and test data contributed to neither gene identity nor model development.

\begin{figure}[H]
\centering
\includegraphics[width=\textwidth]{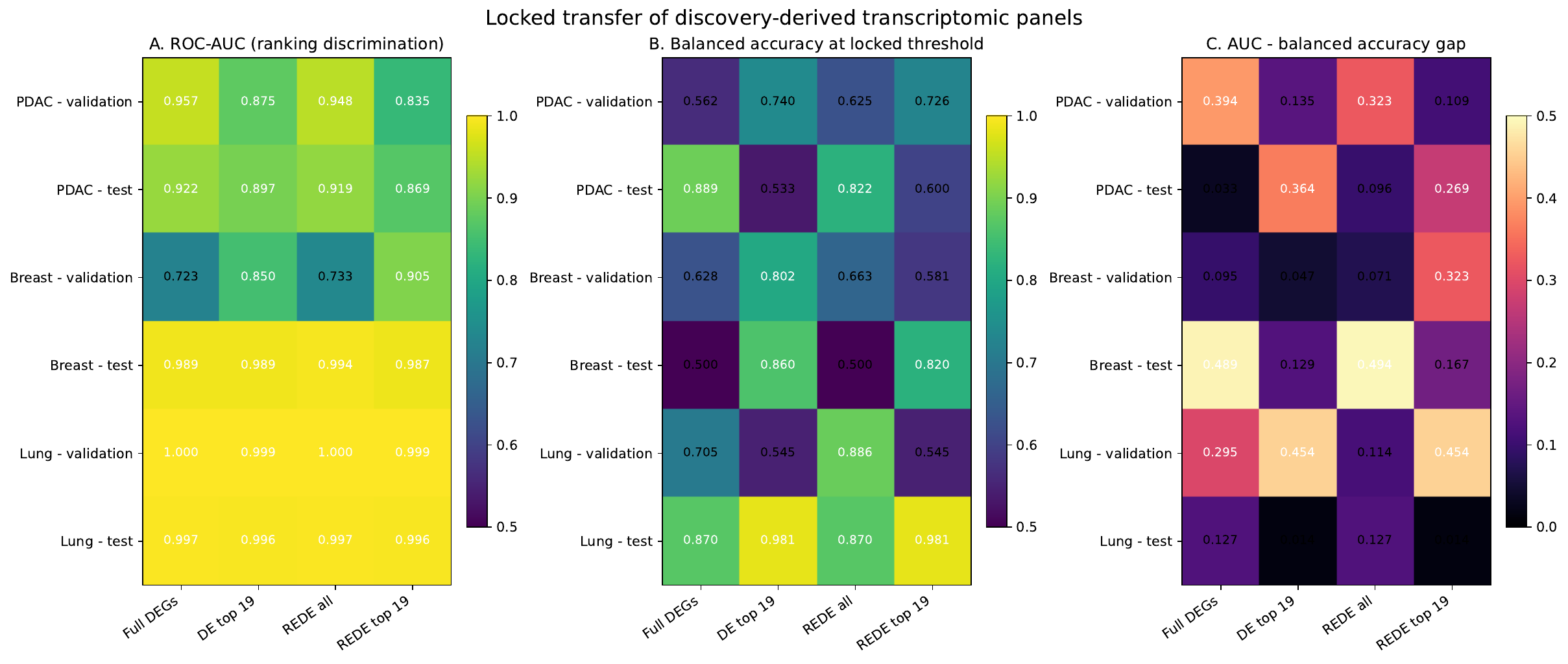}
\caption{\textbf{Locked transfer of discovery-derived transcriptomic panels.} (A) ROC-AUC measures ranking discrimination after the complete discovery-derived model was applied unchanged. (B) Balanced accuracy uses the probability threshold selected from grouped out-of-fold discovery predictions. (C) The difference between ROC-AUC and balanced accuracy exposes panels that preserve score ordering but fail at the locked operating point. Full DEGs, all discovery-cohort FDR genes; DE top 19, the strongest discovery DEGs by absolute moderated t-statistic; REDE all, the complete same-direction two-fold intersection; REDE top 19, genes ranked by the minimum fold-specific absolute t-statistic.}
\label{fig:diagnostic_transfer}
\end{figure}

The compact panels retained substantial external ranking discrimination (Figure~\ref{fig:diagnostic_transfer}; Table~\ref{tab:diagnostic_summary}). In PDAC, conventional and \redetwofold top-19 panels achieved validation ROC-AUCs of 0.875 and 0.835 and test ROC-AUCs of 0.897 and 0.869. In breast cancer, the corresponding values were 0.850 and 0.905 in validation and 0.989 and 0.987 in test. In lung cancer, the two compact strategies contained the same genes and achieved ROC-AUC 0.999 in validation and 0.996 in test. Compact panels therefore preserved useful cross-cohort ranking information despite containing fewer than 20 genes, but neither conventional DE ranking nor \redetwofold ranking was uniformly superior.

Locked classification thresholds revealed a different pattern. The PDAC compact panels had test sensitivities of only 0.067 and 0.200 despite ROC-AUCs near 0.9. The breast full-DEG and full-\redetwofold models had test ROC-AUCs of 0.989 and 0.994 but classified every test sample as tumor, producing specificity 0 and balanced accuracy 0.500. Conversely, the lung compact panels had validation ROC-AUC 0.999 but specificity 0.091 at the discovery-derived threshold; the same locked panels reached sensitivity 0.963, specificity 1.000, and balanced accuracy 0.981 in the external test. Across compact panels, balanced accuracy ranged from 0.545 to 0.802 in validation and from 0.533 to 0.981 in test. Thus, cross-cohort preservation of score ordering did not ensure transfer of the operating threshold.

Every discovery DEG was also evaluated as an individual marker using its discovery effect direction and a discovery-derived threshold. Leading DE-ranked genes showed the same distinction. PDAC COL10A1 retained validation and test ROC-AUCs of 0.880 and 0.882 but had zero test sensitivity at its locked threshold. Breast EZH2 achieved ROC-AUCs of 0.822 and 0.981, whereas lung FABP4 achieved 0.966 and 0.995 but test sensitivity was 0.519. Complete individual-gene results are provided as machine-readable supplementary data. Strong differential-expression evidence and strong ranking discrimination therefore did not guarantee a portable binary diagnostic rule.

\begin{table}[H]
\centering
\caption{External discrimination and locked-threshold balanced accuracy of discovery-derived panels.}
\label{tab:diagnostic_summary}
\scriptsize
\resizebox{0.98\textwidth}{!}{%
\begin{tabular}{llrrrrr}
\toprule
Cancer & Discovery-derived panel & Genes & Validation AUC & Validation balanced accuracy & Test AUC & Test balanced accuracy \\
\midrule
PDAC & Full discovery DEGs & 13,038 & 0.957 & 0.562 & 0.922 & 0.889 \\
PDAC & Discovery DE top 19 & 19 & 0.875 & 0.740 & 0.897 & 0.533 \\
PDAC & All REDE-2Fold genes & 6,257 & 0.948 & 0.625 & 0.919 & 0.822 \\
PDAC & REDE-2Fold top 19 & 19 & 0.835 & 0.726 & 0.869 & 0.600 \\
Breast & Full discovery DEGs & 5,796 & 0.723 & 0.628 & 0.989 & 0.500 \\
Breast & Discovery DE top 19 & 19 & 0.850 & 0.802 & 0.989 & 0.860 \\
Breast & All REDE-2Fold genes & 4,511 & 0.733 & 0.663 & 0.994 & 0.500 \\
Breast & REDE-2Fold top 19 & 19 & 0.905 & 0.581 & 0.987 & 0.820 \\
Lung & Full discovery DEGs & 6,394 & 1.000 & 0.705 & 0.997 & 0.870 \\
Lung & Discovery DE top 19 & 19 & 0.999 & 0.545 & 0.996 & 0.981 \\
Lung & All REDE-2Fold genes & 4,238 & 1.000 & 0.886 & 0.997 & 0.870 \\
Lung & REDE-2Fold top 19 & 19 & 0.999 & 0.545 & 0.996 & 0.981 \\
\bottomrule
\end{tabular}%
}
\end{table}

\subsection{The same hierarchy emerged despite disease-specific differences}

The three cancer experiments differed meaningfully. Lung cancer showed the strongest exact DEG overlap, the highest confirmation of large discovery effects, and near-perfect tissue-discrimination AUC. Breast cancer showed the lowest broad gene confirmation and the largest contrast between high test AUC and failed all-gene operating thresholds. PDAC showed particularly unstable discovery top ranks, high pathway replication, and a marked loss of compact-panel sensitivity in the external test. These differences likely reflect cohort size, tissue composition, platform, variance structure, expression-scale transport, and disease biology.

Nevertheless, the ordering of inferential stability was consistent. Broad list length varied substantially. Exact membership was moderate. Extreme top ranks were fragile. Signed effects were more coherent than ranks, and large discovery effects confirmed more often than broad FDR discoveries. Pathway programs transferred more frequently than broad gene lists. At the predictive layer, ranking discrimination could remain high while the locked threshold, sensitivity, specificity, and predictive values shifted substantially. The reproducibility hierarchy therefore extends beyond gene discovery: association, prioritization, biological-program transfer, score discrimination, and operating-point transport are distinct claims requiring distinct evidence.

\section{Discussion}

This nine-cohort study shows that transcriptomic reproducibility cannot be represented by one overlap statistic, one ROC-AUC, or one binary label. Across three cancers, the same paired contrast and analytic procedure produced DEG lists covering between one fifth and three quarters of the tested transcriptome. Exact agreement and extreme ranks were unstable, while signed effects, large independently confirmed effects, and pathways were more transferable. A one-time two-fold cross-validation enriched external confirmation relative to the full FDR list, although matched-size controls showed that most gain reflected selection of a smaller, stronger evidence set. The locked diagnostic extension then exposed another separation: compact panels frequently retained high external AUC, but sensitivity, specificity, and balanced accuracy at the discovery threshold could fail. The central finding is therefore a hierarchy of reproducibility from molecular association to operating-point transport, not a failure of transcriptomics.

\subsection{A DEG list is a decision boundary, not a molecular object}

A thresholded DEG list is frequently discussed as though it were directly observed. In reality, it is a discretization of effect estimates and uncertainty. The number of calls depends on sample size, within-pair variance, empirical-Bayes moderation, and the full distribution of $P$ values. This explains why the broad FDR burden varied so strongly while large-effect counts followed a different pattern. A cohort with many precise moderate effects can produce a much longer FDR list than a cohort containing a similar number of large effects but greater variance.

The practical implication is that DEG counts cannot be interpreted as a measure of biological complexity or signal strength without examining the effect distribution. Nor should a gene just below FDR 0.05 be treated as biologically opposite to one just above it. Reporting only a list discards the continuous evidence needed to understand cross-cohort differences. A single-cohort DEG list provides evidence only of within-cohort association. It does not establish a reproducible gene ranking, a transferable biomarker panel, or a generalizable molecular signature; those claims require independent confirmation. Complete effect estimates, uncertainty, adjusted $P$ values, and rank behavior should remain part of the primary result.

\subsection{Why top-gene narratives are especially vulnerable}

The instability of extreme ranks is particularly consequential because experimental follow-up often begins with the first few genes in a table. The top of a genome-wide ranking is an extreme-order statistic. When hundreds or thousands of genes participate in correlated tumor-associated programs, small changes in variance, cellular admixture, probe behavior, or sample composition can reorder the leaders without erasing the underlying biology. This phenomenon has been described previously for cancer signatures \citep{eindor2005outcome,eindor2006thousands}, and the present study shows it across three paired tumor-normal settings.

Biological plausibility does not solve this problem. The leading PDAC stromal genes, breast cell-cycle genes, and lung tissue-associated genes were coherent within their respective discovery cohorts. Their plausibility may justify further study, but it does not make their numerical rank a transferable property. A claim such as ``the most important DEG'' requires evidence that the prioritization is robust to independent cohorts, alternative preprocessing choices, and reasonable ranking definitions. Without that evidence, the honest statement is that the gene was highly ranked in one analysis. Selecting top-ranked genes for functional validation while presenting their order as biologically stable is scientifically unsound unless the ranking itself is independently confirmed.

\subsection{Effect direction and large-effect confirmation provide stronger evidence}

The effect analyses prevent an overly pessimistic reading. Genome-wide correlations were moderate, and agreement became stronger within selected evidence layers. Among genes significant in both members of a pair, 85.6\% to 98.7\% had the same direction. Most importantly, restricting discovery to \abslogfc at least 1 increased confirmation in both independent cohorts to 50.1\% to 84.3\%. This cross-cancer effect-magnitude gradient is one of the most actionable findings.

Large effects are not automatically causal, clinically useful, or free of confounding. They can still reflect tissue composition or technical properties. However, a large signed effect that confirms independently has a stronger empirical basis for mechanistic prioritization than a small effect selected mainly by precision or an unstable rank. Gene-level follow-up should therefore prioritize the joint evidence of magnitude, direction, independent confirmation, and biological context rather than discovery rank alone.

\subsection{Discrimination can transfer while the diagnostic rule fails}

The diagnostic analysis demonstrates why a gene list cannot become a biomarker panel through terminology alone. ROC-AUC asks whether tumor samples tend to receive higher scores than non-tumor samples across all possible thresholds. Sensitivity and specificity ask what happens at one chosen threshold. A platform- or cohort-dependent shift in the score distribution can preserve ranking and therefore AUC while moving both classes across the fixed decision boundary. This is exactly what occurred for the breast all-gene models, the PDAC compact panels, and the lung compact panels in validation.

This distinction is not a technical footnote. A locked threshold is part of the diagnostic rule, and PPV and NPV additionally depend on prevalence. Prediction-model guidance therefore separates discrimination, calibration, and clinical operating characteristics and requires all development decisions to be isolated from external evaluation data \citep{pepe2003statistical,collins2015tripod,steyerberg2019clinical,wolff2019probast}. Reporting only the best external AUC would have hidden clinically consequential errors in this study. Conversely, poor performance at one transported threshold does not erase useful ranking information; it indicates that the score is not directly portable as a binary decision rule.

The compact-panel results also reject a simplistic conclusion that fewer genes are always better or that internally stable genes are automatically more diagnostic. Compact panels were competitive and sometimes superior, especially in breast validation and lung test, but full panels were more robust at the PDAC test operating point. \redetwofold improved breast validation AUC relative to the conventional top 19, whereas conventional DE ranking was stronger in PDAC, and both compact strategies were identical in lung cancer. Feature stability, discrimination, calibration, and threshold transport should therefore be evaluated separately rather than collapsed into one claim of biomarker validity.

\subsection{REDE-2Fold imports machine-learning validation logic into molecular feature selection}

In machine learning, k-fold cross-validation tests whether a model or feature-selection process remains useful when fitted and evaluated on different subsets of the data. \redetwofold transfers this computer-science principle to differential expression. The output is not a prediction score. It is a cross-validated molecular feature set: genes independently selected from both halves of the available cohort with the same effect direction. We use the term \emph{bio-machine learning} for this type of integration, in which validation logic developed for computational learning systems is used to strengthen biological feature discovery.

The external results support a precise interpretation. Across all three cancers, the \redetwofold intersection transferred better than the complete discovery FDR list. This demonstrates that requiring a gene to survive two independent patient subsets can remove cohort-dependent calls. The matched-size controls are equally important. They performed almost identically to the cross-validated FDR intersection, and the absolute-\logfc control was stronger in PDAC. \redetwofold should therefore not be presented as a universally superior gene-ranking algorithm. Its main contribution is procedural: it forces the analyst to demonstrate internal feature stability rather than reporting a single complete-cohort list without perturbation.

When only one cohort exists, we recommend \redetwofold as a minimum internal validation step. The random seed, fold allocation, pair preservation, model, thresholds, and intersection rule must be reported. A gene retained by this procedure has stronger internal evidence than a gene selected only once, but both folds share the same recruitment process, laboratory conditions, platform, preprocessing, and population. In addition, a single fixed split can be sensitive to the particular allocation of patients. \redetwofold reduces the single-cohort fallacy but cannot eliminate it; independent validation remains the evidential standard for claims of transferability.

\subsection{Pathways are more transferable, but not automatically universal}

Pathway analysis replicated more successfully than the broad discovery DEG list in every cancer. Aggregating correlated genes reduces dependence on the identity and order of individual members, allowing a biological program to remain detectable when different cohorts assign the strongest statistics to different genes. The recurrence of cell-cycle and growth programs across all nine cohorts supports this interpretation.

The lower pathway replication in breast cancer (52.2\%) may reflect greater molecular heterogeneity across the discovery, validation, and external-test cohorts, including differences in age and receptor or subtype composition reported by the source studies, together with the use of three distinct microarray platforms \citep{quigley2017age,pauni2010malaysian,chang2018gas7}. This result is informative rather than anomalous: pathway replication, although more stable than broad gene-level membership in these analyses, is not automatic and still depends on biological consistency across cohorts.

Pathway stability must still be interpreted carefully. Hallmark sets contain fewer hypotheses than gene-level analysis and deliberately compress overlapping biology, which makes replication easier. A replicated pathway does not imply that every leading-edge gene is shared, that the pathway is causal, or that its direction is the same across diseases. UV Response Dn, interferon signaling, TNF-alpha signaling, and KRAS signaling illustrated this point: internal reproducibility within a cancer coexisted with cross-cancer directional differences. Pathway-level reporting should therefore include normalized enrichment scores, direction, significance in every cohort, and leading-edge composition where mechanistic claims are made.

\subsection{Confirmatory testing should be distinguished from repeated discovery}

A recurring ambiguity in validation studies is whether the independent cohort performs a new genome-wide discovery analysis or tests a prespecified family from discovery. The two procedures have different multiplicity burdens and answer different questions. Requiring a gene to pass genome-wide FDR in every cohort is stringent and can be appropriate for claiming repeated de novo discovery. Applying multiplicity correction across the declared discovery family is closer to confirmatory inference. Neither procedure should omit the direction requirement.

The similarity between confirmatory and genome-wide percentages in this study indicates that many confirming genes carried strong independent evidence, especially in lung cancer. Nevertheless, the distinction remains essential for transparent reporting. Terms such as ``validated DEG'' or ``replicated biomarker'' should state the tested family, the correction denominator, the required direction, the effect-size criterion, and whether the independent data were used for any model or threshold tuning.

\subsection{REDE: a seven-level Reproducible Expression Differential Evaluation framework}

The results support \rede - a seven-level Reproducible Expression Differential Evaluation framework - for transcriptome-wide studies that make claims beyond one cohort:

\begin{center}
\fbox{\begin{minipage}{0.93\textwidth}
\textbf{REDE seven-level reproducibility report}
\begin{enumerate}
\item \textbf{Preprocessing validity:} document scale, transformation, annotation, probe aggregation, missingness, and the rationale for any harmonization.
\item \textbf{Within-cohort evidence:} report signed effect sizes, uncertainty, FDR values, DEG burden, and the full tested gene space.
\item \textbf{Threshold robustness:} show how exact membership changes across reasonable FDR and effect-magnitude cutoffs.
\item \textbf{Rank and internal selection stability:} quantify top-k overlap and effect-rank correlation; when only one dataset is available, apply a declared internal procedure such as \redetwofold rather than presenting one complete-cohort gene table as stable.
\item \textbf{Independent gene confirmation:} prespecify the discovery family, multiplicity correction, direction requirement, and effect criterion.
\item \textbf{Program-level transfer:} evaluate pathway direction and significance while retaining the distinction between pathway replication and leading-edge gene replication.
\item \textbf{Predictive and diagnostic transfer:} lock the feature set, preprocessing, model, hyperparameters, and threshold before independent evaluation; report discrimination, calibration-sensitive measures, sensitivity, specificity, predictive values, and prevalence.
\end{enumerate}
\end{minipage}}
\end{center}

\rede is intended as both an analysis workflow and a reporting checklist. A manuscript that reports these levels can make claims proportional to its evidence. A single cohort establishes association in that cohort. \redetwofold can add an internal cross-validation certificate when an external dataset is unavailable, but it remains below independent confirmation in the claim-evidence ladder. Independent confirmatory evidence supports replication of a declared gene family. Repeated direction and magnitude strengthen prioritization. A locked external model can establish transport of ranking discrimination, but a biomarker decision rule additionally requires a stable operating point, calibration in the intended population, and a clinically appropriate sampling design. Differential expression alone does not provide that evidence.

\begin{table}[H]
\centering
\caption{REDE claim-evidence ladder for differential-expression studies.}
\label{tab:claim_ladder}
\scriptsize
\begin{tabularx}{0.99\textwidth}{>{\raggedright\arraybackslash}p{0.17\textwidth} >{\raggedright\arraybackslash}p{0.29\textwidth} >{\raggedright\arraybackslash}p{0.23\textwidth} >{\raggedright\arraybackslash}X}
\toprule
Claim & Minimum evidence & Recommended analysis & Language justified without that evidence \\
\midrule
Differentially expressed in one cohort & Prespecified contrast, valid scale, appropriate design, multiplicity control, and effect estimate & Full effect table, DEG burden, uncertainty, and scale audit & Associated with the contrast in the analyzed cohort \\
Internally cross-validated candidate set & One declared patient-level two-fold split, independent DE in each fold, same-direction intersection, and exact fold reporting & \redetwofold with fixed seed, fold assignments, overlap, and effect comparison & Stable across one internal partition, not externally replicated \\
Replicated gene-level effect & Independent cohort, prespecified tested family, explicit multiplicity denominator, concordant direction, and reported magnitude & Confirmatory FDR, effect comparison, and forest plot & Not yet established as independently replicated \\
Reproducible prioritization & Stable effect and rank behavior across independent cohorts and reasonable analysis choices & Top-k overlap, rank correlation, and ranking sensitivity & Highly ranked in the discovery analysis \\
Replicated biological program & Independent pathway significance with concordant enrichment direction, plus leading-edge inspection for mechanistic claims & NES comparison, direction table, and leading-edge overlap & Enriched in the discovery cohort \\
Transferable prediction score & Locked features and model with leakage-free independent evaluation of discrimination and calibration-sensitive metrics & External ROC-AUC, PR-AUC, Brier score, log loss, and score-distribution assessment & A discovery-derived score requiring external validation \\
Diagnostic decision rule & Locked threshold, sensitivity, specificity, predictive values at target prevalence, and evaluation in the intended clinical sampling setting & External operating-point evaluation, calibration or prespecified updating, and decision analysis & Tissue-classification proof of concept, not a clinical diagnostic assay \\
\bottomrule
\end{tabularx}
\end{table}

\subsection{Strengths and limitations}

This study has several strengths. It includes nine independent datasets and three cancer types rather than relying on one disease or one cohort pair. Every analysis used complete matched tumor and non-tumor pairs and a patient-adjusted model. The role of each cohort was fixed, the gene space was held constant within cancer, and the cohorts were never pooled. The scale audit explicitly identified three datasets requiring transformation and protected against accidental double transformation. Reproducibility was evaluated across complementary estimands, including prespecified confirmation, pathway transfer, a prospectively specified one-time \redetwofold analysis, and a locked diagnostic-transfer experiment derived only from each discovery cohort. Matched-size full-cohort controls prevented the external-confirmation gain from being attributed uncritically to cross-validation alone. Finally, the source archive contains all six original analysis notebooks, the \redetwofold notebook, the diagnostic-validation notebook and locked predictions, exact summary tables, manifests, integrated figures, and a portable implementation of \rede with example inputs and automated tests.

Several limitations should be considered. First, the datasets are historical microarrays with study-specific normalization, platforms, annotations, procurement protocols, and clinical composition. These differences are part of the external-reproducibility problem but cannot be fully decomposed using public metadata. Second, adjacent or matched non-tumor tissue is not identical to healthy donor tissue, and field effects may differ by cancer and cohort. Third, whole-tissue expression reflects variable malignant, stromal, immune, vascular, and normal-cell fractions. The analysis quantifies transfer of the observed tumor-normal contrast, not cell-intrinsic causality. Fourth, probe-to-symbol aggregation can hide transcript-specific effects, and restricting each cancer to a common gene intersection may exclude platform-specific genes. Fifth, the study uses one primary model and one curated pathway collection. Alternative differential-expression engines, annotation resources, and gene-set databases could alter numerical results. Sixth, the three cancer experiments use different gene spaces, so the cross-cancer synthesis is descriptive rather than a pooled meta-analysis of identical genes. Seventh, \redetwofold intentionally used one fixed random split rather than repeated resampling; its exact membership can therefore depend on that allocation, and the analysis should be interpreted as a reproducible minimum internal cross-validation rather than a full stability-selection estimate. Finally, the new predictive task distinguishes resected tumor from adjacent or non-tumor tissue. It is not a screening study, does not evaluate blood-based detection or preoperative clinical diagnosis, and should not be interpreted as evidence of patient-level clinical utility. The cohorts are small for high-dimensional prediction, uncertainty around external metrics is therefore material, and no survival, recurrence, treatment-response, or follow-up outcomes were available for prognostic analysis.

Future work should extend the same audit to RNA sequencing, prospective multi-center cohorts, single-cell pseudobulk analyses, and proteogenomic data. It should also compare conventional moderated-statistic rankings with stability-aware prioritization, hierarchical models, and explicitly reproducible feature-selection procedures. The essential design principle should remain unchanged: the independent cohort must be protected from selection and tuning, and every claim should be matched to the level of evidence actually tested.

\section{Conclusion}

Across nine cohorts and three cancers, transcriptomic evidence formed a reproducibility hierarchy. Binary DEG burden, exact membership, and top-gene rank were strongly cohort dependent. Signed effects were more coherent, large discovery effects confirmed more often, and biological programs transferred more consistently. \redetwofold showed how a technique analogous to machine-learning cross-validation can strengthen molecular feature selection when only one dataset is available, although matched-size controls showed that it is an internal stability certificate rather than a universally superior ranking rule.

The locked tissue-classification analysis extended the hierarchy. Panels containing only 19 discovery-derived genes retained high external AUC in all three cancers, demonstrating that compact transcriptomic scores can preserve cross-cohort discrimination. However, several models with excellent AUC showed unacceptable sensitivity or specificity at the discovery-derived threshold. To avoid the single-cohort fallacy, neither a significant DEG list nor a high AUC from an adaptively evaluated model should be presented as a transferable diagnostic signature. The expanded seven-level \rede framework separates within-cohort association, internal stability, independent molecular confirmation, pathway transfer, predictive discrimination, and operating-point transport. Each claim requires its own protected validation design.

\section{Materials and methods}

\subsection{Study design and data sources}

Processed expression data and sample metadata were obtained from the NCBI Gene Expression Omnibus \citep{edgar2002geo}. Three disease-specific experiments were defined before differential-expression analysis. For PDAC, GSE15471 was discovery, GSE16515 validation, and GSE28735 external test \citep{badea2008combined,pei2009fkbp51,zhang2012dpep1}. For breast cancer, GSE70947 was discovery, GSE15852 validation, and GSE109169 external test \citep{quigley2017age,pauni2010malaysian,chang2018gas7}. For lung cancer, GSE32863 was discovery, GSE18842 validation, and GSE7670 external test \citep{selamat2012methylation,sanchezpalencia2011biomarkers,chen2009vegfa}. The differential-expression analyses retained only complete one-to-one matched tumor and non-tumor pairs and were designed as three parallel replication experiments rather than a pooled pan-cancer analysis. The subsequent diagnostic-transfer analysis preserved the same cohort roles but used all processed samples with a valid tumor or non-tumor label. This added 20 tumor-only GSE16515 samples to locked validation; all discovery feature selection and model development remained restricted to GSE15471.

\subsection{Expression-scale audit and gene-level preprocessing}

Submitted processed values were audited before gene-level aggregation using global and sample-level ranges and quantiles. Datasets with plausible log2 distributions were preserved. GSE16515, GSE15852, and GSE7670 contained intensity-scale values and were transformed using $\log_2(x+1)$. The pipeline avoided double transformation when values were already log2-like and recorded every decision in preprocessing manifests.

Probe identifiers were mapped to platform-provided gene symbols. Unmapped probes and records without a usable symbol were removed. When multiple probes mapped to the same gene symbol, expression was aggregated by the median within each sample. Technical replicate arrays in GSE15471 were averaged within patient and tissue class after scale correction and gene aggregation. For GSE70947, partially observed probes were retained through probe-to-gene aggregation; genes that remained incomplete in any sample after aggregation were removed. No expression value was imputed in any cohort. Within each cancer, the three matrices were restricted to their common gene-symbol intersection, yielding 18,034, 7,701, and 10,549 genes for PDAC, breast cancer, and lung cancer. Submitted within-cohort normalization was retained. No cross-cohort normalization, feature scaling, z-score standardization, or batch correction was performed.

\subsection{Paired differential-expression analysis}

Each cohort was analyzed independently in Python using InMoose's implementation of limma methods \citep{smyth2004linear,ritchie2015limma,colange2025inmoose}. For gene $g$, patient $i$, and tissue $j$, the model was
\[
Y_{gij}=\alpha_{gi}+\beta_g T_{ij}+\epsilon_{gij},
\]
where $Y_{gij}$ is log2 expression, $\alpha_{gi}$ is a gene-specific patient fixed effect, and $T_{ij}$ equals 1 for tumor and 0 for non-tumor. The coefficient $\beta_g$ estimates the mean paired tumor-minus-normal \logfc. The design was implemented as \texttt{0 + C(patient\_id) + target}. Limma-trend empirical-Bayes moderation was applied using intensity-dependent prior variance and standard, non-robust hyperparameter estimation. Coefficients were checked numerically against direct within-patient mean differences.

Two-sided gene-level $P$ values were adjusted using the Benjamini-Hochberg procedure \citep{benjamini1995controlling}. The primary DEG definition was adjusted $P<0.05$. A secondary descriptive definition additionally required \abslogfc at least 1. This magnitude filter was used to evaluate effect-strength sensitivity and was not treated as a separate inferential error-control procedure.

\subsection{Exact set membership and threshold sensitivity}

For each within-cancer cohort pair, exact thresholded-set agreement was quantified with the Jaccard index,
\[
J(A,B)=\frac{|A\cap B|}{|A\cup B|},
\]
and the overlap coefficient,
\[
O(A,B)=\frac{|A\cap B|}{\min(|A|,|B|)}.
\]
Three-way membership patterns were also enumerated. Threshold sensitivity was evaluated at FDR cutoffs of 0.01 and 0.05 and absolute \logfc cutoffs of 0, 0.5, 1.0, and 1.5. This analysis assessed whether list-overlap conclusions depended on one chosen boundary.

\subsection{Effect and rank reproducibility}

Effect-size agreement was summarized using Spearman and Pearson correlations of cohort-specific \logfc values. Direction concordance was the proportion of compared genes for which the signs of the two effects agreed. Metrics were calculated across all common genes, discovery-selected subsets, and overlapping significant subsets where appropriate.

Rank agreement was evaluated for the top 10, 25, 50, 100, 250, 500, and 1,000 genes ordered by absolute moderated t-statistic. The shared fraction was the intersection divided by $k$. Under independent random selection of two lists of size $k$ from $N$ genes, the expected intersection is $k^2/N$ and the expected shared fraction is $k/N$. Random expectations were used descriptively; no rank-overlap hypothesis test was used.

\subsection{Independent confirmation of discovery genes}

The discovery FDR list in each cancer was treated as a prespecified family in validation and external test. Within each independent cohort, Benjamini-Hochberg correction was applied only across the selected discovery genes. Confirmatory replication required confirmatory FDR below 0.05 and the same effect direction as discovery. Replication in both required these criteria to be met in validation and external test.

A second, more stringent definition required each discovery gene to pass the independent cohort's full genome-wide FDR threshold and have concordant direction. The same two procedures were applied to the discovery subset satisfying both FDR below 0.05 and \abslogfc at least 1. Independent confirmation of this subset did not require \abslogfc to exceed 1 again unless exact large-effect set membership was being evaluated.

\subsection{REDE-2Fold bio-machine-learning analysis}

For each cancer, only the fixed discovery cohort was used to construct the \redetwofold feature set. Patients were randomly divided once into two approximately equal non-overlapping folds with NumPy's deterministic random-number generator and random seed 3. Matched tumor and normal or non-tumor samples from the same patient were retained in the same fold. The resulting partitions contained 18 and 18 PDAC pairs, 74 and 74 breast-cancer pairs, and 29 and 28 lung-cancer pairs. No repeated splitting, resampling, model selection, or tuning against validation or external-test results was performed.

The same patient-adjusted InMoose limma-trend model, tested gene universe, FDR cutoff, and effect-direction convention used in the primary analysis were applied independently to both discovery folds. A gene entered the primary \redetwofold set when it had Benjamini-Hochberg FDR below 0.05 in both folds and concordant \logfc sign. A secondary large-effect set additionally required \abslogfc at least 1 in both folds. The fold-specific results, patient assignments, same-direction intersections, and between-fold Jaccard indices were retained as explicit outputs.

The selected gene family was frozen before external evaluation. Within validation and external test, Benjamini-Hochberg correction was applied across the prespecified \redetwofold genes, and confirmation required confirmatory FDR below 0.05 and the same direction as the fold-consensus effect. Confirmation in both required the criterion in both independent cohorts. To distinguish cross-validated stability from the effect of selecting fewer genes, two full-discovery controls were constructed with exactly the same number of genes as the primary \redetwofold set: one ranked by absolute moderated t-statistic and one by absolute \logfc. The full discovery FDR list and full discovery FDR-plus-large-effect list were also retained as reference strategies.

\subsection{Training-only diagnostic transfer analysis}

The predictive extension was defined as tumor-versus-non-tumor tissue classification. It was termed diagnostic transfer for brevity but was not designed as a clinical screening or patient-diagnosis study. For each cancer, candidate genes were selected exclusively from the fixed discovery cohort. The conventional full panel contained every gene with discovery FDR below 0.05. The conventional compact panel contained the 19 genes with the largest absolute moderated t-statistics. The full \redetwofold panel contained all same-direction genes passing FDR below 0.05 in both discovery folds. Its compact panel contained the 19 genes with the largest minimum absolute t-statistic across the two folds. A maximum of 19 genes was prespecified before external performance was examined as a practical ceiling below 20 features, rather than selected by optimization against validation or test results. When fewer than 19 eligible genes are available, the procedure uses all eligible genes; all three cancers had at least 19. No validation- or test-cohort statistic contributed to gene selection or ranking.

Each candidate gene was first evaluated individually. Its score was multiplied by the sign of the discovery-cohort \logfc so that larger values indicated tumor. A threshold was selected in discovery by maximizing Youden's $J$ statistic, with deterministic tie-breaking by distance to the upper-left ROC point, and was applied unchanged in validation and test.

Multigene panels used a pipeline comprising training-fitted standardization and L2-regularized logistic regression. The inverse regularization parameter $C$ was selected from $\{0.001,0.01,0.1,1,10,100\}$ using five-fold GroupKFold cross-validation in the discovery cohort with ROC-AUC as the selection metric. Patient identifiers defined groups so that paired tumor and non-tumor samples remained in the same fold. Grouped out-of-fold discovery probabilities were then generated with the selected pipeline, and the probability threshold was chosen by Youden's $J$ from those out-of-fold predictions. The pipeline was refitted on the full discovery cohort and applied without feature reselection, rescaling, hyperparameter tuning, recalibration, or threshold adjustment to validation and external test.

Performance measures were ROC-AUC, precision-recall AUC, sensitivity, specificity, PPV, NPV, accuracy, balanced accuracy, F1 score, Matthews correlation coefficient, positive and negative likelihood ratios, diagnostic odds ratio, Brier score, and log loss \citep{pepe2003statistical,steyerberg2019clinical}. PPV and NPV were interpreted relative to the observed cohort prevalence. Training metrics for multigene panels used grouped out-of-fold probabilities; all independent metrics used predictions from the locked full-discovery model.

\subsection{Hallmark gene set enrichment analysis}

Genes were ranked within each cohort by signed moderated t-statistic. Preranked gene set enrichment analysis was performed with GSEApy 1.1.9 using 1,000 permutations, random seed 3, minimum gene-set size 15, maximum size 500, and the MSigDB Hallmark 2020 collection \citep{subramanian2005gsea,liberzon2015hallmark,fang2023gseapy}. A pathway was significant at FDR below 0.05. A discovery pathway replicated in an independent cohort when it was significant there and its normalized enrichment score had the same sign. Replication in both required these conditions in validation and external test.

\subsection{Cross-cancer synthesis}

All inferential analyses were conducted within cancer because the platforms, common-gene spaces, and biological diseases differed. Cross-cancer results were synthesized descriptively by comparing the range and ordering of the same prespecified metrics. The \redetwofold analysis was likewise evaluated separately within each disease, with cross-cancer comparison restricted to set sizes and external confirmation rates. No pooled effect estimate, cross-cancer $P$ value, or formal meta-analysis was calculated. Medians and ranges across the nine within-cancer cohort comparisons are descriptive summaries of the parallel experiments.

\subsection{Software and reproducibility}

The differential-expression analyses used Python 3.10.16, InMoose 0.8.1, GSEApy 1.1.9, Patsy 1.0.1, statsmodels 0.14.5, SciPy 1.15.3, pandas 2.3.2, NumPy 2.2.6, and Matplotlib 3.10.5. The diagnostic-transfer analysis additionally used scikit-learn 1.7.2. The complete executable workflows are publicly available at \href{https://github.com/Bluesman79/REDE-transcriptomic-reproducibility}{github.com/Bluesman79/REDE-transcriptomic-reproducibility}. The exact software release accompanying this study, \rede v0.2.1, is permanently archived by Zenodo under DOI \href{https://doi.org/10.5281/zenodo.21774116}{10.5281/zenodo.21774116} \citep{angelakis2026rede}. The repository contains the three verified preprocessing notebooks, three disease-specific DEG notebooks, the outputs-only \redetwofold notebook, the unified three-cancer diagnostic-validation notebook, the cross-cancer synthesis script, the portable \rede workflow, pinned environments, automated tests, and synthetic examples. Raw GEO data, generated study outputs, manuscript files, and saved notebook outputs are intentionally excluded from the software repository; the public workflows regenerate the analytical outputs locally from the source datasets. The six molecular levels of the \rede framework are implemented in a portable command-line workflow and Colab-ready quick-start notebook that accept two paired expression matrices or two precomputed differential-expression tables. The seventh level, predictive and diagnostic transfer, is implemented in the unified three-cancer diagnostic-validation notebook because it requires sample-level outcomes, locked model development, and independent predictions. Predictive models were tuned only by grouped cross-validation within discovery and were applied without modification to both independent cohorts.

\section*{Data availability}

All expression datasets are publicly available from NCBI GEO under accessions GSE15471, GSE16515, GSE28735, GSE70947, GSE15852, GSE109169, GSE32863, GSE18842, and GSE7670. No new primary data were generated. The numerical results supporting the main conclusions are reported in the article and its supplementary tables and figures, and the released code regenerates the complete derived outputs from the public source datasets. The software repository does not redistribute the original GEO expression files or generated study outputs.

\section*{Code availability}

The complete executable analysis code is available at \href{https://github.com/Bluesman79/REDE-transcriptomic-reproducibility}{github.com/Bluesman79/REDE-transcriptomic-reproducibility}. The version used for this manuscript is \rede v0.2.1 and is permanently archived at Zenodo under DOI \href{https://doi.org/10.5281/zenodo.21774116}{10.5281/zenodo.21774116} \citep{angelakis2026rede}. The repository includes the verified preprocessing and disease-specific DEG notebooks, \path{code/04_REDE_2Fold_Outputs_Only_Three_Cancers.ipynb}, \path{code/05_DEG_Diagnostic_Validation_Three_Cancers.ipynb}, \path{code/rede_workflow.py}, the cross-cancer figure-generation script, a Colab-ready quick-start notebook, synthetic inputs, documentation, automated tests, and pinned environment specifications. The \redetwofold notebook deterministically reconstructs the one-time patient-level split using random seed 3, and the diagnostic-transfer notebook performs training-only gene selection and locked independent evaluation.

\section*{Ethics statement}

This study is a secondary analysis of deidentified public datasets and involved no new participant recruitment or intervention. Ethical approval and consent procedures were the responsibility of the original data-generating studies.

\section*{Author contributions}

A.A. conceived the study, designed the methodology, implemented and integrated the analyses, interpreted the results, prepared the figures, and wrote the manuscript.

\section*{Competing interests}

The author declares no competing interests.

\section*{Funding}

No project-specific funding was used for this secondary computational analysis.

\section*{Acknowledgments}

The author thanks the investigators and participants responsible for generating and publicly depositing the nine GEO datasets analyzed in this study.

\bibliographystyle{plainnat}
\bibliography{references}

\clearpage
\section*{Supplementary information}
\setcounter{table}{0}
\setcounter{figure}{0}
\renewcommand{\thetable}{S\arabic{table}}
\renewcommand{\thefigure}{S\arabic{figure}}

\subsection*{Supplementary Table S1: Expression-scale audit}

\begin{longtable}{lllrrrr}
\caption{Final expression distributions after verified scale handling.}\label{tab:s_scale}\\
\toprule
Cancer and cohort & Scale action & Pairs & Minimum & Median & 99th percentile & Maximum \\
\midrule
\endfirsthead
\toprule
Cancer and cohort & Scale action & Pairs & Minimum & Median & 99th percentile & Maximum \\
\midrule
\endhead
PDAC GSE15471 & Preserve log2 & 36 & 2.16 & 5.90 & 11.16 & 14.64 \\
PDAC GSE16515 & $\log_2(x+1)$ & 16 & 1.82 & 5.00 & 11.64 & 15.98 \\
PDAC GSE28735 & Preserve log2 & 45 & 1.13 & 4.71 & 8.85 & 12.58 \\
Breast GSE70947 & Preserve log2 & 148 & -1.44 & 8.62 & 15.21 & 18.39 \\
Breast GSE15852 & $\log_2(x+1)$ & 43 & 0.52 & 8.82 & 11.94 & 13.99 \\
Breast GSE109169 & Preserve log2 & 25 & 1.77 & 7.58 & 11.35 & 13.73 \\
Lung GSE32863 & Preserve log2 & 57 & 6.35 & 7.64 & 13.43 & 15.99 \\
Lung GSE18842 & Preserve log2 & 44 & 2.68 & 5.90 & 12.13 & 14.62 \\
Lung GSE7670 & $\log_2(x+1)$ & 27 & 0.14 & 7.86 & 13.18 & 16.96 \\
\bottomrule
\end{longtable}

\subsection*{Supplementary Table S2: Pairwise reproducibility metrics}

\begin{landscape}
\begin{table}[H]
\centering
\caption{Pairwise reproducibility under the primary FDR definition.}
\label{tab:s_pairwise}
\scriptsize
\resizebox{0.96\linewidth}{!}{%
\begin{tabular}{llrrrrrr}
\toprule
Cancer & Cohort pair & FDR Jaccard & Direction concordance among shared FDR genes & Spearman effect & All-gene direction & Top-25 shared & Top-1,000 shared \\
\midrule
PDAC & Discovery vs validation & 0.252 & 96.2\% & 0.650 & 78.1\% & 0.0\% & 13.6\% \\
PDAC & Discovery vs external test & 0.403 & 88.8\% & 0.563 & 72.5\% & 0.0\% & 19.8\% \\
PDAC & Validation vs external test & 0.357 & 98.7\% & 0.727 & 78.4\% & 32.0\% & 45.4\% \\
Breast & Discovery vs validation & 0.249 & 88.5\% & 0.474 & 64.4\% & 4.0\% & 41.7\% \\
Breast & Discovery vs external test & 0.505 & 90.2\% & 0.664 & 71.7\% & 0.0\% & 42.1\% \\
Breast & Validation vs external test & 0.229 & 85.6\% & 0.407 & 57.6\% & 12.0\% & 28.4\% \\
Lung & Discovery vs validation & 0.547 & 86.2\% & 0.574 & 70.4\% & 20.0\% & 41.0\% \\
Lung & Discovery vs external test & 0.425 & 95.9\% & 0.644 & 73.1\% & 28.0\% & 50.0\% \\
Lung & Validation vs external test & 0.419 & 95.1\% & 0.650 & 73.4\% & 20.0\% & 49.5\% \\
\bottomrule
\end{tabular}%
}
\end{table}
\end{landscape}

\subsection*{Supplementary Table S3: Discovery-gene replication}

\begin{table}[H]
\centering
\caption{Replication of discovery-defined gene families in both independent cohorts.}
\label{tab:s_replication}
\small
\begin{tabular}{llrrrr}
\toprule
Cancer & Discovery family & Selected & Confirmatory, n (\%) & Genome-wide, n (\%) \\
\midrule
PDAC & FDR & 13,038 & 2,623 (20.1) & 2,521 (19.3) \\
PDAC & FDR and large effect & 952 & 593 (62.3) & 512 (53.8) \\
Breast & FDR & 5,796 & 901 (15.5) & 860 (14.8) \\
Breast & FDR and large effect & 819 & 410 (50.1) & 369 (45.1) \\
Lung & FDR & 6,394 & 2,526 (39.5) & 2,419 (37.8) \\
Lung & FDR and large effect & 681 & 574 (84.3) & 562 (82.5) \\
\bottomrule
\end{tabular}
\end{table}

\subsection*{Supplementary Table S4: Hallmark pathway replication}

\begin{table}[H]
\centering
\caption{Replication of discovery-significant Hallmark pathways.}
\label{tab:s_gsea}
\small
\resizebox{0.96\textwidth}{!}{%
\begin{tabular}{lrrrr}
\toprule
Cancer & Discovery significant & Replicated in validation & Replicated in test & Replicated in both, n (\%) \\
\midrule
PDAC & 36 & 33 & 33 & 32 (88.9) \\
Breast & 23 & 17 & 16 & 12 (52.2) \\
Lung & 32 & 26 & 22 & 22 (68.8) \\
\bottomrule
\end{tabular}%
}
\end{table}

\clearpage

\begin{landscape}
\subsection*{Supplementary Table S5: REDE-2Fold external confirmation and matched controls}
\scriptsize
\setlength{\tabcolsep}{4pt}
\begin{longtable}{llrrrr}
\caption{External confirmation of REDE-2Fold and reference discovery strategies. Percentages are calculated relative to the number selected by each strategy.}\label{tab:s_rede2fold}\\
\toprule
Cancer & Discovery strategy & Selected & Confirmed validation, n (\%) & Confirmed test, n (\%) & Confirmed both, n (\%) \\
\midrule
\endfirsthead
\toprule
Cancer & Discovery strategy & Selected & Confirmed validation, n (\%) & Confirmed test, n (\%) & Confirmed both, n (\%) \\
\midrule
\endhead
PDAC & Full discovery FDR & 13,038 & 3,591 (27.5) & 5,413 (41.5) & 2,623 (20.1) \\
PDAC & Matched top $|t|$ & 6,257 & 2,205 (35.2) & 3,059 (48.9) & 1,618 (25.9) \\
PDAC & Matched top $|\log_2\mathrm{FC}|$ & 6,257 & 2,577 (41.2) & 3,385 (54.1) & 2,084 (33.3) \\
PDAC & REDE-2Fold FDR intersection & 6,257 & 2,220 (35.5) & 3,047 (48.7) & 1,618 (25.9) \\
PDAC & Full FDR and large effect & 952 & 617 (64.8) & 750 (78.8) & 593 (62.3) \\
PDAC & REDE-2Fold FDR and large effect & 536 & 362 (67.5) & 444 (82.8) & 352 (65.7) \\
Breast & Full discovery FDR & 5,796 & 1,390 (24.0) & 3,006 (51.9) & 901 (15.5) \\
Breast & Matched top $|t|$ & 4,511 & 1,313 (29.1) & 2,667 (59.1) & 899 (19.9) \\
Breast & Matched top $|\log_2\mathrm{FC}|$ & 4,511 & 1,312 (29.1) & 2,619 (58.1) & 897 (19.9) \\
Breast & REDE-2Fold FDR intersection & 4,511 & 1,307 (29.0) & 2,643 (58.6) & 895 (19.8) \\
Breast & Full FDR and large effect & 819 & 456 (55.7) & 684 (83.5) & 410 (50.1) \\
Breast & REDE-2Fold FDR and large effect & 719 & 407 (56.6) & 621 (86.4) & 373 (51.9) \\
Lung & Full discovery FDR & 6,394 & 4,070 (63.7) & 3,056 (47.8) & 2,526 (39.5) \\
Lung & Matched top $|t|$ & 4,238 & 2,995 (70.7) & 2,526 (59.6) & 2,141 (50.5) \\
Lung & Matched top $|\log_2\mathrm{FC}|$ & 4,238 & 2,924 (69.0) & 2,494 (58.8) & 2,095 (49.4) \\
Lung & REDE-2Fold FDR intersection & 4,238 & 2,995 (70.7) & 2,511 (59.2) & 2,127 (50.2) \\
Lung & Full FDR and large effect & 681 & 613 (90.0) & 610 (89.6) & 574 (84.3) \\
Lung & REDE-2Fold FDR and large effect & 580 & 531 (91.6) & 522 (90.0) & 499 (86.0) \\
\bottomrule
\end{longtable}
\end{landscape}

\subsection*{Supplementary Table S6: REDE reporting checklist}

\begin{table}[H]
\centering
\caption{Operational checklist for a complete REDE report.}
\label{tab:s_rede}
\small
\begin{tabularx}{0.97\textwidth}{>{\raggedright\arraybackslash}p{0.08\textwidth} >{\raggedright\arraybackslash}p{0.24\textwidth} >{\raggedright\arraybackslash}X}
\toprule
Level & Domain & Minimum reported output \\
\midrule
1 & Preprocessing validity & Input scale, transformation, annotation, aggregation, missingness, and harmonization rationale \\
2 & Within-cohort evidence & Signed effects, uncertainty, FDR, DEG burden, and tested gene universe \\
3 & Threshold robustness & Membership and burden across reasonable FDR and effect-size cutoffs \\
4 & Rank and internal stability & Top-k overlap, rank correlation, ranking sensitivity, and REDE-2Fold when only one dataset is available \\
5 & Independent confirmation & Prespecified family, multiplicity denominator, direction, magnitude, and independent confirmation rate \\
6 & Program-level transfer & Pathway significance, normalized enrichment direction, and leading-edge composition where interpreted \\
7 & Predictive and diagnostic transfer & Locked features, preprocessing, model, threshold, discrimination, calibration-sensitive metrics, sensitivity, specificity, predictive values, and evaluated prevalence \\
\bottomrule
\end{tabularx}
\end{table}

\subsection*{Supplementary Table S7: Diagnostic-transfer cohort composition}

\begin{table}[H]
\centering
\caption{Samples used for locked tumor-versus-non-tumor tissue classification.}
\label{tab:s_diagnostic_cohorts}
\small
\begin{tabular}{lllrrrr}
\toprule
Cancer & Role & GEO & Samples & Non-tumor & Tumor & Patients \\
\midrule
PDAC & Discovery & GSE15471 & 72 & 36 & 36 & 36 \\
PDAC & Validation & GSE16515 & 52 & 16 & 36 & 36 \\
PDAC & External test & GSE28735 & 90 & 45 & 45 & 45 \\
Breast & Discovery & GSE70947 & 296 & 148 & 148 & 148 \\
Breast & Validation & GSE15852 & 86 & 43 & 43 & 43 \\
Breast & External test & GSE109169 & 50 & 25 & 25 & 25 \\
Lung & Discovery & GSE32863 & 114 & 57 & 57 & 57 \\
Lung & Validation & GSE18842 & 88 & 44 & 44 & 44 \\
Lung & External test & GSE7670 & 54 & 27 & 27 & 27 \\
\bottomrule
\end{tabular}
\end{table}

\clearpage
\begin{landscape}
\subsection*{Supplementary Table S8: Complete locked external panel performance}
\scriptsize
\setlength{\tabcolsep}{3pt}
\begin{longtable}{lllrrrrrrrrrr}
\caption{Performance of every discovery-derived panel in validation and external test. Threshold-dependent metrics use the unchanged discovery-derived threshold. PPV and NPV reflect the prevalence in each evaluated cohort.}\label{tab:s_diagnostic_performance}\\
\toprule
Cancer & Evaluation & Panel & Genes & AUC & PR-AUC & Sens. & Spec. & PPV & NPV & Bal. acc. & F1 & MCC \\
\midrule
\endfirsthead
\toprule
Cancer & Evaluation & Panel & Genes & AUC & PR-AUC & Sens. & Spec. & PPV & NPV & Bal. acc. & F1 & MCC \\
\midrule
\endhead
PDAC & Validation & Full discovery DEGs & 13,038 & 0.957 & 0.983 & 1.000 & 0.125 & 0.720 & 1.000 & 0.562 & 0.837 & 0.300 \\
PDAC & External test & Full discovery DEGs & 13,038 & 0.922 & 0.939 & 0.867 & 0.911 & 0.907 & 0.872 & 0.889 & 0.886 & 0.779 \\
PDAC & Validation & Discovery DE top 19 & 19 & 0.875 & 0.926 & 0.917 & 0.562 & 0.825 & 0.750 & 0.740 & 0.868 & 0.525 \\
PDAC & External test & Discovery DE top 19 & 19 & 0.897 & 0.913 & 0.067 & 1.000 & 1.000 & 0.517 & 0.533 & 0.125 & 0.186 \\
PDAC & Validation & All REDE-2Fold genes & 6,257 & 0.948 & 0.979 & 1.000 & 0.250 & 0.750 & 1.000 & 0.625 & 0.857 & 0.433 \\
PDAC & External test & All REDE-2Fold genes & 6,257 & 0.919 & 0.933 & 0.711 & 0.933 & 0.914 & 0.764 & 0.822 & 0.800 & 0.661 \\
PDAC & Validation & REDE-2Fold top 19 & 19 & 0.835 & 0.894 & 0.889 & 0.562 & 0.821 & 0.692 & 0.726 & 0.853 & 0.481 \\
PDAC & External test & REDE-2Fold top 19 & 19 & 0.869 & 0.877 & 0.200 & 1.000 & 1.000 & 0.556 & 0.600 & 0.333 & 0.333 \\
Breast & Validation & Full discovery DEGs & 5,796 & 0.723 & 0.714 & 0.907 & 0.349 & 0.582 & 0.789 & 0.628 & 0.709 & 0.308 \\
Breast & External test & Full discovery DEGs & 5,796 & 0.989 & 0.990 & 1.000 & 0.000 & 0.500 & -- & 0.500 & 0.667 & 0.000 \\
Breast & Validation & Discovery DE top 19 & 19 & 0.850 & 0.897 & 0.744 & 0.860 & 0.842 & 0.771 & 0.802 & 0.790 & 0.609 \\
Breast & External test & Discovery DE top 19 & 19 & 0.989 & 0.988 & 0.760 & 0.960 & 0.950 & 0.800 & 0.860 & 0.844 & 0.735 \\
Breast & Validation & All REDE-2Fold genes & 4,511 & 0.733 & 0.753 & 0.767 & 0.558 & 0.635 & 0.706 & 0.663 & 0.695 & 0.333 \\
Breast & External test & All REDE-2Fold genes & 4,511 & 0.994 & 0.994 & 1.000 & 0.000 & 0.500 & -- & 0.500 & 0.667 & 0.000 \\
Breast & Validation & REDE-2Fold top 19 & 19 & 0.905 & 0.928 & 0.977 & 0.186 & 0.545 & 0.889 & 0.581 & 0.700 & 0.266 \\
Breast & External test & REDE-2Fold top 19 & 19 & 0.987 & 0.987 & 1.000 & 0.640 & 0.735 & 1.000 & 0.820 & 0.847 & 0.686 \\
Lung & Validation & Full discovery DEGs & 6,394 & 1.000 & 1.000 & 0.409 & 1.000 & 1.000 & 0.629 & 0.705 & 0.581 & 0.507 \\
Lung & External test & Full discovery DEGs & 6,394 & 0.997 & 0.997 & 0.741 & 1.000 & 1.000 & 0.794 & 0.870 & 0.851 & 0.767 \\
Lung & Validation & Discovery DE top 19 & 19 & 0.999 & 0.999 & 1.000 & 0.091 & 0.524 & 1.000 & 0.545 & 0.688 & 0.218 \\
Lung & External test & Discovery DE top 19 & 19 & 0.996 & 0.996 & 0.963 & 1.000 & 1.000 & 0.964 & 0.981 & 0.981 & 0.964 \\
Lung & Validation & All REDE-2Fold genes & 4,238 & 1.000 & 1.000 & 0.773 & 1.000 & 1.000 & 0.815 & 0.886 & 0.872 & 0.793 \\
Lung & External test & All REDE-2Fold genes & 4,238 & 0.997 & 0.997 & 0.741 & 1.000 & 1.000 & 0.794 & 0.870 & 0.851 & 0.767 \\
Lung & Validation & REDE-2Fold top 19 & 19 & 0.999 & 0.999 & 1.000 & 0.091 & 0.524 & 1.000 & 0.545 & 0.688 & 0.218 \\
Lung & External test & REDE-2Fold top 19 & 19 & 0.996 & 0.996 & 0.963 & 1.000 & 1.000 & 0.964 & 0.981 & 0.981 & 0.964 \\
\bottomrule
\end{longtable}
\end{landscape}

\clearpage
\begin{landscape}
\subsection*{Supplementary Table S9: Compact discovery-derived gene panels}
\scriptsize
\begin{longtable}{llp{0.72\linewidth}}
\caption{Genes in the conventional and REDE-2Fold compact panels. All identities and ranks were fixed using discovery data only.}\label{tab:s_compact_genes}\\
\toprule
Cancer & Panel & Genes in training-derived rank order \\
\midrule
\endfirsthead
\toprule
Cancer & Panel & Genes in training-derived rank order \\
\midrule
\endhead
PDAC & Discovery DE top 19 & COL10A1, SULF1, INHBA, PMEPA1, NOX4, EDNRA, AEBP1, FAM19A5, THBS2, FN1, ANKRD50, DACT1, LTBP1, CTHRC1, BGN, MXRA8, SLPI, VCAN, COL1A2 \\
PDAC & REDE-2Fold top 19 & COL10A1, SULF1, EDNRA, PMEPA1, INHBA, DACT1, FAM19A5, ANKRD50, FN1, AEBP1, SHISA2, LTBP1, TIMP1, BGN, SLPI, PDGFRB, COL8A2, LOXL1, CDH11 \\
Breast & Discovery DE top 19 & EZH2, RAD54L, KIAA0101, CORO2B, BUB1B, CKS2, TYMS, SQLE, CENPM, DTL, NUP210, CAT, SPC25, ECHDC3, SLC20A1, HADH, MYOM1, TOP2A, KIF2C \\
Breast & REDE-2Fold top 19 & RAD54L, CAT, EZH2, KIAA0101, CORO2B, BUB1B, SQLE, HADH, DTL, TYMS, ECHDC3, CKS2, NDN, BIRC5, NDC80, CDKN2C, CENPF, MYOM1, ACADL \\
Lung & Discovery DE top 19 & FABP4, FMO2, FAM107A, CLEC3B, CAV1, ADH1B, LDB2, CLEC1A, TCF21, SPOCK2, AGER, PECAM1, TEK, HBB, GHR, STX11, CA4, MMRN1, FHL1 \\
Lung & REDE-2Fold top 19 & FABP4, FMO2, CLEC3B, CAV1, LDB2, ADH1B, FAM107A, CLEC1A, TCF21, AGER, HBB, SPOCK2, PECAM1, STX11, MMRN1, FHL1, CA4, GHR, TEK \\
\bottomrule
\end{longtable}
\end{landscape}

\subsection*{Supplementary Figure S1: Cohort-specific volcano plots}

\begin{figure}[H]
\centering
\begin{subfigure}[t]{0.32\textwidth}\includegraphics[width=\textwidth]{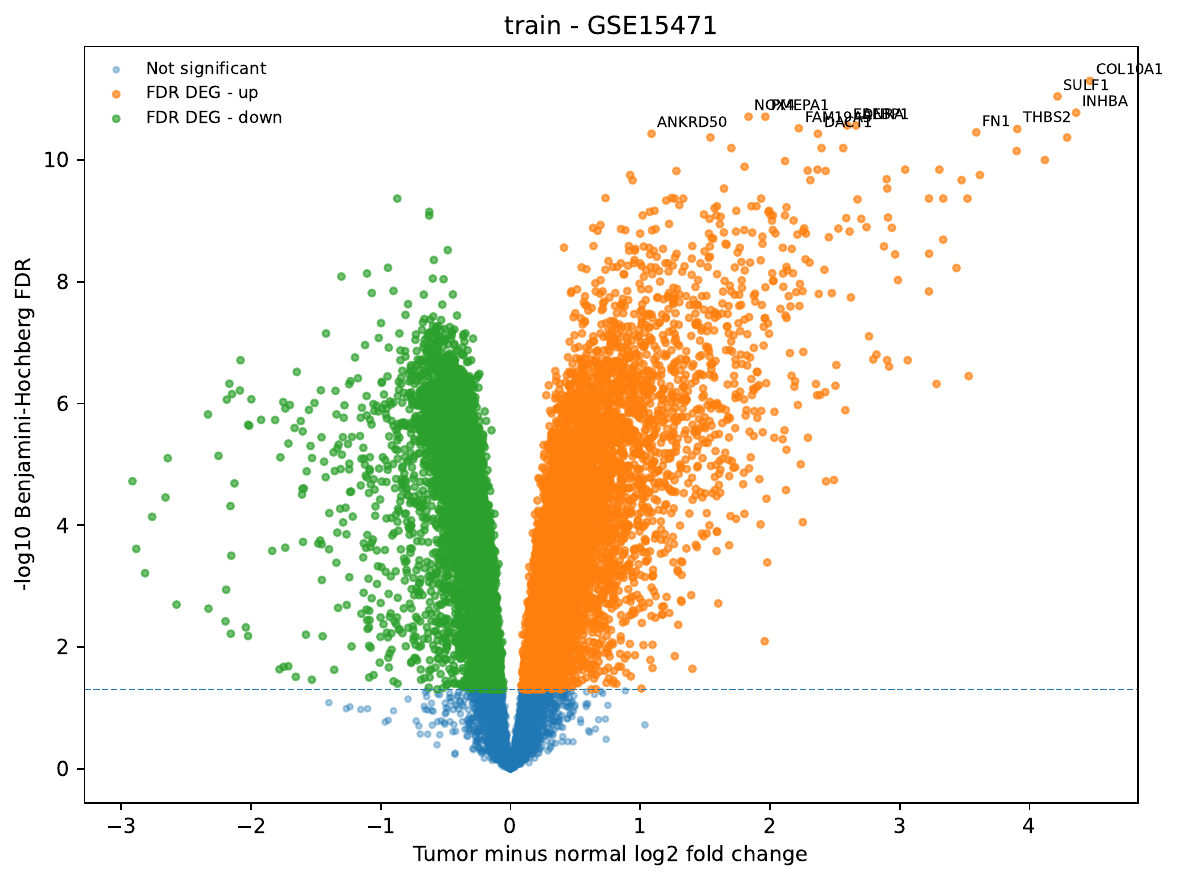}\caption{PDAC discovery}\end{subfigure}
\begin{subfigure}[t]{0.32\textwidth}\includegraphics[width=\textwidth]{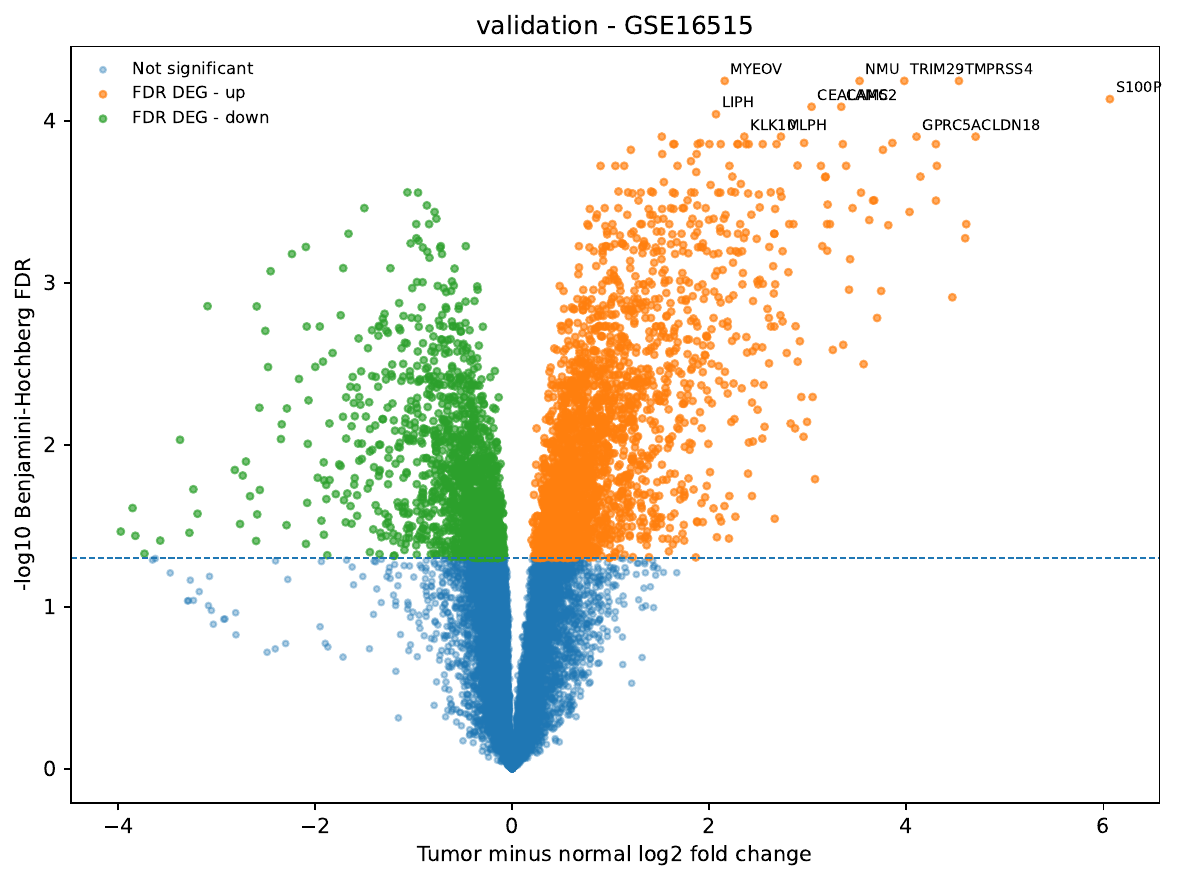}\caption{PDAC validation}\end{subfigure}
\begin{subfigure}[t]{0.32\textwidth}\includegraphics[width=\textwidth]{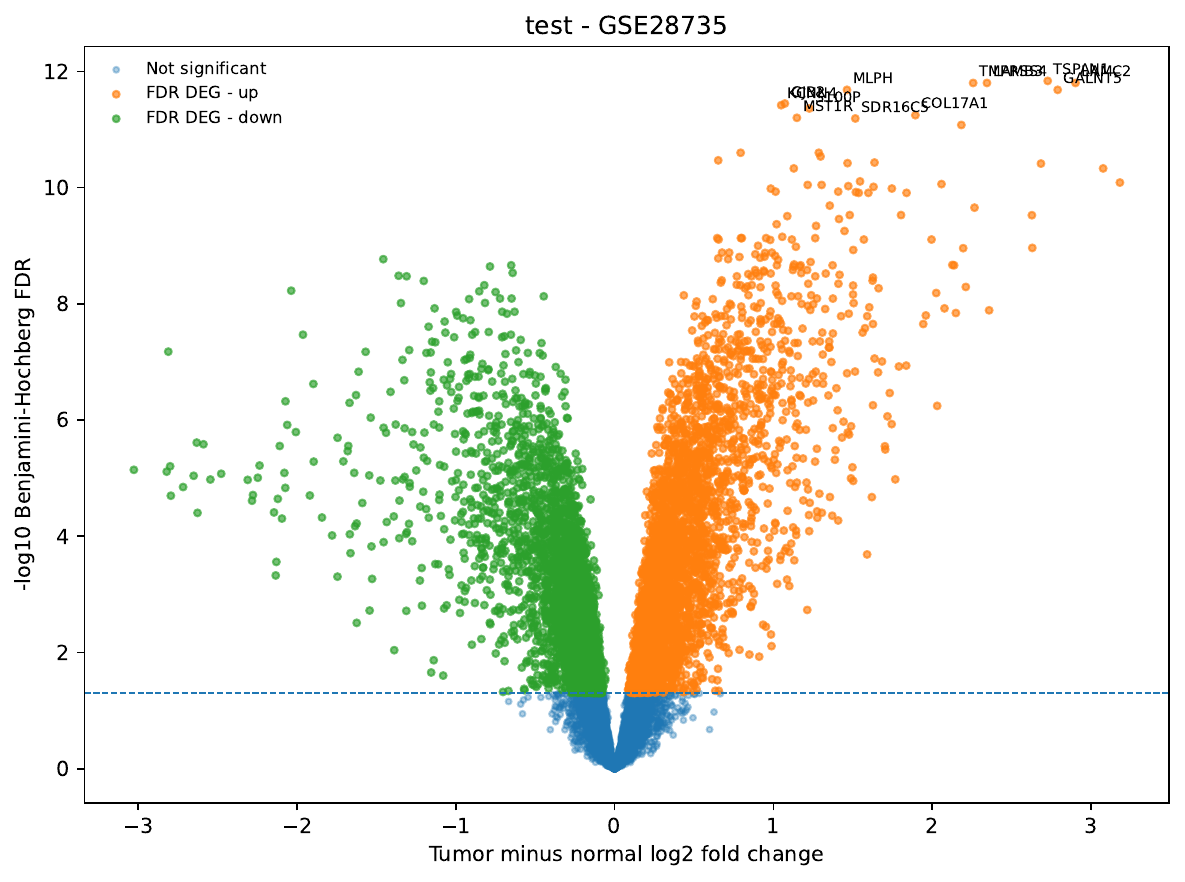}\caption{PDAC external test}\end{subfigure}

\vspace{0.8em}
\begin{subfigure}[t]{0.32\textwidth}\includegraphics[width=\textwidth]{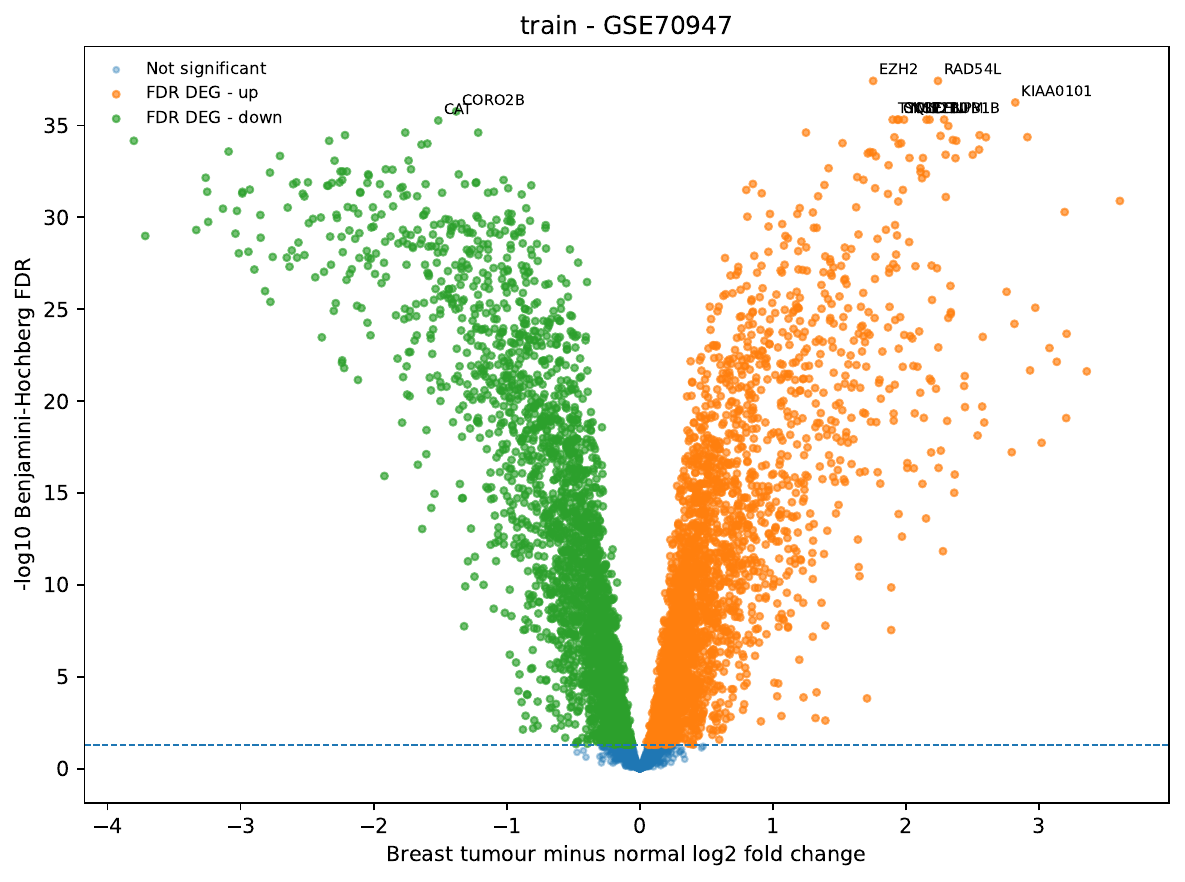}\caption{Breast discovery}\end{subfigure}
\begin{subfigure}[t]{0.32\textwidth}\includegraphics[width=\textwidth]{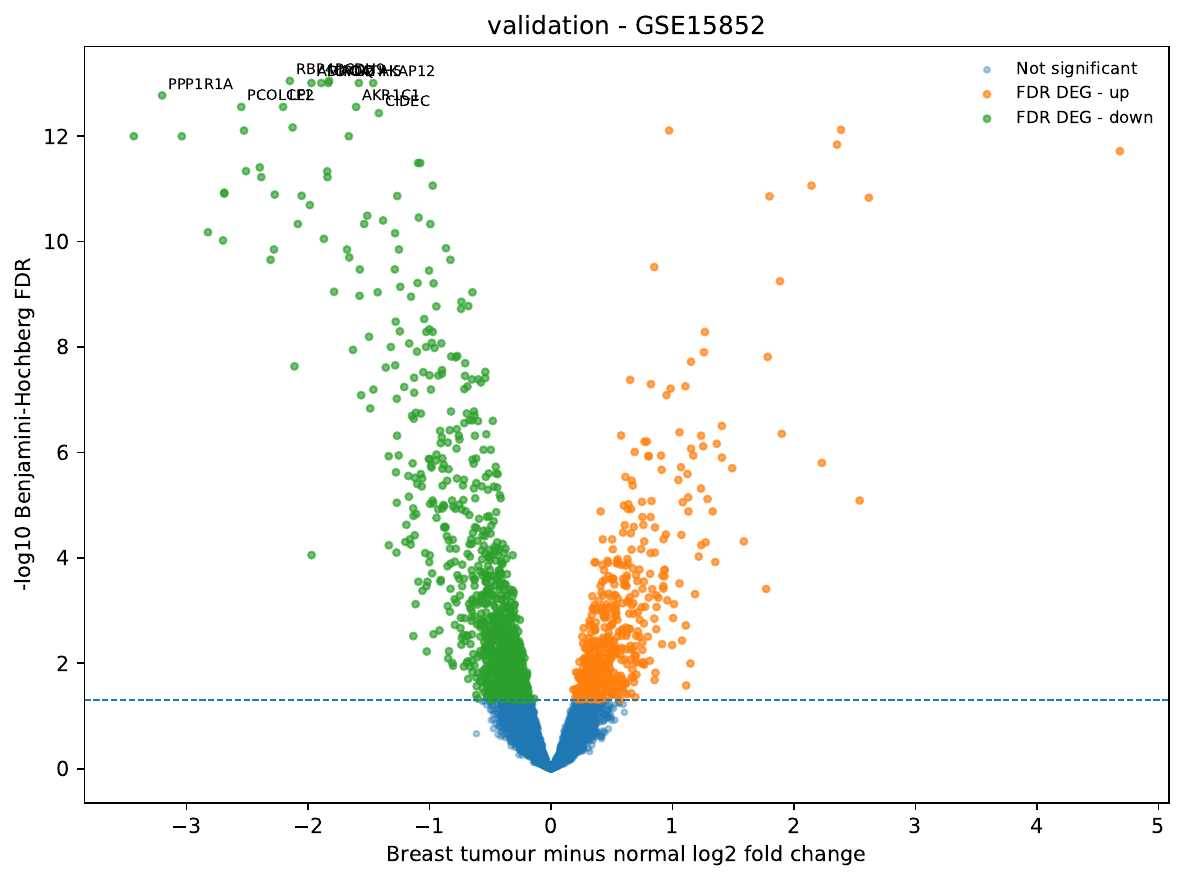}\caption{Breast validation}\end{subfigure}
\begin{subfigure}[t]{0.32\textwidth}\includegraphics[width=\textwidth]{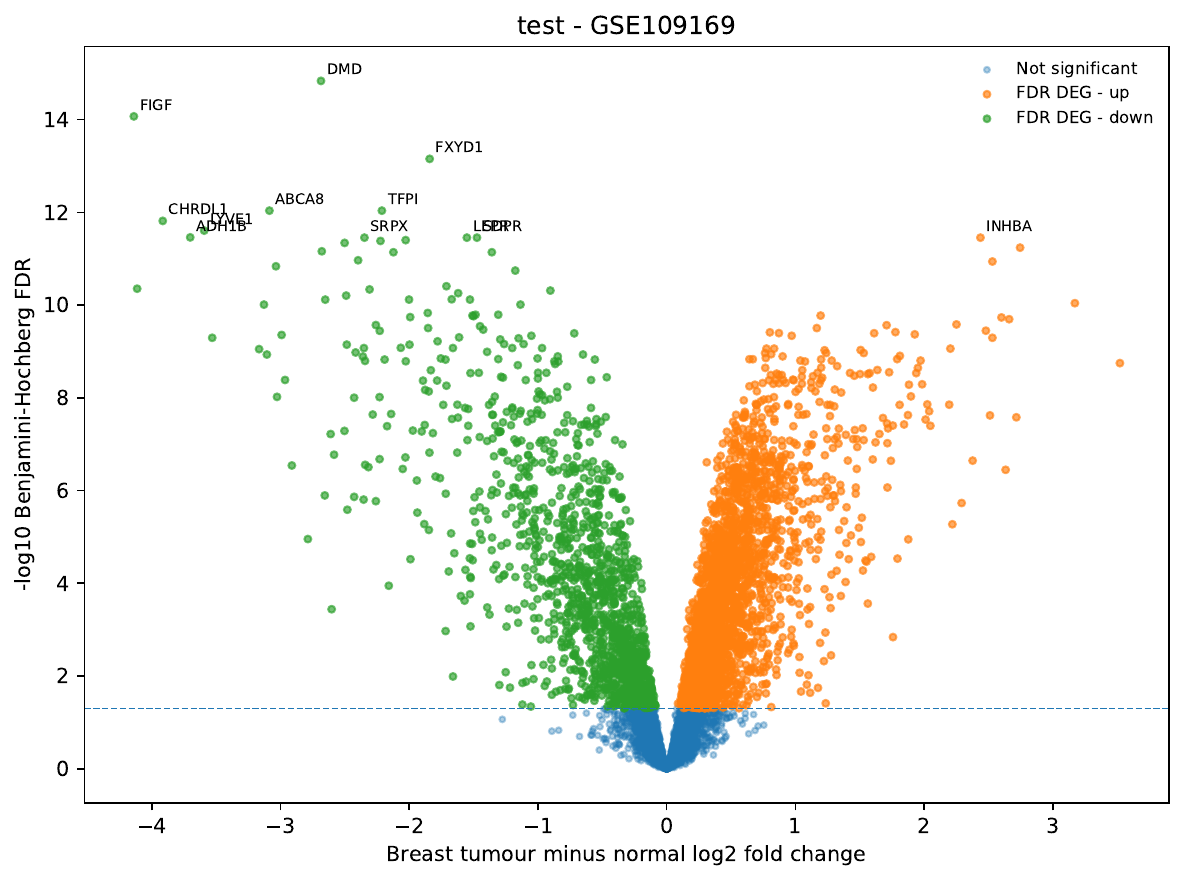}\caption{Breast external test}\end{subfigure}

\vspace{0.8em}
\begin{subfigure}[t]{0.32\textwidth}\includegraphics[width=\textwidth]{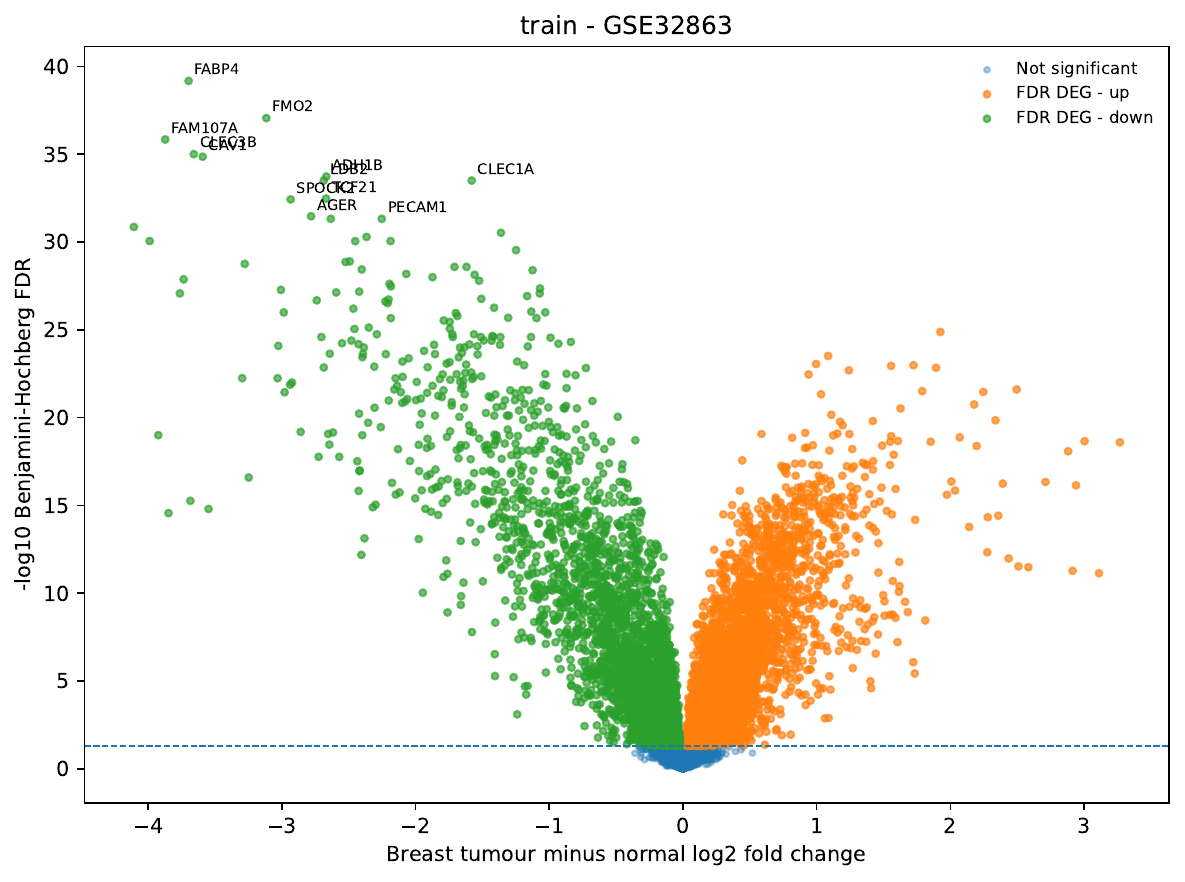}\caption{Lung discovery}\end{subfigure}
\begin{subfigure}[t]{0.32\textwidth}\includegraphics[width=\textwidth]{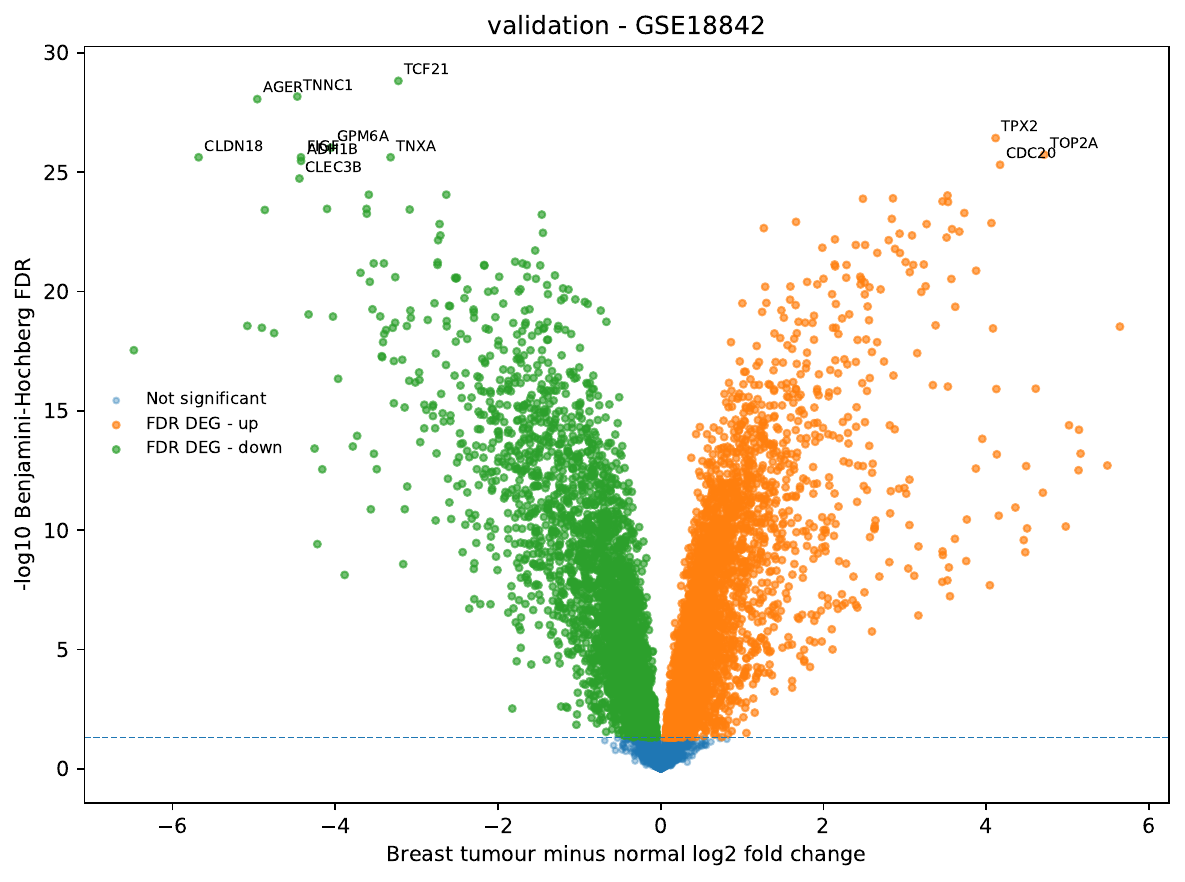}\caption{Lung validation}\end{subfigure}
\begin{subfigure}[t]{0.32\textwidth}\includegraphics[width=\textwidth]{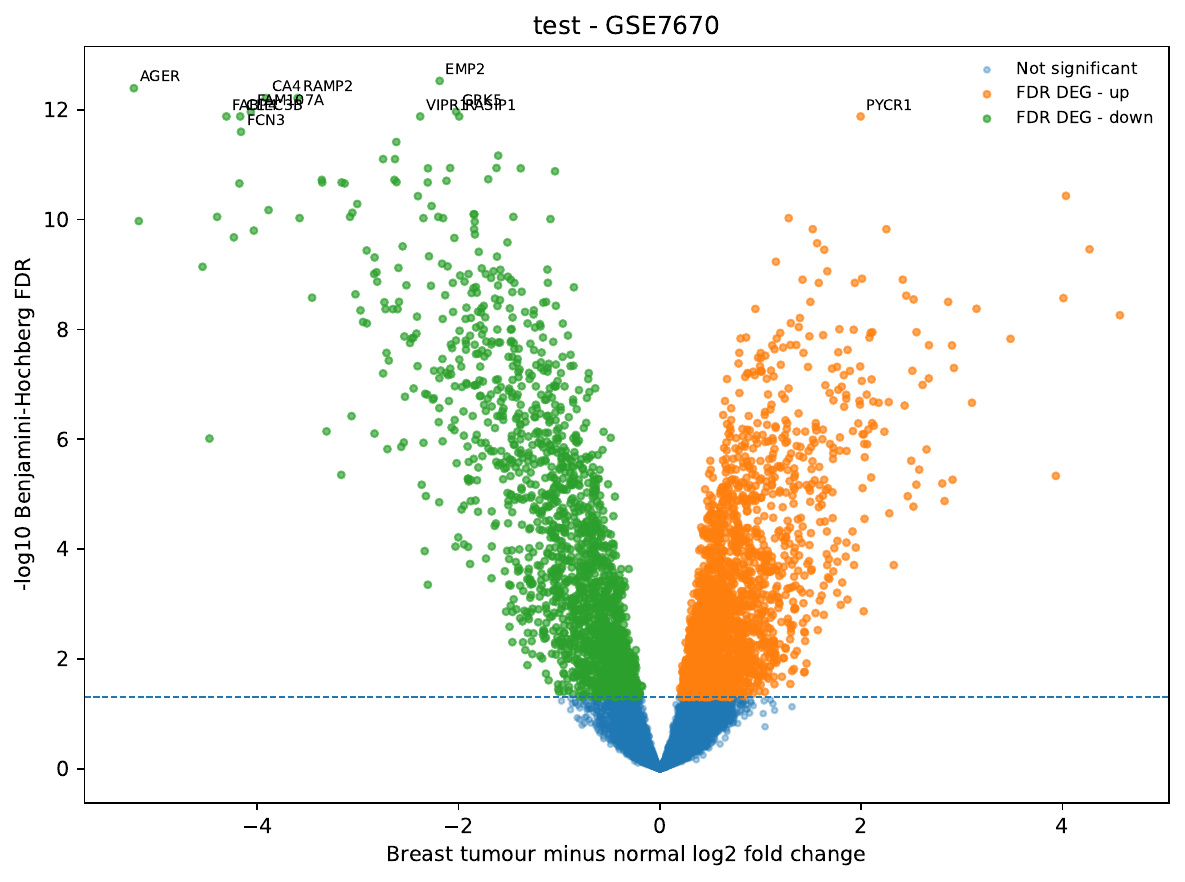}\caption{Lung external test}\end{subfigure}
\caption{\textbf{Volcano plots for all nine paired differential-expression analyses.} Each panel displays cohort-specific tumor-minus-normal effects and statistical evidence.}
\label{fig:s_volcano}
\end{figure}

\clearpage
\subsection*{Supplementary Figure S2: Complete top-k overlap curves}

\begin{figure}[H]
\centering
\includegraphics[width=0.92\textwidth]{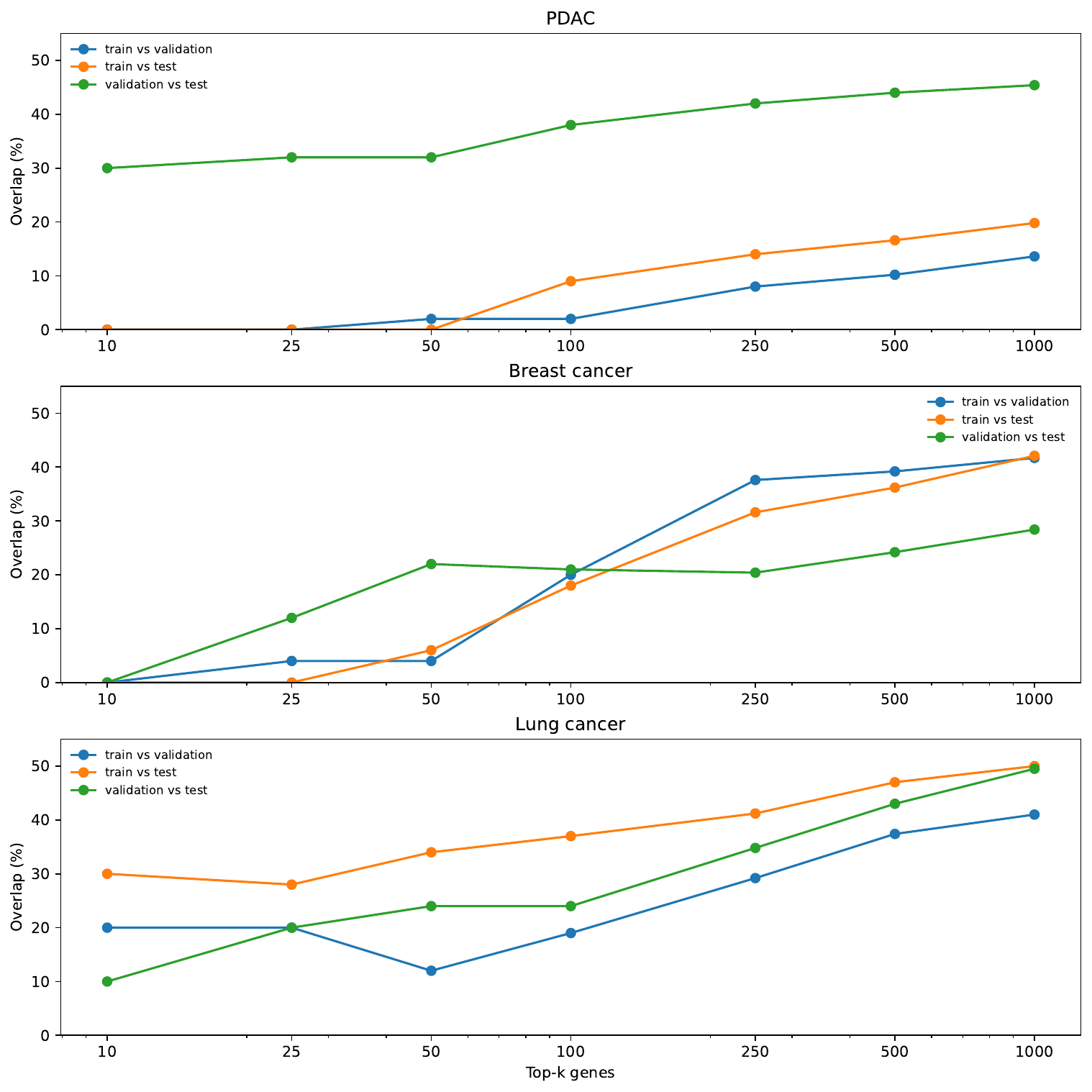}
\caption{\textbf{Top-k ranked-gene overlap for all within-cancer cohort comparisons.} Genes were ordered by absolute moderated t-statistic. Exact leading-rank agreement was weakest at small $k$ and increased as broader portions of the ranking were considered. Dashed reference curves show random-selection expectations where available.}
\label{fig:s_rank}
\end{figure}

\clearpage
\subsection*{Supplementary Figure S3: Disease-specific DEG-set agreement}

\begin{figure}[H]
\centering
\begin{subfigure}[t]{0.32\textwidth}\includegraphics[width=\textwidth]{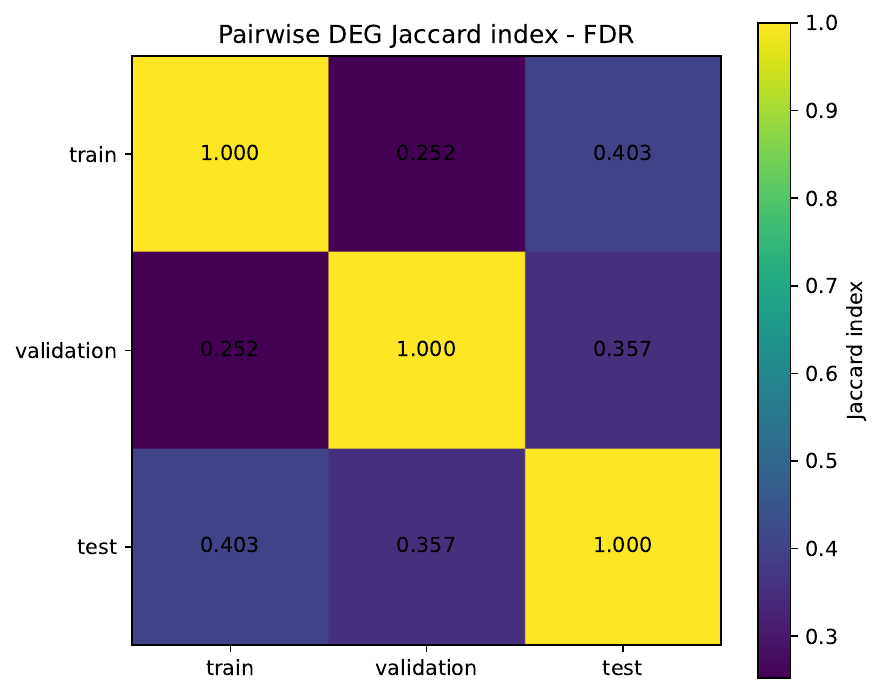}\caption{PDAC, FDR definition}\end{subfigure}
\begin{subfigure}[t]{0.32\textwidth}\includegraphics[width=\textwidth]{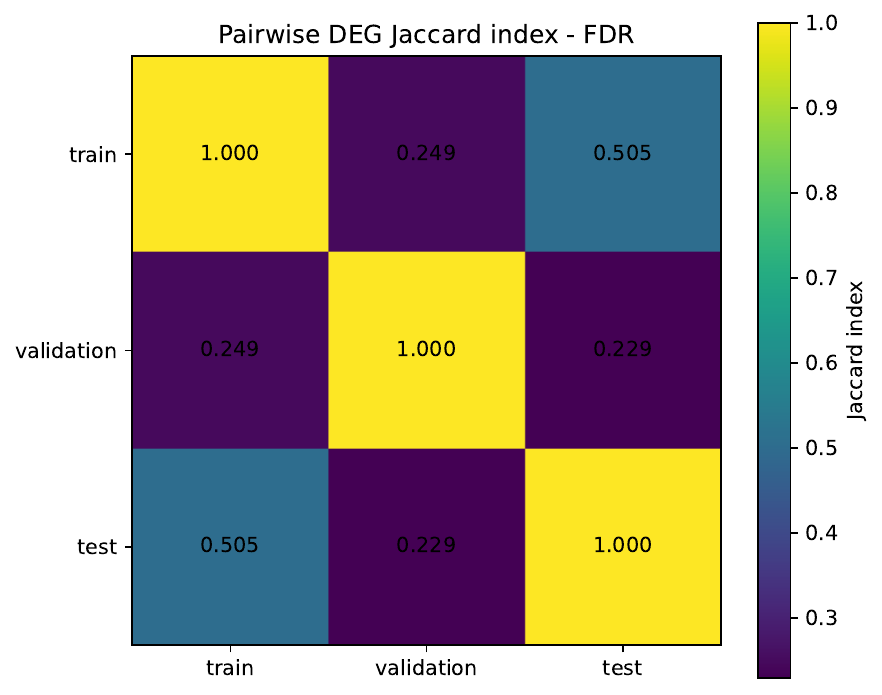}\caption{Breast, FDR definition}\end{subfigure}
\begin{subfigure}[t]{0.32\textwidth}\includegraphics[width=\textwidth]{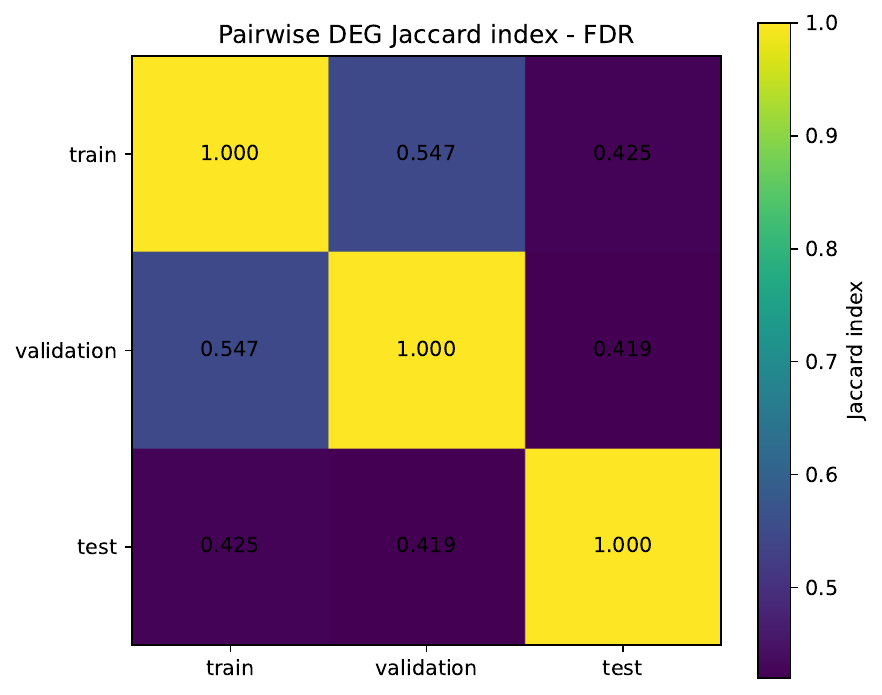}\caption{Lung, FDR definition}\end{subfigure}

\vspace{0.7em}
\begin{subfigure}[t]{0.32\textwidth}\includegraphics[width=\textwidth]{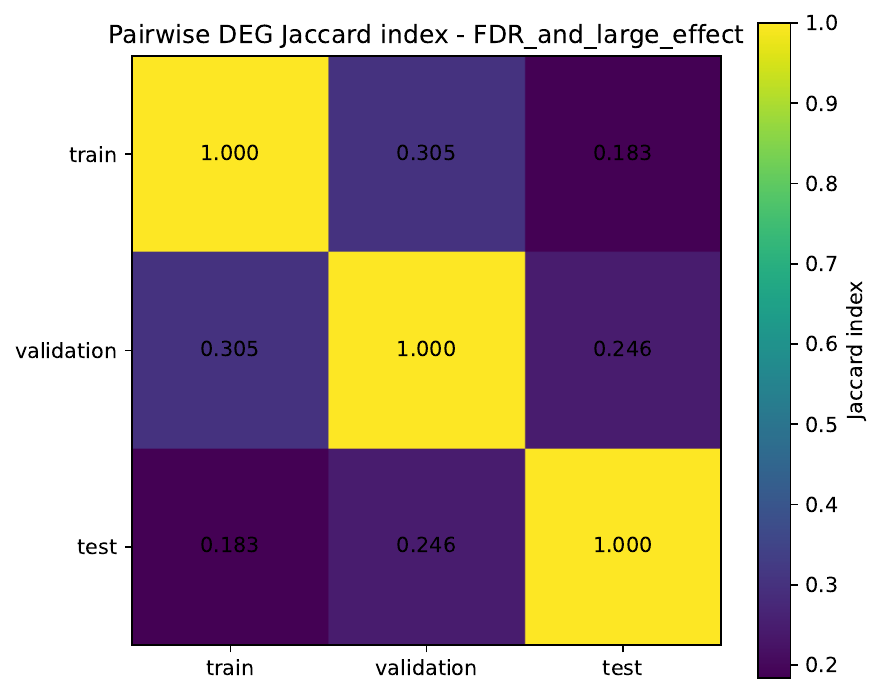}\caption{PDAC, FDR and large effect}\end{subfigure}
\begin{subfigure}[t]{0.32\textwidth}\includegraphics[width=\textwidth]{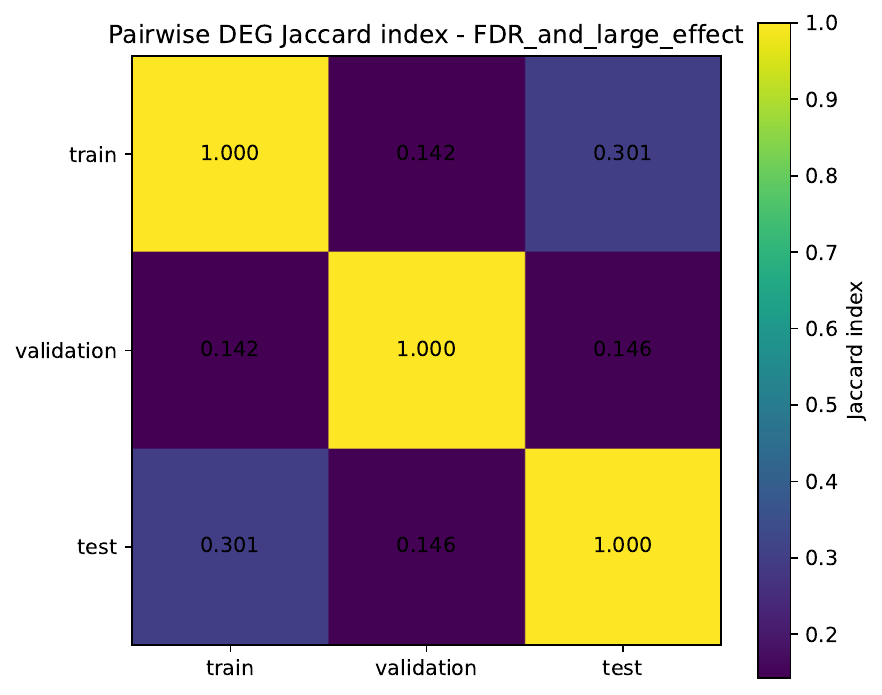}\caption{Breast, FDR and large effect}\end{subfigure}
\begin{subfigure}[t]{0.32\textwidth}\includegraphics[width=\textwidth]{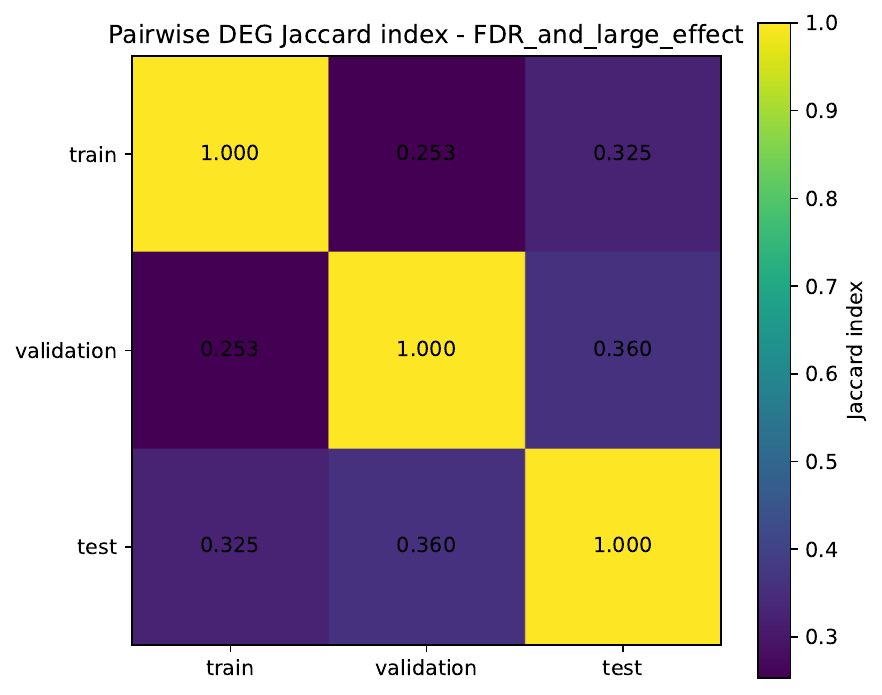}\caption{Lung, FDR and large effect}\end{subfigure}
\caption{\textbf{Exact DEG-set agreement within each cancer.} Heatmaps show pairwise Jaccard indices for broad FDR lists and for the secondary subset additionally requiring $|\log_2\mathrm{FC}|\geq1$. Effect filtering did not produce uniformly high exact-list agreement.}
\label{fig:s_threshold}
\end{figure}

\clearpage
\begin{landscape}
\subsection*{Supplementary Figure S4: Disease-specific Hallmark reproducibility}

\begin{figure}[H]
\centering
\begin{subfigure}[t]{0.32\linewidth}\includegraphics[width=\textwidth]{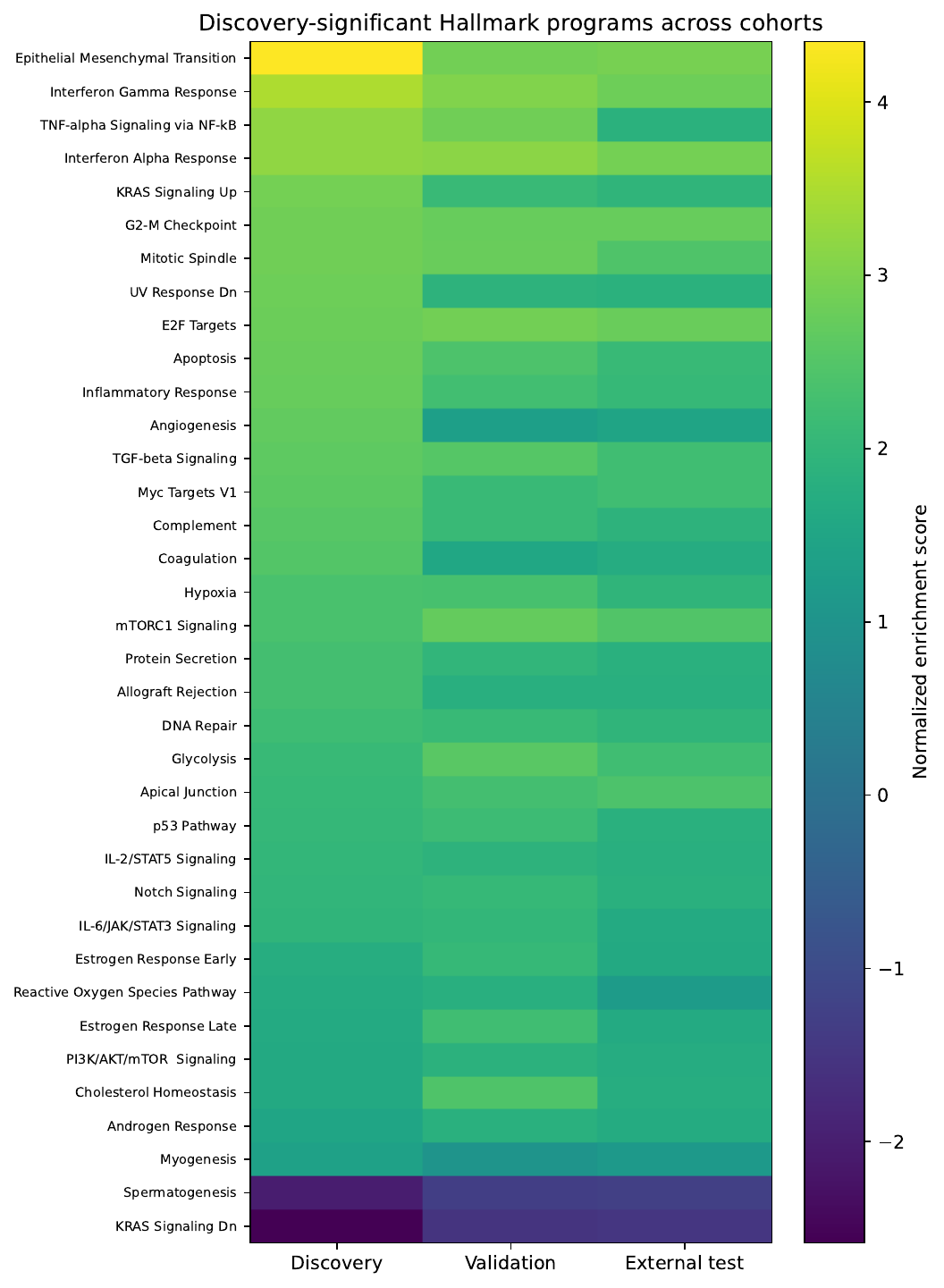}\caption{PDAC}\end{subfigure}
\begin{subfigure}[t]{0.32\linewidth}\includegraphics[width=\textwidth]{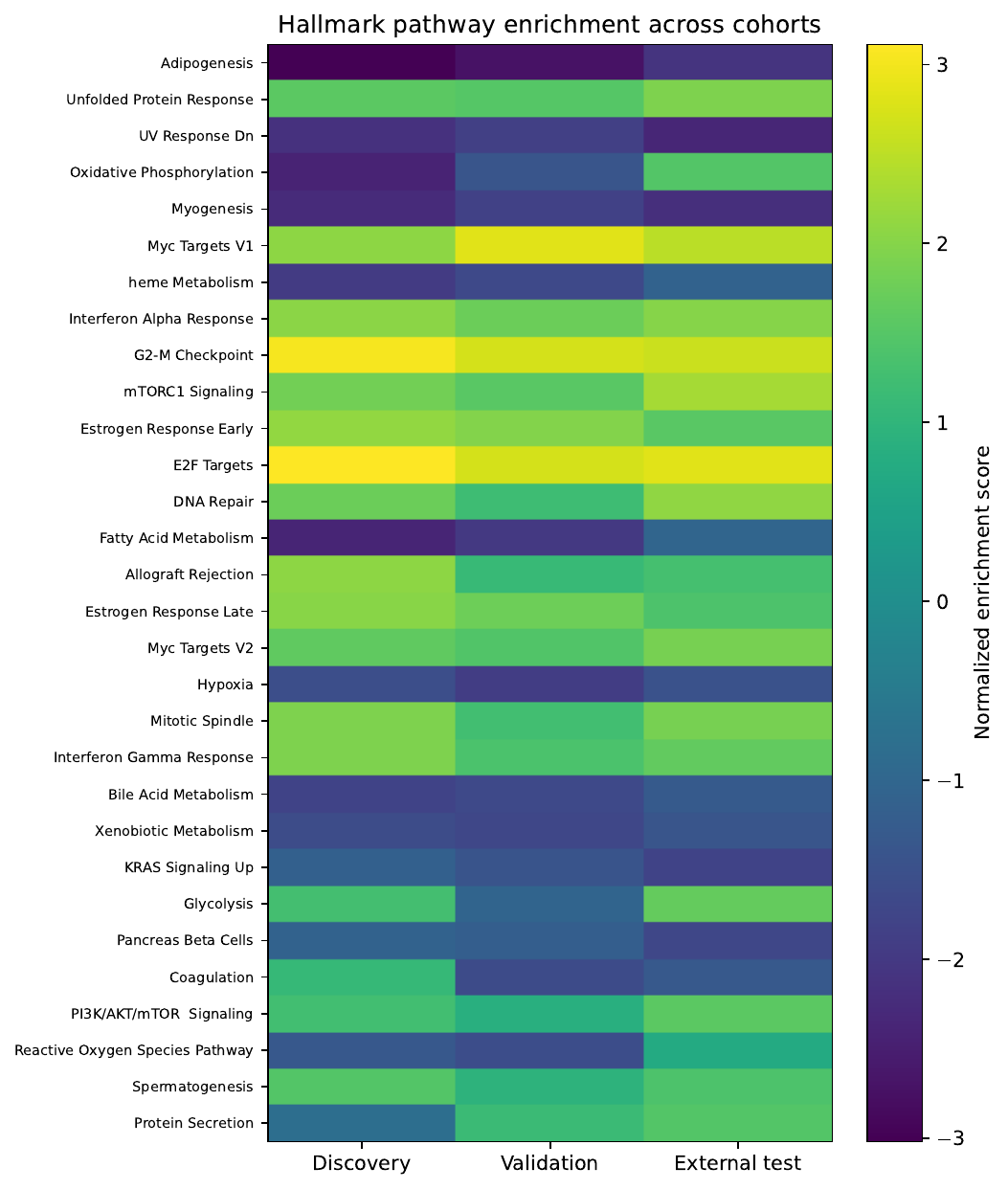}\caption{Breast cancer}\end{subfigure}
\begin{subfigure}[t]{0.32\linewidth}\includegraphics[width=\textwidth]{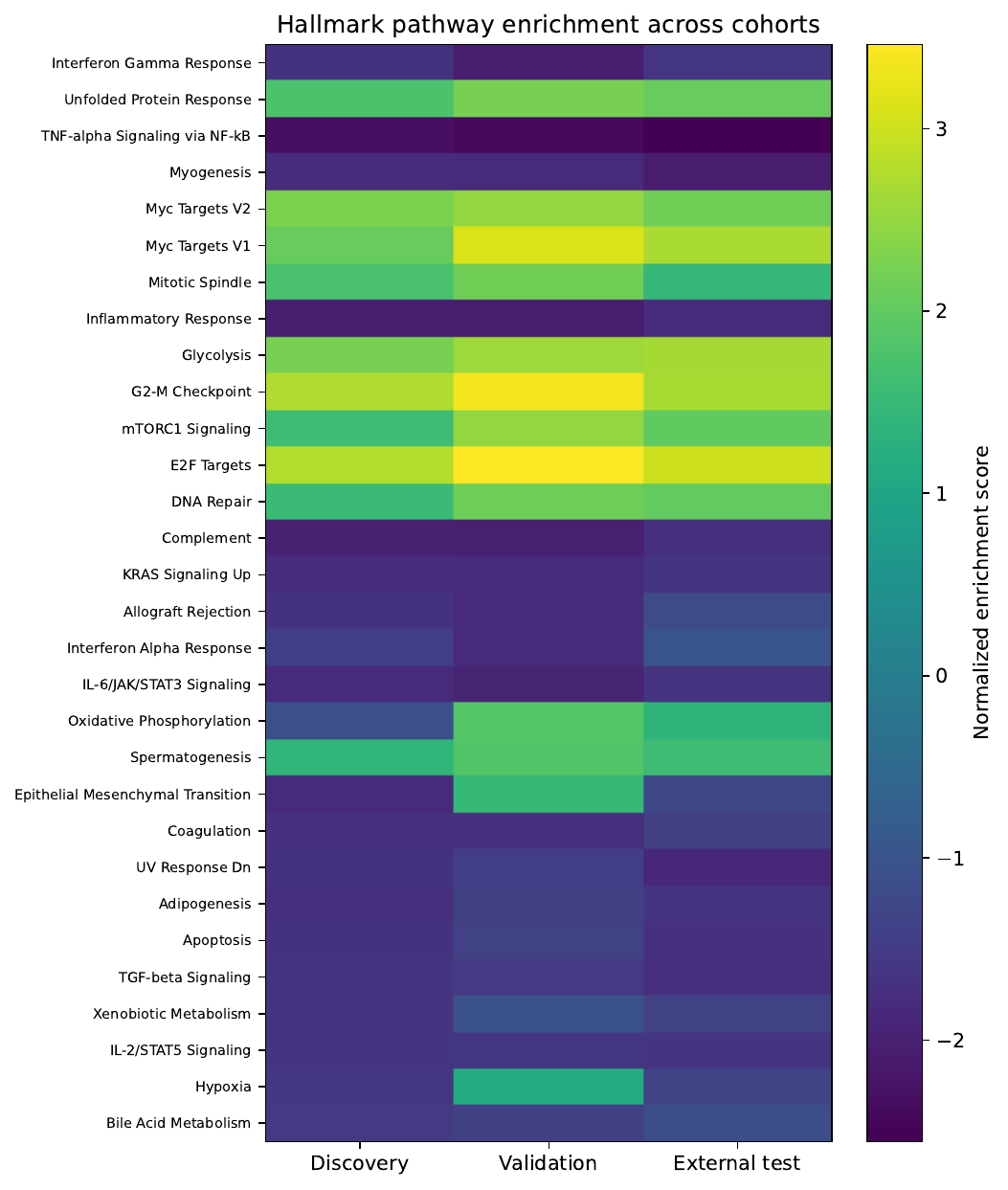}\caption{Lung cancer}\end{subfigure}
\caption{\textbf{Disease-specific Hallmark normalized enrichment scores.} Pathway patterns were reproducible within each cancer while several programs differed in direction across cancers.}
\label{fig:s_gsea}
\end{figure}
\end{landscape}

\clearpage
\subsection*{Supplementary Figure S5: Discovery-leading genes across independent cohorts}

\begin{figure}[H]
\centering
\begin{subfigure}[t]{0.32\textwidth}\includegraphics[width=\textwidth]{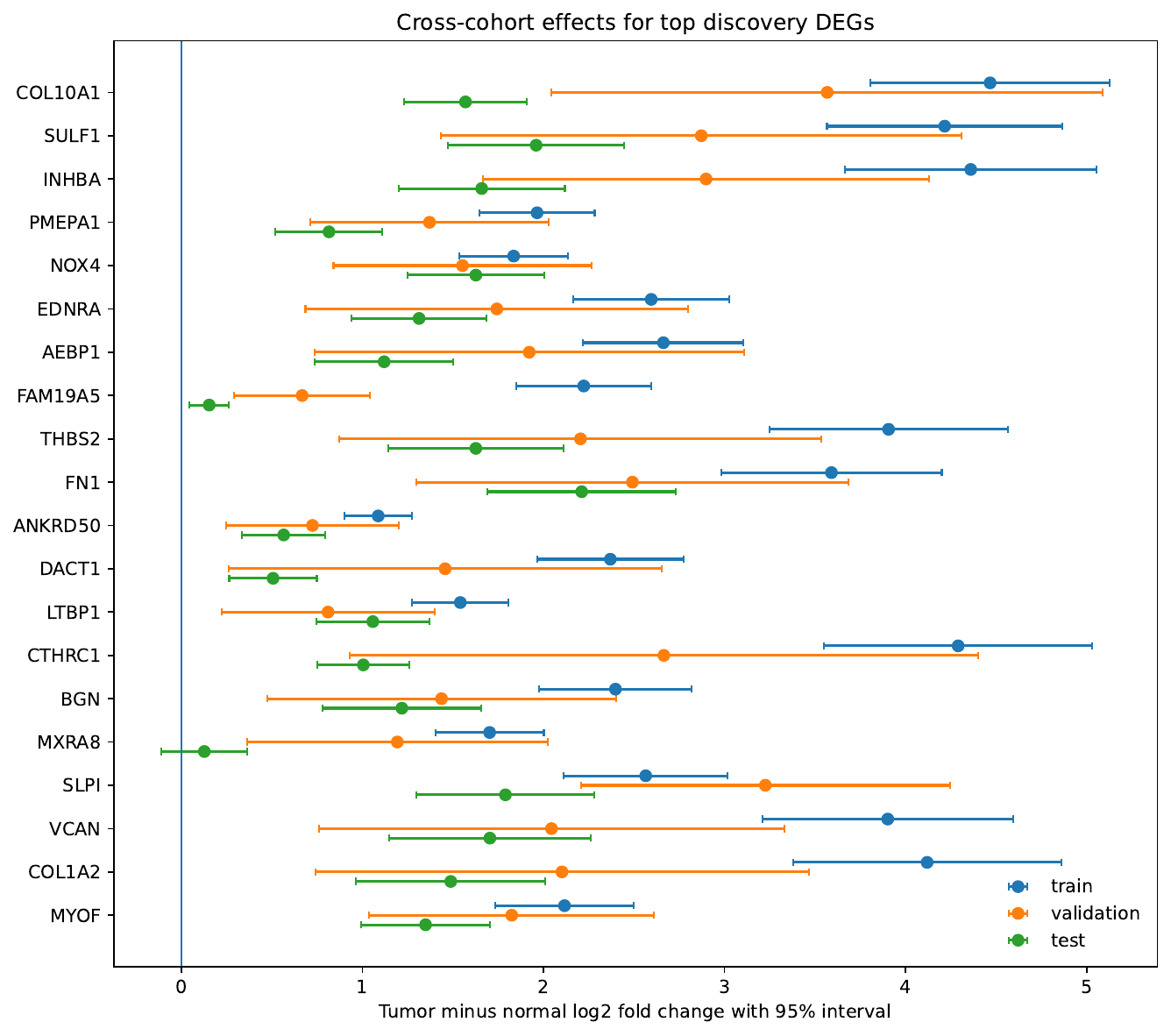}\caption{PDAC}\end{subfigure}
\begin{subfigure}[t]{0.32\textwidth}\includegraphics[width=\textwidth]{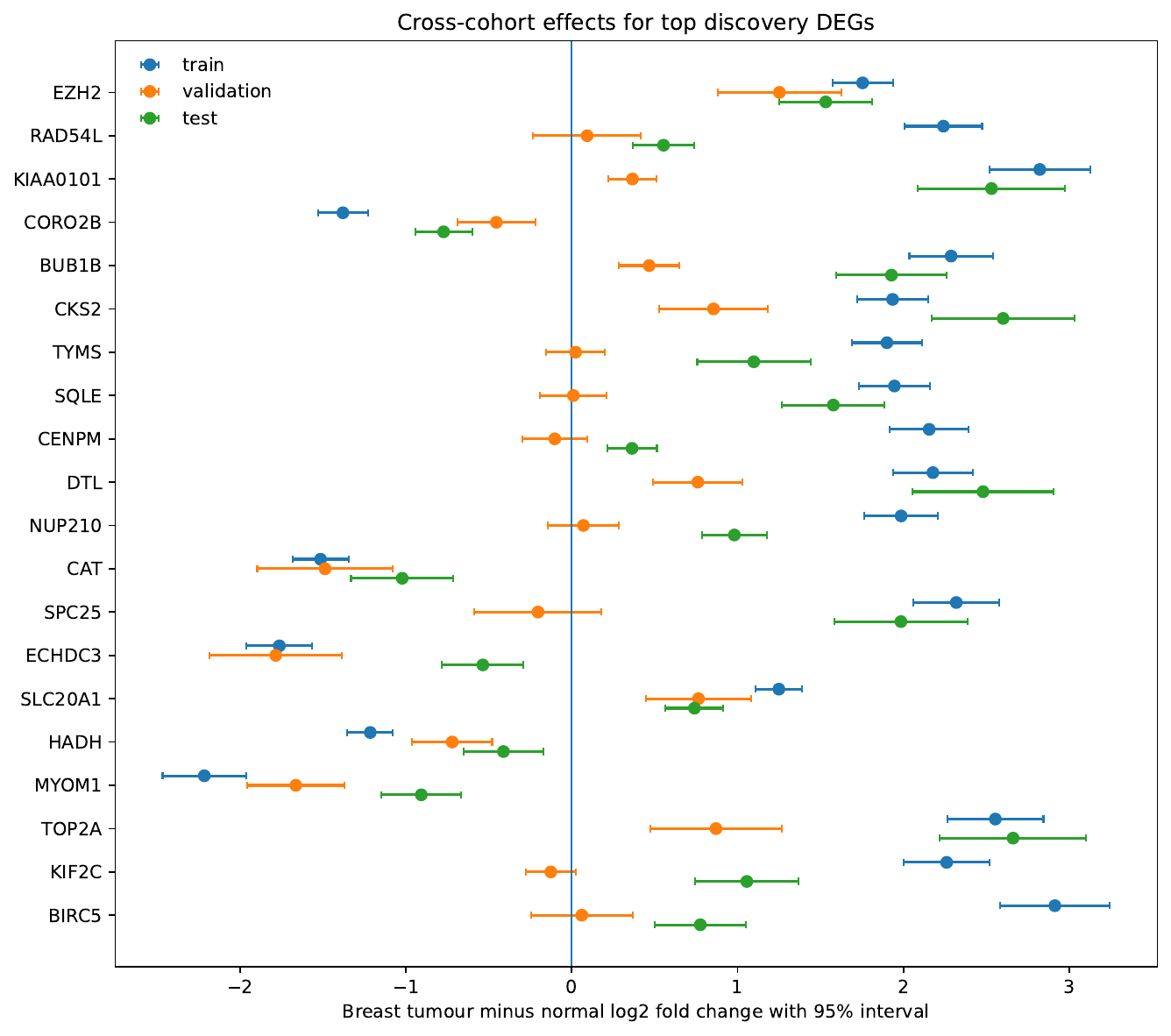}\caption{Breast cancer}\end{subfigure}
\begin{subfigure}[t]{0.32\textwidth}\includegraphics[width=\textwidth]{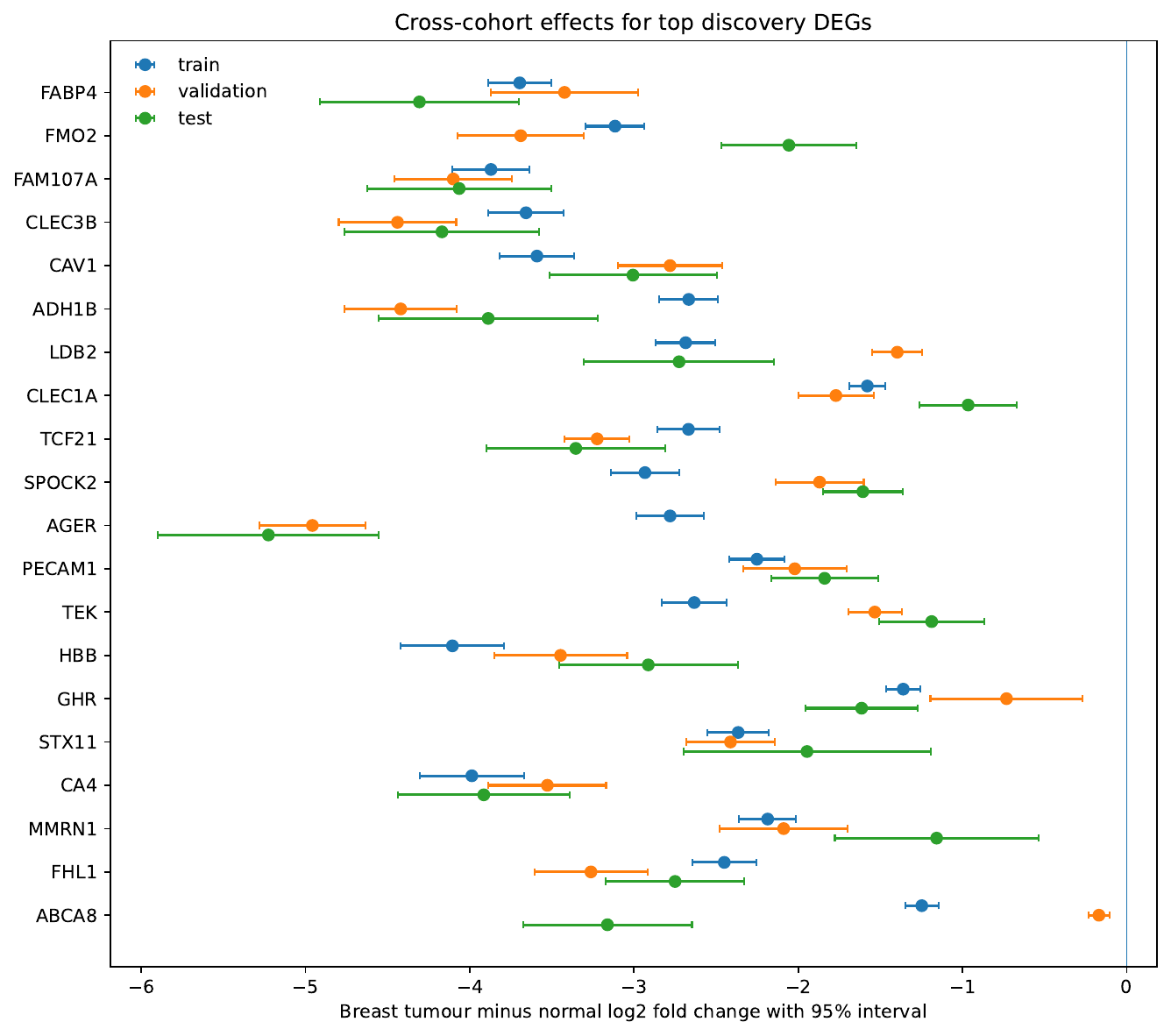}\caption{Lung cancer}\end{subfigure}
\caption{\textbf{Cross-cohort effects for leading discovery DEGs.} Forest plots display cohort-specific effect estimates and uncertainty for genes prioritized by the discovery analysis. Biological plausibility and strong discovery rank did not guarantee identical rank or magnitude in independent cohorts.}
\label{fig:s_forest}
\end{figure}

\clearpage
\begin{landscape}
\subsection*{Supplementary Figure S6: Disease-specific REDE-2Fold strategy comparisons}

\begin{figure}[H]
\centering
\begin{subfigure}[t]{0.32\linewidth}\includegraphics[width=\textwidth]{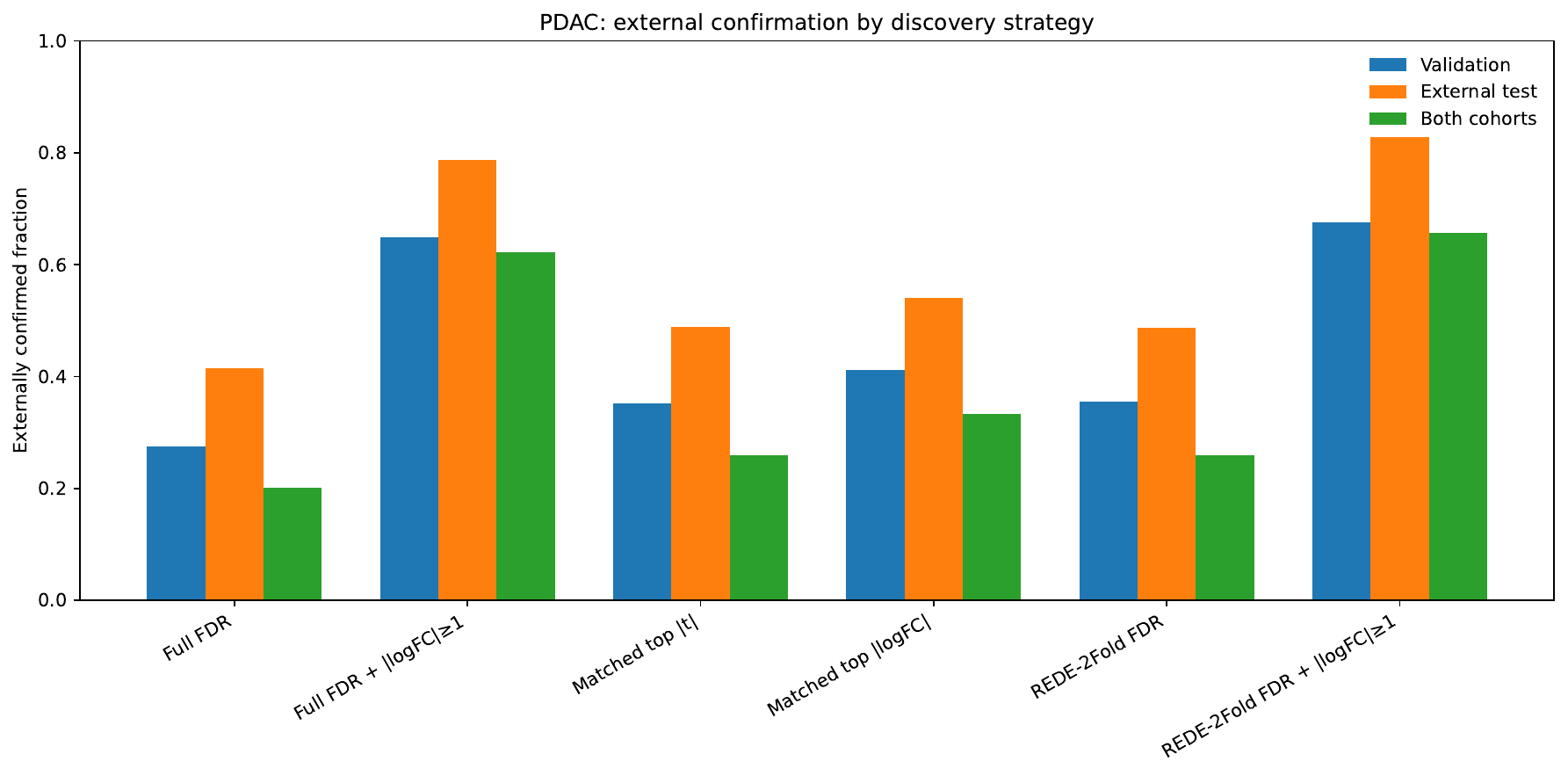}\caption{PDAC}\end{subfigure}
\begin{subfigure}[t]{0.32\linewidth}\includegraphics[width=\textwidth]{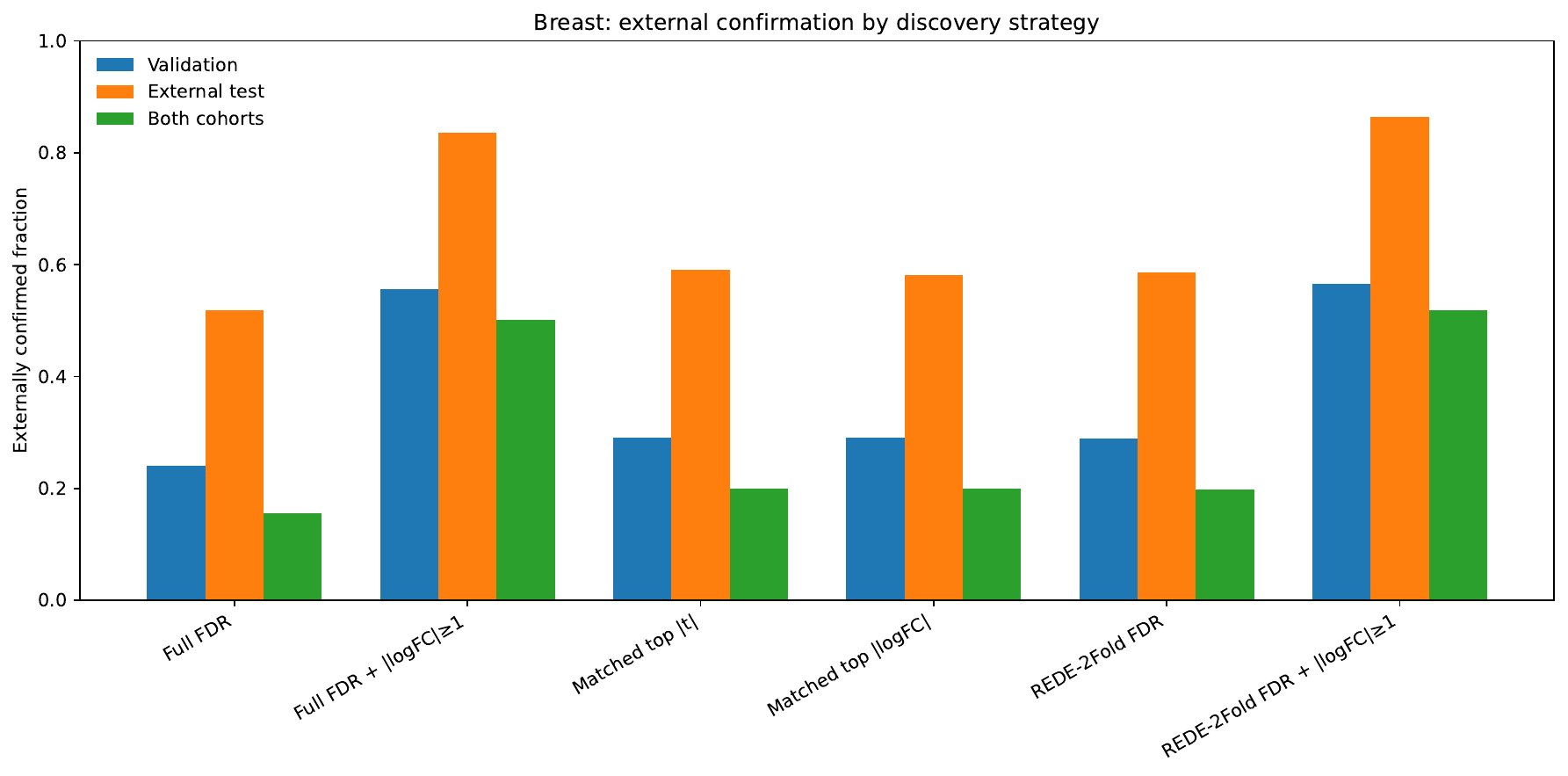}\caption{Breast cancer}\end{subfigure}
\begin{subfigure}[t]{0.32\linewidth}\includegraphics[width=\textwidth]{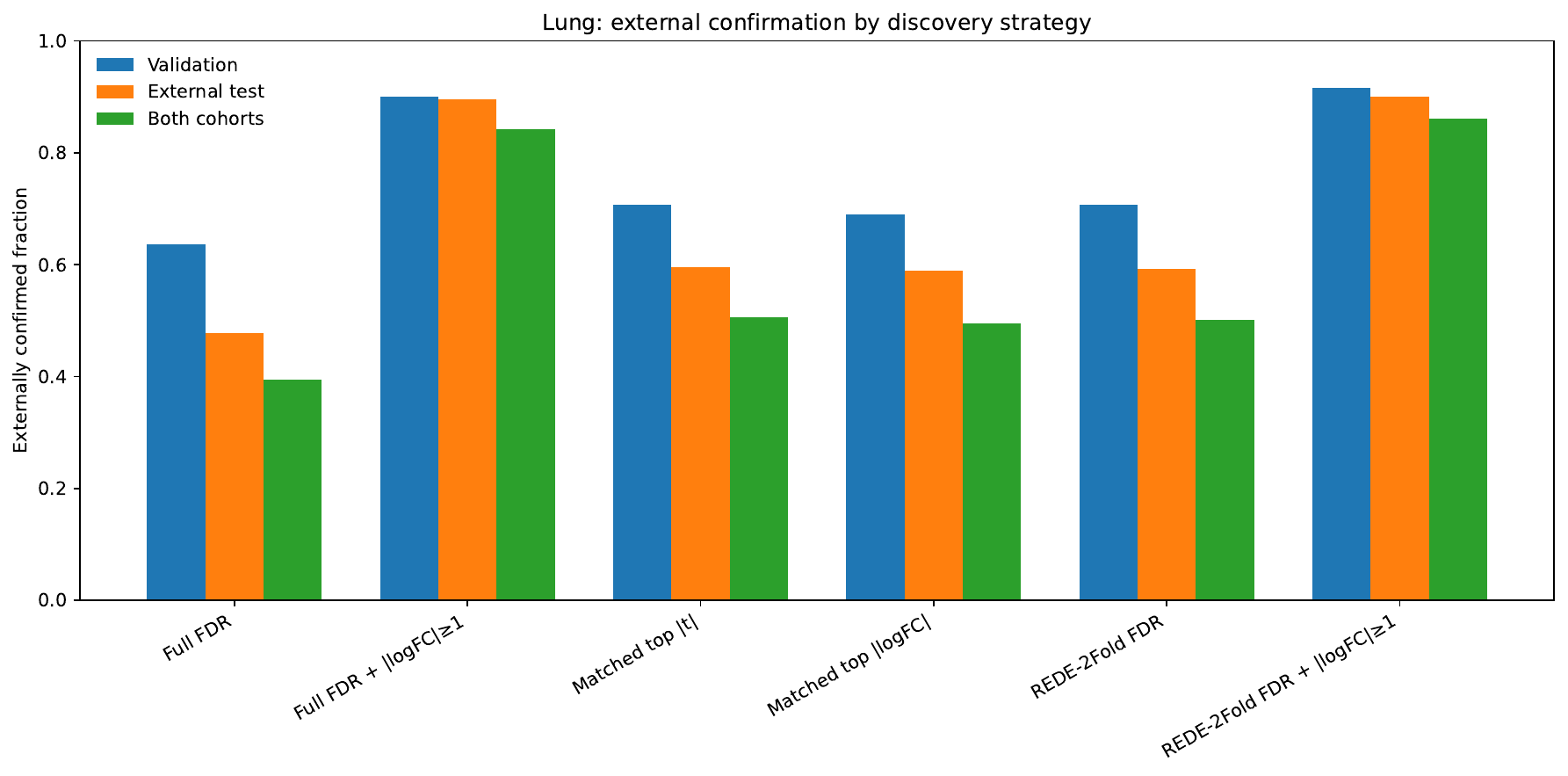}\caption{Lung cancer}\end{subfigure}
\caption{\textbf{Validation, external-test, and joint confirmation by discovery strategy.} Each panel compares the complete discovery FDR list, the large-effect subset, matched-size full-cohort evidence rankings, and the corresponding \redetwofold intersections.}
\label{fig:s_rede2fold}
\end{figure}
\end{landscape}

\clearpage
\subsection*{Supplementary Figure S7: Individual compact-panel gene discrimination}

\begin{figure}[H]
\centering
\begin{subfigure}[t]{0.32\textwidth}\includegraphics[width=\textwidth]{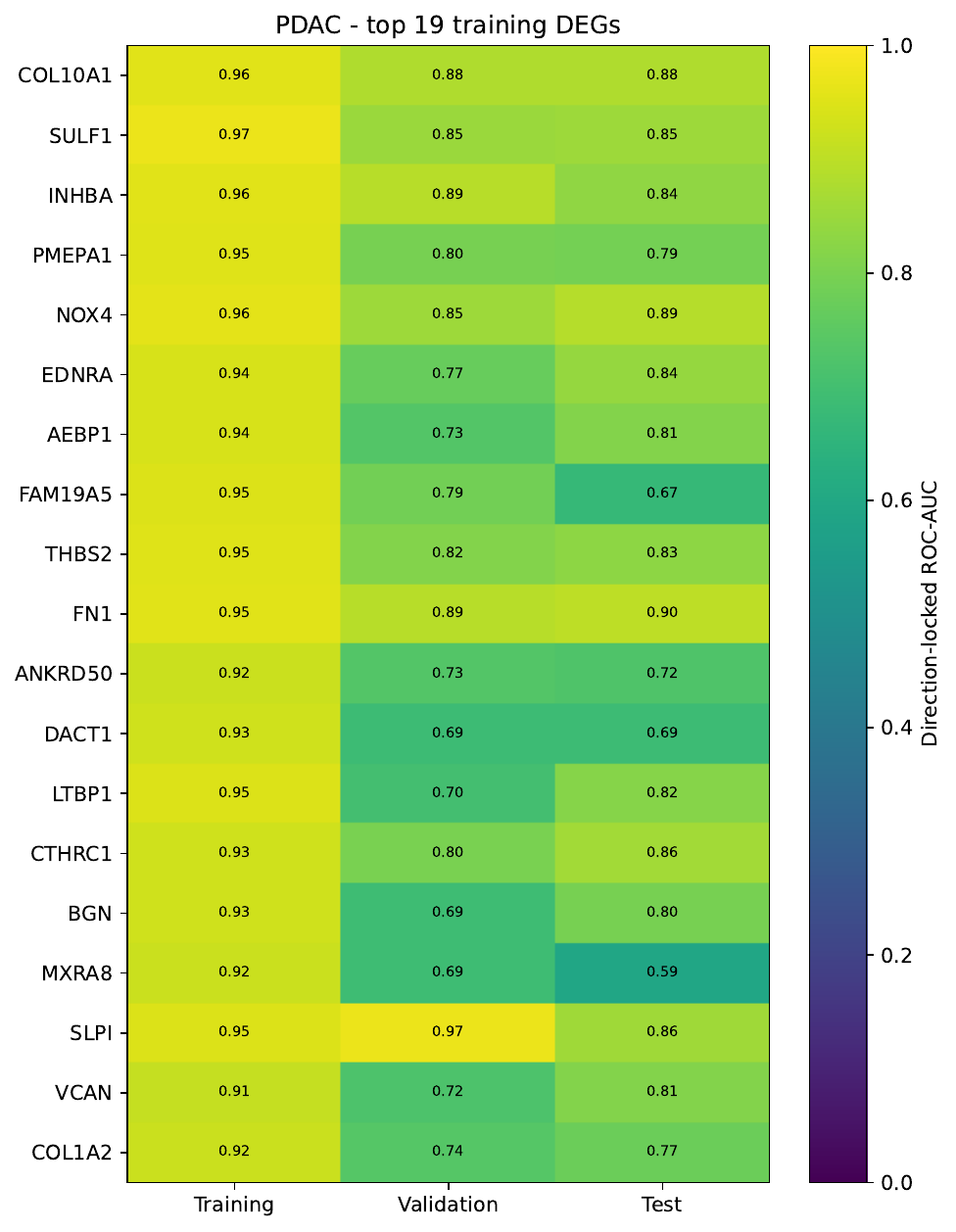}\caption{PDAC}\end{subfigure}
\begin{subfigure}[t]{0.32\textwidth}\includegraphics[width=\textwidth]{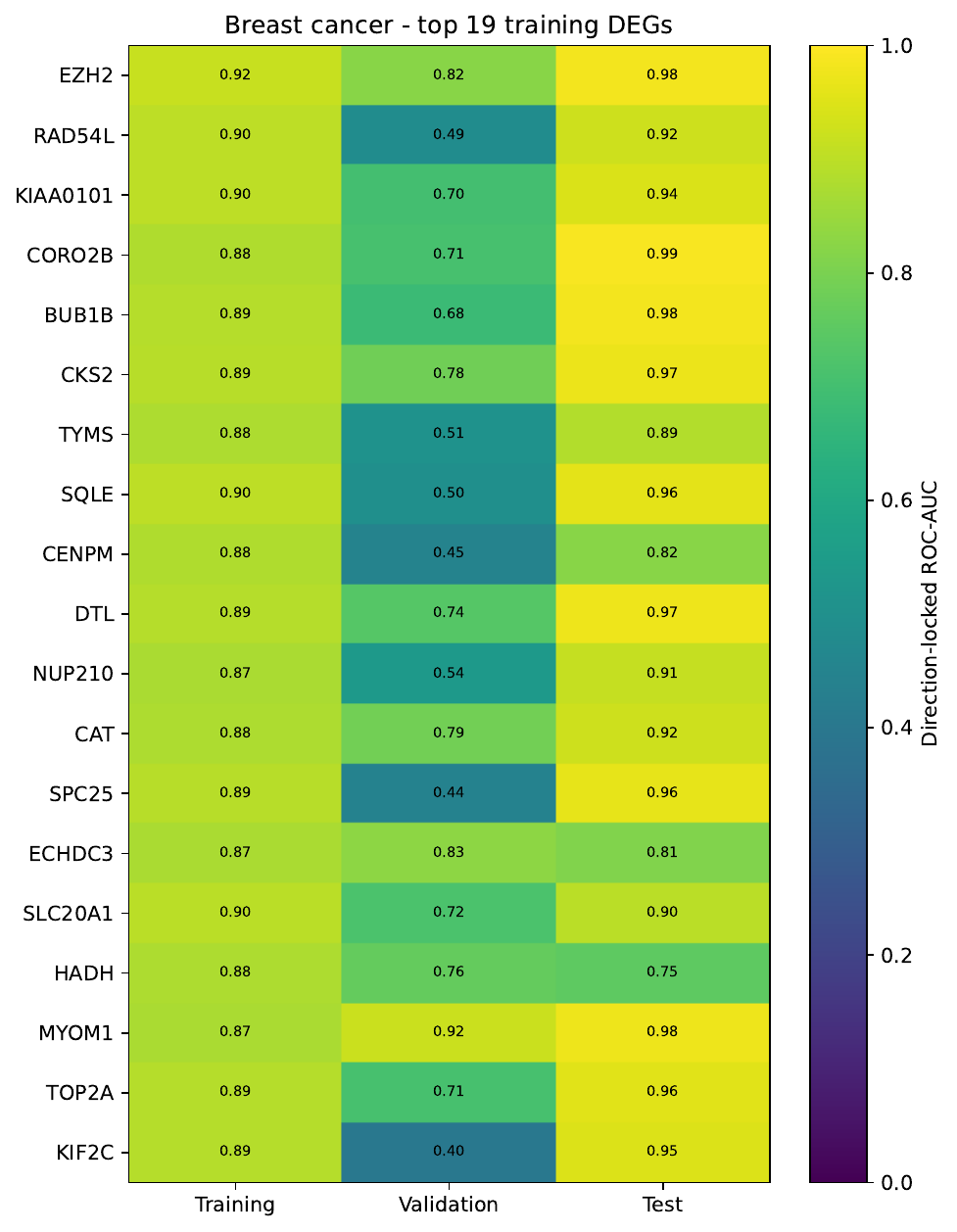}\caption{Breast cancer}\end{subfigure}
\begin{subfigure}[t]{0.32\textwidth}\includegraphics[width=\textwidth]{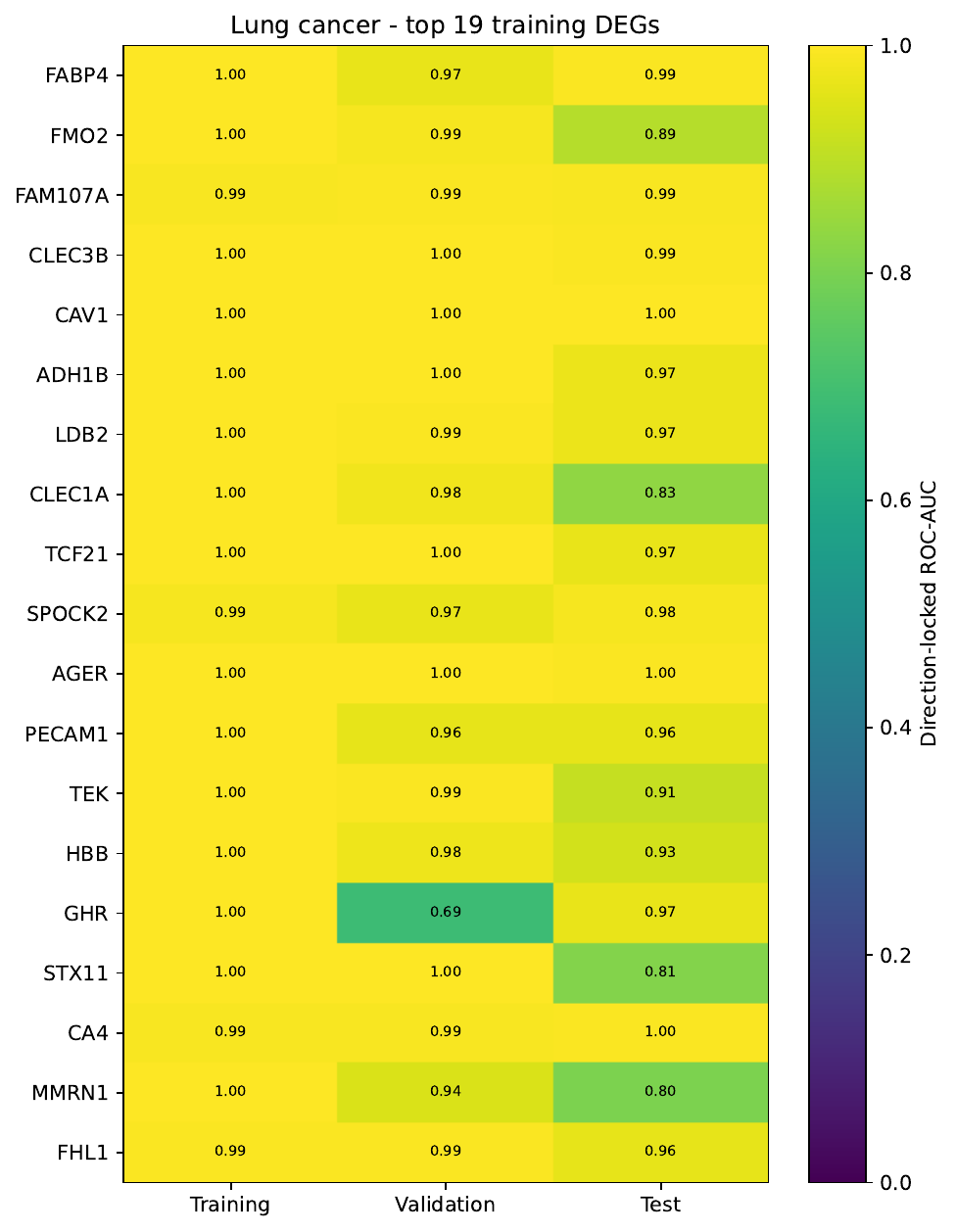}\caption{Lung cancer}\end{subfigure}
\caption{\textbf{ROC-AUC of individual genes belonging to either training-derived compact panel.} Gene orientation was fixed by the discovery effect direction. The heatmaps show that individual-gene discrimination and training differential-expression rank are related but not equivalent.}
\label{fig:s_individual_diagnostic}
\end{figure}

\clearpage
\begin{landscape}
\subsection*{Supplementary Figure S8: External ROC curves for all panel strategies}

\begin{figure}[H]
\centering
\begin{subfigure}[t]{0.32\linewidth}\includegraphics[width=\textwidth]{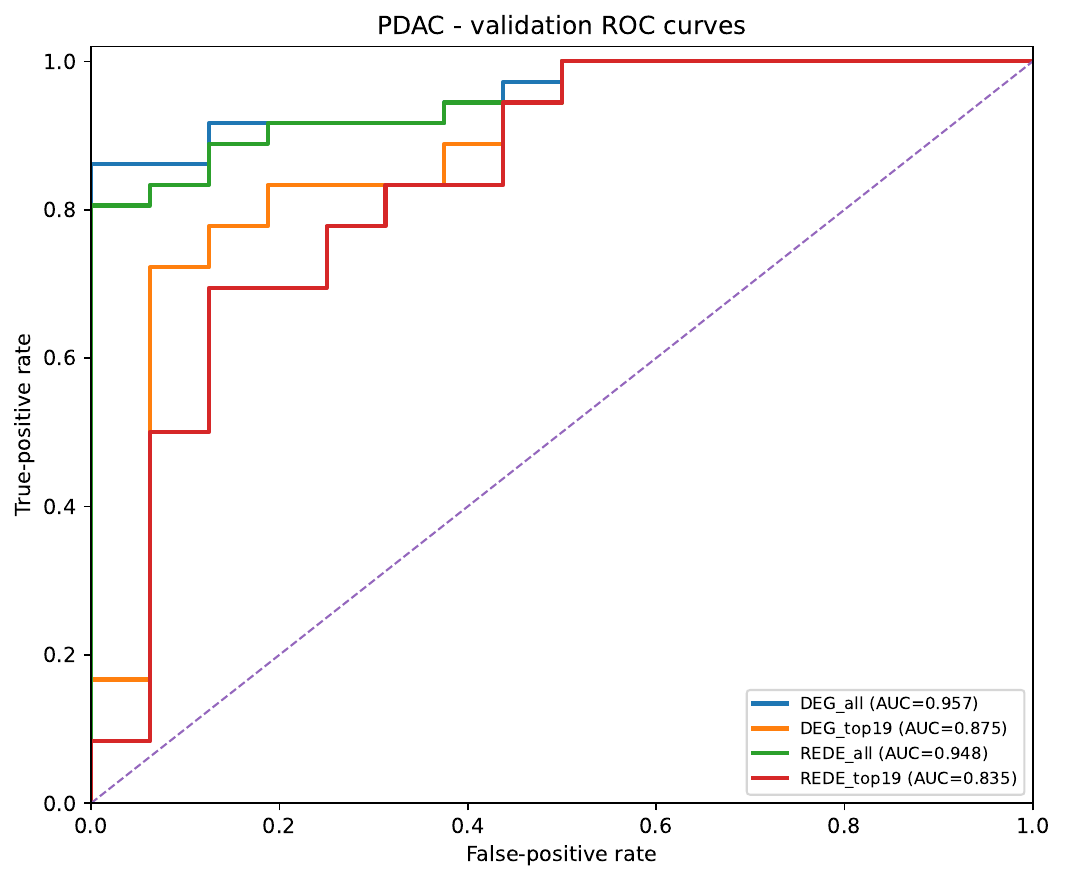}\caption{PDAC validation}\end{subfigure}
\begin{subfigure}[t]{0.32\linewidth}\includegraphics[width=\textwidth]{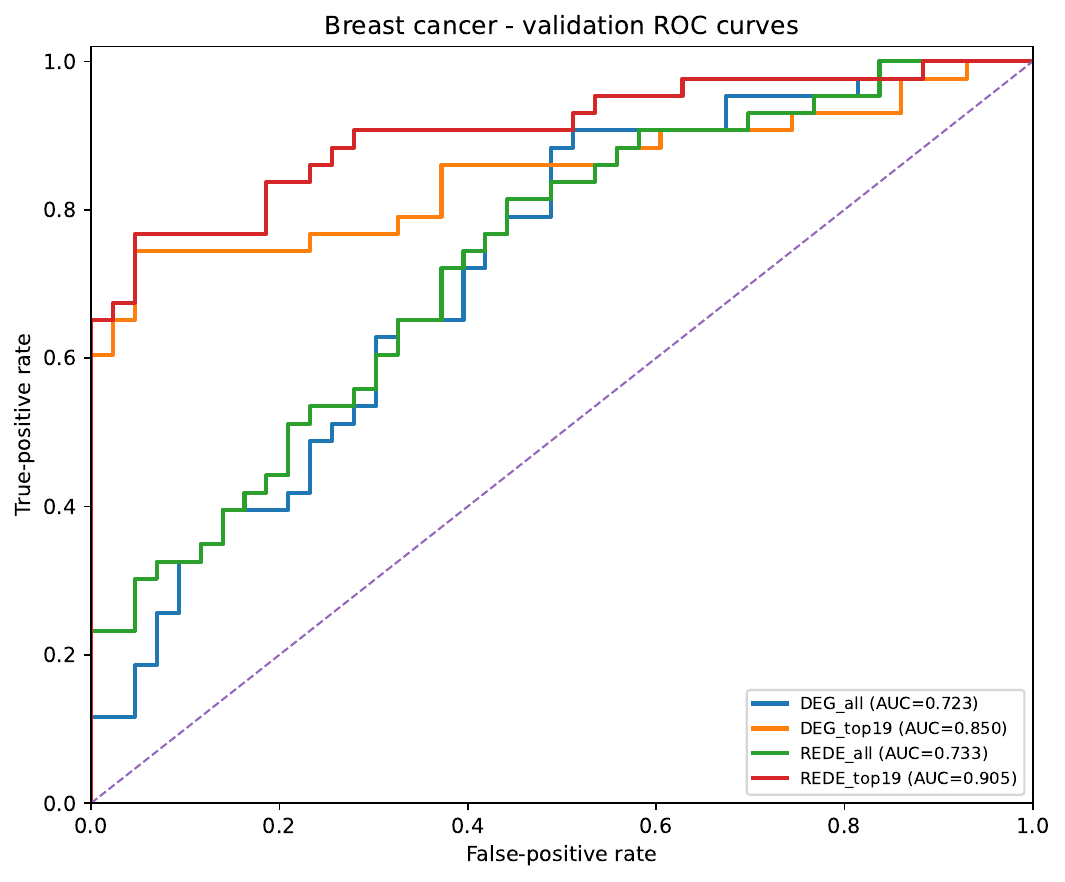}\caption{Breast validation}\end{subfigure}
\begin{subfigure}[t]{0.32\linewidth}\includegraphics[width=\textwidth]{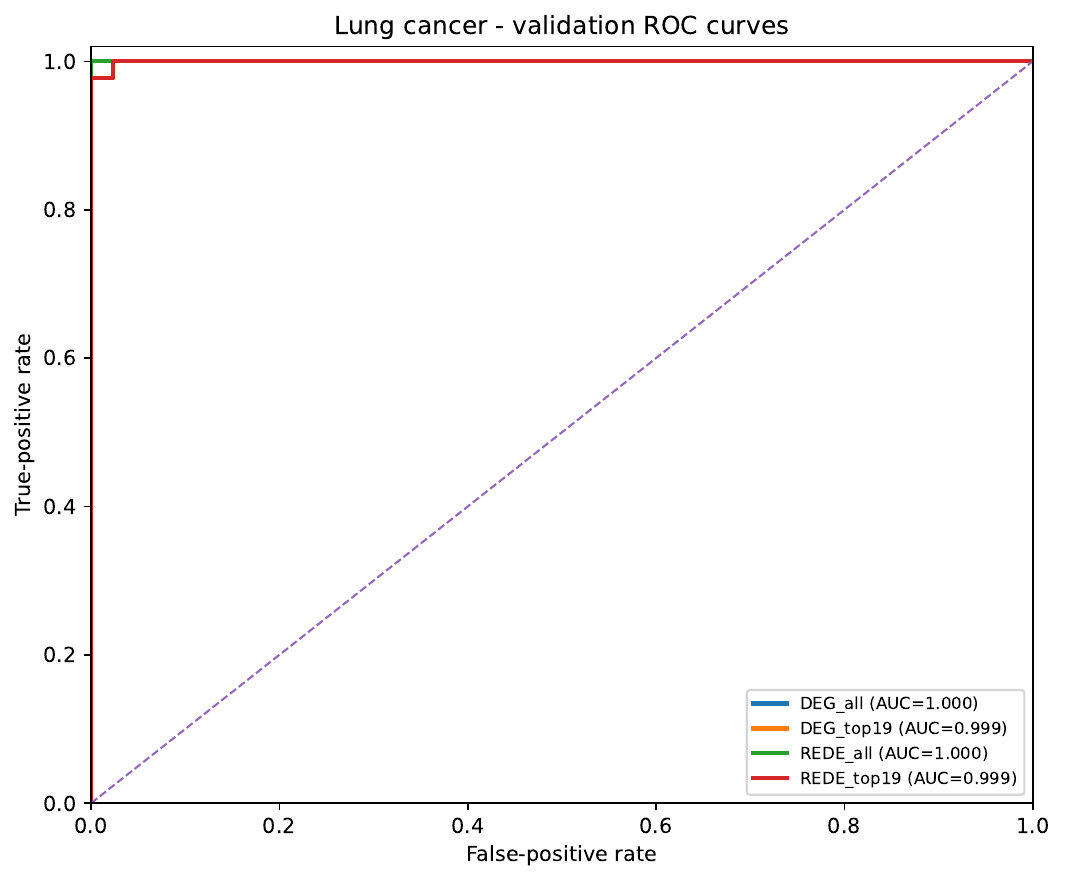}\caption{Lung validation}\end{subfigure}

\vspace{0.6em}
\begin{subfigure}[t]{0.32\linewidth}\includegraphics[width=\textwidth]{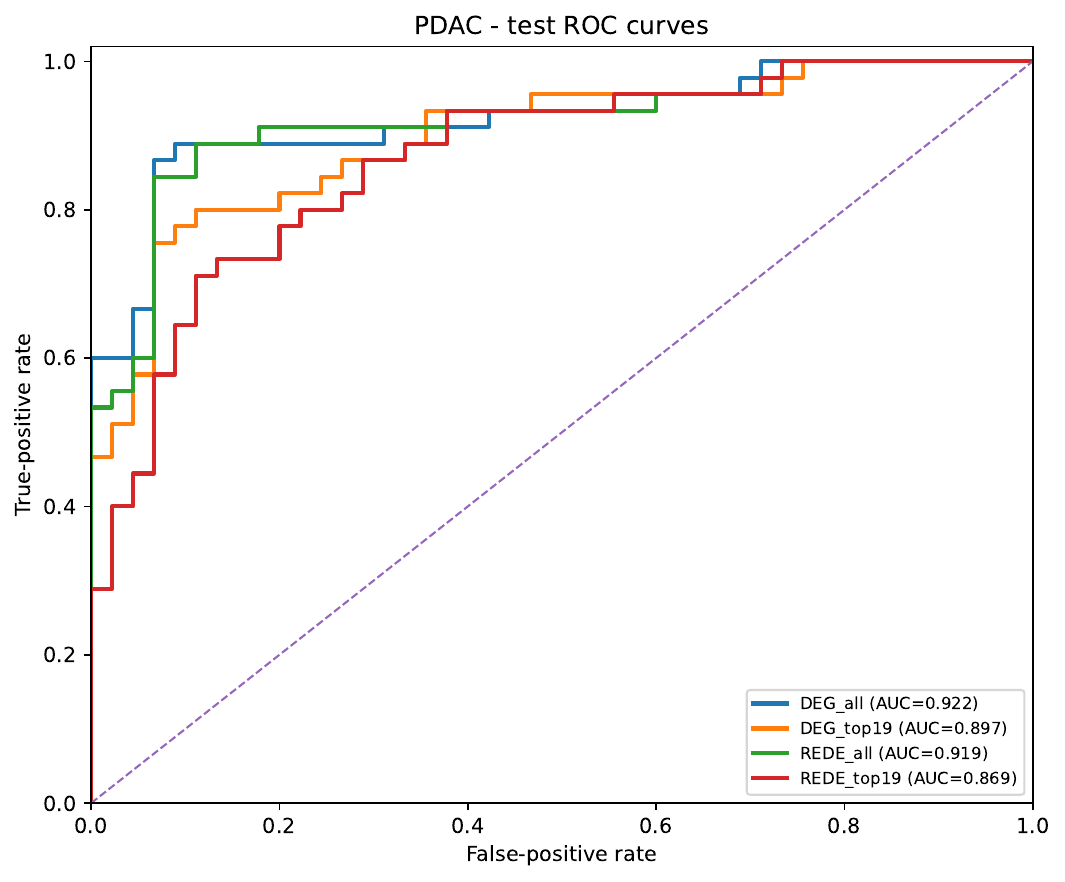}\caption{PDAC external test}\end{subfigure}
\begin{subfigure}[t]{0.32\linewidth}\includegraphics[width=\textwidth]{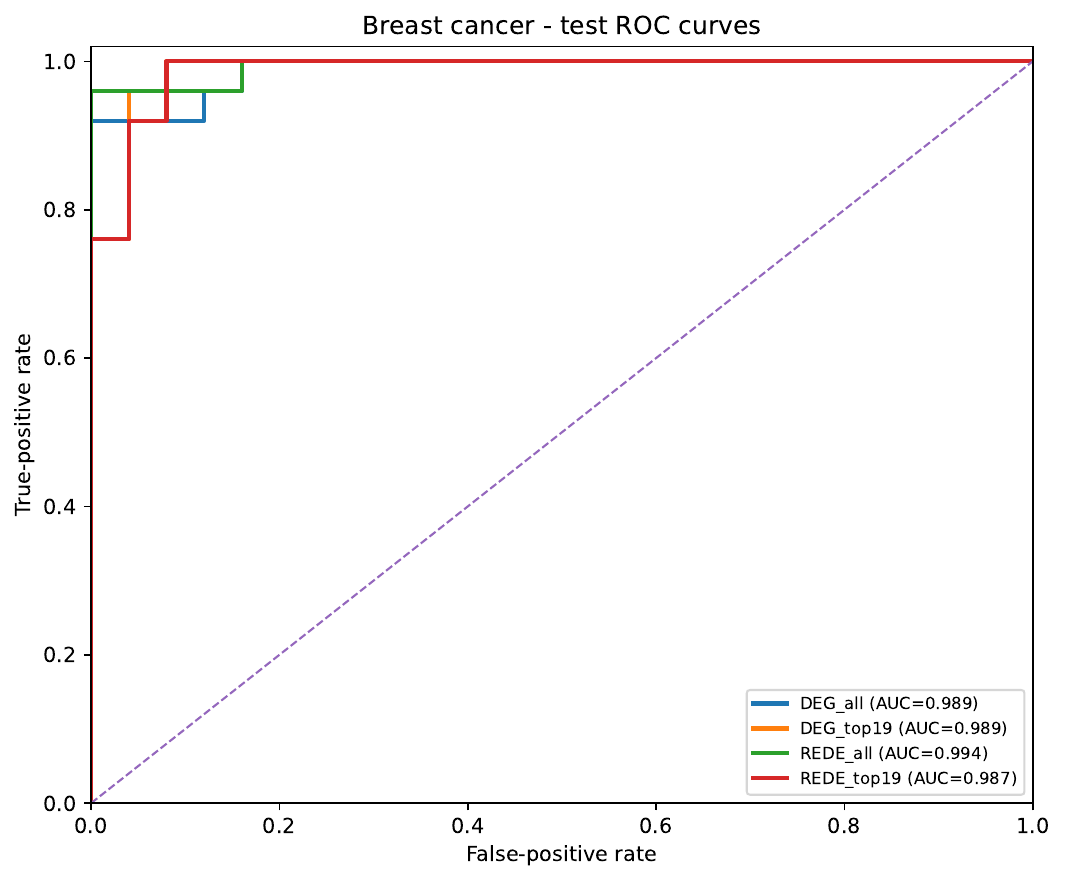}\caption{Breast external test}\end{subfigure}
\begin{subfigure}[t]{0.32\linewidth}\includegraphics[width=\textwidth]{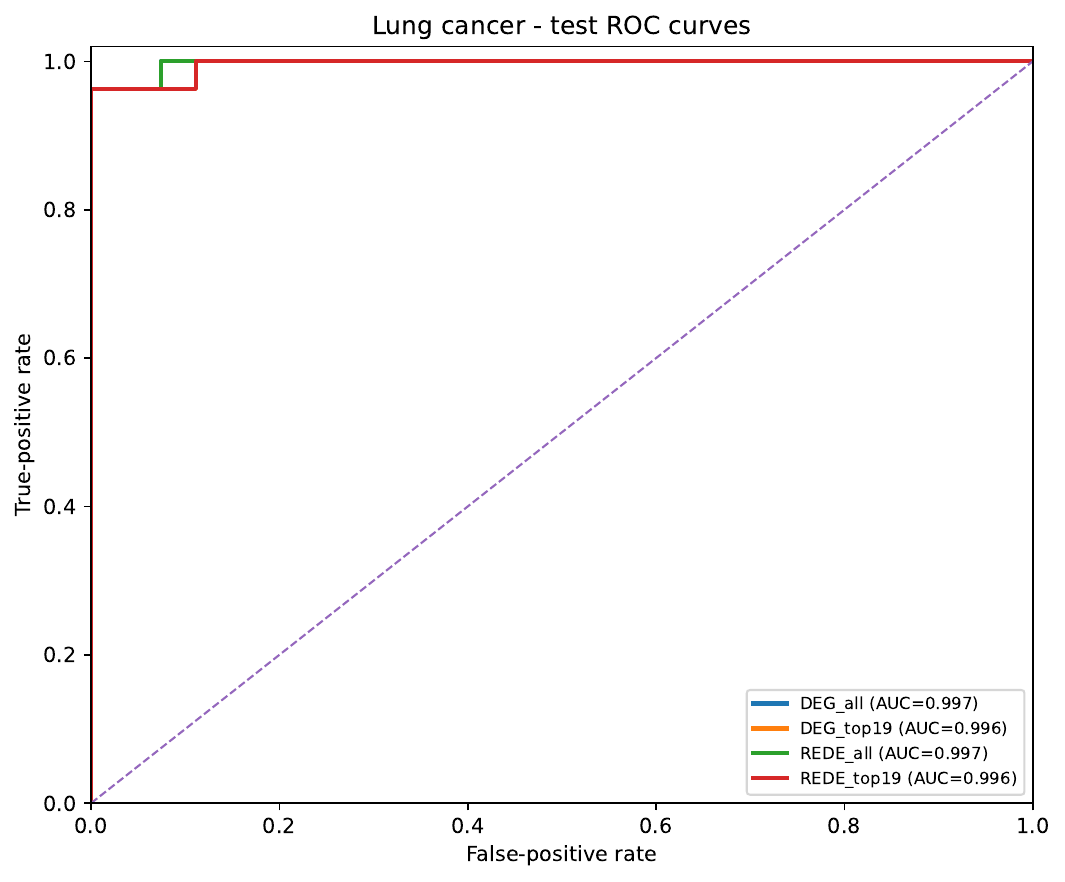}\caption{Lung external test}\end{subfigure}
\caption{\textbf{ROC curves from locked validation and external-test predictions.} Curves compare full discovery DEGs, conventional DE top 19, all REDE-2Fold genes, and REDE-2Fold top 19. ROC curves assess ranking discrimination and do not display the transported decision threshold.}
\label{fig:s_diagnostic_roc}
\end{figure}
\end{landscape}

\end{document}